\documentclass[structabstract]{aa}  
\usepackage{graphicx}
\usepackage{amsmath,amssymb}
\usepackage{longtable,lscape}
\usepackage{txfonts}
\begin{document}

\def\Arrow{\mathop{\longrightarrow}\limits}
\def\Harpoons{\mathop{\rightleftharpoons}\limits}

   \title{LABOCA 870 $\mu$m dust continuum mapping of selected infrared-dark 
cloud regions in the Galactic plane\thanks{This publication is based on data acquired with the Atacama Pathfinder EXperiment (APEX) under programme 087.F-9315(A,B). APEX is a collaboration between the Max-Planck-Institut f\"{u}r Radioastronomie, the European Southern Observatory, and the Onsala Space Observatory.}}

   \author{O. Miettinen}

 \offprints{O. Miettinen}

   \institute{Department of Physics, P.O. Box 64, FI-00014 University of Helsinki, Finland\\ \email{oskari.miettinen@helsinki.fi}}

   \date{Received ; accepted}

\authorrunning{Miettinen}
\titlerunning{LABOCA mapping of IRDCs}

  \abstract
   {Imaging surveys of dust continuum emission at (sub)millimetre wavelengths 
provide a powerful tool to study molecular clouds and the early stages of star 
formation.}
   {Through submm dust continuum mapping, we attempt to search for genuine 
infrared dark clouds (IRDCs) and precursors to massive stars and stellar 
clusters in the Galactic plane, and to determine their basic physical 
properties.}
   {We have mapped four selected about $0\fdg5 \times 0\fdg5$-sized fields 
containing \textit{Spitzer} 8-$\mu$m dark regions with APEX/LABOCA at 870 
$\mu$m. Selected positions in the fields were observed in C$^{17}$O$(2-1)$ 
to obtain kinematic information. The obtained LABOCA maps are used in 
conjunction with the \textit{Spitzer} IR images.} 
   {The total number of clumps identified in this survey is 91, out of which 
40 (44\%) appear dark at 8 and 24 $\mu$m. The remaining clumps are associated 
with mid-IR emission. Seven clumps associated with extended-like 4.5 $\mu$m 
emission are candidate extended green objects (EGOs). Filamentary dust 
``ridges'' were found towards the \textit{Spitzer} bubbles N10/11 in one of 
our fields. The relative number of IR-dark and IR-bright clumps suggest that 
the duration of the former stage is about $1.6\times10^5$ yr. The mass 
distribution of the total sample of clumps, and that separately constructed 
for the IR-dark and IR-bright clumps, could be fitted at the high-mass end 
with the power-law function ${\rm d}N/{\rm d}\log M \propto M^{-\Gamma}$, where 
$\Gamma \simeq 0.7\ldots 0.8$. The C$^{17}$O observation positions appear to 
be dominated by non-thermal motions, and the data also revealed some potential 
sites of strong CO depletion. In G11.36+0.80, which is the best example of a 
filamentary IRDC in our sample, the clumps appear to be 
gravitationally bound. The fragmentation of the filament can be understood in 
terms of a ``sausage''-type fluid instability, in agreement with the results 
for other IRDCs. The fragmentation and the CO depletion timescales in G11.36 
appear to be very similar to each other.}
   {Many of the identified clumps are massive enough to allow high-mass star 
formation, and some of them already show clear signposts of that. In the 
N10/11 bubble environment, the morphology of the detected dust emission 
conforms to the triggered high-mass star formation in the system. The clump 
mass distributions are similar to those found for diffuse CO clumps, and can 
be explained by the action of supersonic turbulence. The formation 
of filamentary IRDCs might be caused by converging turbulent flows, and the 
same process may play a role in exciting the fluid perturbations responsible 
for the fragmentation of the clouds into clumps.}


   \maketitle
%

\section{Introduction}

Understanding the origin of high-mass ($M>8$ M$_{\sun}$) stars, particularly 
the first steps in the formation process, is probably one of the greatest 
challenges of modern astrophysics. Since the discovery of the so-called 
infrared dark clouds, or IRDCs (\cite{perault1996}; \cite{egan1998}), 
ample evidence has been gathered concerning their important role in the 
earliest stages of Galactic high-mass star formation 
(e.g., \cite{rathborne2006}; \cite{beuther2007}; \cite{chambers2009}; 
\cite{battersby2010}; \cite{zhang2011}, and many other works). In particular, 
studies of IRDCs have the potential to help to understand the inital 
conditions of high-mass star and stellar cluster formation, which is 
necessary in order to constrain, or even distinguish, between different 
theoretical views\footnote{At least in some high-mass star-forming 
regions, however, other factors, such as interactions between the cluster 
members, stellar feedback, and external forces, may be more important for the 
source evolution than the initial conditions of the parent cloud.}.

Thermal dust continuum emission at far-infrared (FIR) and (sub)millimetre 
wavelengths provides a powerful observational tool to search and study the 
densest parts of IRDCs. Dust conti\-nuum imaging with bolometer cameras can 
be used to distinguish the real IRDCs, i.e., cold dense molecular clouds, 
from the mi\-nima in the Galactic mid-infrared (MIR) background radiation, 
which may look like candidate IRDCs (\cite{wilcock2012}). Optically thin dust 
emission also provides a probe of the basic phy\-sical properties of dense 
clouds, such as the column density of molecular hydrogen and the mass of 
the cloud. This information is needed to learn the physical conditions 
prevailing in the precursor regions of stellar clusters and high-mass stars.

In this paper, we present the results of our submm dust continuum observations 
at 870 $\mu$m of four selected regions in the Galactic plane, each of which 
contain IRDCs. Throughout the paper, we will use the term ``clump'' to refer 
to sources whose typical radii, masses, and mean densities are, respectively, 
$\sim0.2-1$ pc, $\sim10^2-10^3$ M$_{\sun}$, and $10^3-10^4$ cm$^{-3}$ 
(cf. \cite{bergin2007}). The rest of the present paper 
is orga\-nised as follows. Observations and data reduction are 
described in Sect.~2. Observational results are presented in Sect.~3. 
Analysis and its results are presented in Sect.~4, and in Sect.~5 we discuss 
the obtained results. In Sect.~6, we summarise the results and draw our main 
conclusions.

\section{Observations and data reduction}

\subsection{Archival data from the Spitzer Space Telescope}

In this study, we use the \textit{Spitzer} (\cite{werner2004}) IR
data taken as part of the GLIMPSE (\cite{benjamin2003}; \cite{churchwell2009}) 
and MIPSGAL (\cite{carey2009}) Galactic plane surveys. 
The former survey employed the IRAC instrument operating at 3.6, 4.5, 5.8, 
and 8.0 $\mu$m (\cite{fazio2004}), whereas the latter one used the MIPS 
instrument at 24 and 70 $\mu$m (\cite{rieke2004}). 
The angular resolution of the \textit{Spitzer}-IRAC 
instrument is $1\farcs9$ at 8 $\mu$m, and that of the MIPS instrument is 
$6\arcsec$ at 24 $\mu$m. We note that we have used the data provided by both 
the GLIMPSE I and II surveys, which covered the nominal Galactic longitude 
ranges of $10\degr \leq \vert l \vert \leq 65\degr$ and 
$\vert l \vert \leq 10\degr$, respectively. The data were retrieved from the 
\textit{Spitzer} science archive\footnote{{\tt http://irsa.ipac.caltech.edu/data/SPITZER/docs/ \\
spitzerdataarchives/}}.

The 8 $\mu$m images of the Galactic plane are particularly useful for the 
search of candidate IRDCs. The GLIMPSE 8 $\mu$m band contains the UV-excited 
7.7 and 8.6 $\mu$m PAH (polycyclic aromatic hydrocarbon) features (e.g., 
\cite{draine2003}), which together with emission from warm interstellar dust 
yield a bright MIR background. High columns of cold dust in IRDCs cause 
them to appear as dark absorption features against this background radiation 
field.

\subsection{LABOCA dust continuum mapping}

As a starting point of our study we visually inspected the 
\textit{Spitzer}-GLIMPSE 8-$\mu$m images of the Galactic plane, and chose 
four target fields containing filamentary IR-dark features to be mapped 
in the submm dust continuum emission. Because IRDCs often exhibit 
filamentary shapes and are relatively devoid of (visible) star formation, 
the target sources of this study are likely to represent rather typical IRDCs. 
The selected fields, which all belong to 
the first Galactic quadrant ($0\degr < l < 90 \degr$), are listed in 
Table~\ref{table:laboca}. These target fields were mapped 
with the Large APEX BOlometer CAmera (LABOCA; \cite{siringo2009}) on the
12-m Atacama Pathfinder EXperiment (APEX) telescope at Llano de Chajnantor in 
the Atacama desert of the Chilean Andes (\cite{gusten2006})\footnote{The 
observations presented here are \textit{not} part of the APEX Telescope Large 
Area Survey of the GALaxy (ATLASGAL) conducted with the LABOCA array 
(\cite{schuller2009}; {\tt http://www.mpifr.de/div/atlasgal/index.html}). 
However, the ATLASGAL survey covers the same regions we have mapped. At the 
time of writing, the ATLASGAL data were not yet released into the public 
domain. We note that ATLASGAL has the average $1\sigma$ rms of $\sim50$ 
mJy~beam$^{-1}$ and the resolution of $\sim19\farcs2$.}.
The LABOCA instrument is a multi-channel bolometer array, where 295 
semiconducting composite bolometers are arranged in a series of 
nine concentric hexagons around a central channel. The system operates at a 
central frequency of 345 GHz ($\lambda=870$ $\mu$m) with a bandwidth of about 
60 GHz to match the corresponding atmospheric window. The nominal angular 
resolution of LABOCA is $19\farcs2 \pm 0\farcs3$ (half power beamwidth; HPBW), 
and its total field of view is $11\farcm4$ (about 0.09 pc and 3.3 pc at 1 kpc, 
respectively). 

Our LABOCA observations took place on 19 May 2011, during the UTC time ranges 
of 03:17--06:20 and 07:54--10:50. The observing conditions were very good: 
the atmospheric zenith opacity, as determined using skydip measurements, was
in the range $\tau_{\rm z}=0.10-0.12$, and the amount of precipitable water 
vapour (PWV) was in the range 0.15--0.30 mm. The telescope focus and pointing 
were, respectively, optimised and checked at regular intervals on the planets 
Saturn and Neptune, the Class 0 protostellar core IRAS 16293-2422, the 
ultracompact (UC) H{\scriptsize II} region G10.62−0.38, and the 
massive young stellar object (MYSO) G305.80-0.24 (B13134). The absolute 
calibration uncertainty is estimated to be about 10\%. 

The observations were performed using the on-the-fly (OTF) mapping mode, 
in which the telescope scanned continuously in right ascension (RA) along 
each row. We used a scanning speed of $3\arcmin$ s$^{-1}$ and the step size 
$6\farcs5$ ($\sim1/3$ the beam HPBW) between RA subscans. The step size 
$\lesssim1/3\times$ the beam HPBW is re\-commended to avoid beam broadening. 
The angular sizes of the maps are given in Col.~(2) of Table~\ref{table:laboca} 
(sizes are in the range of $\sim0.23-0.30$ deg$^2$, with a total angular area 
of about 1 deg$^2$). The target fields were mapped three to five times, with 
total on-source integration times in the range 49--79 min [Col.~(4) of 
Table~\ref{table:laboca}]. 

Data reduction was done with the CRUSH-2 (Comprehensive Reduction Utility for 
SHARC-2) (version 2.11-a1) software package\footnote{{\tt 
http://www.submm.caltech.edu/$\sim$sharc/crush/index.htm}} (\cite{kovacs2008}).
We used the pipeline iterations both with the default reduction parameters, 
and also with specifying the 'extended' option, which better preserves 
the extended structures. In the cases of G1.87-0.14, G2.11+0.00, and 
G13.22-0.06 (hereafter, G1.87, etc.), the 'extended'-reduced maps were finally 
chosen for the ana\-lysis because fainter extended structures were clearly 
better recovered compared to the default reduction method. In the case of 
G11.36, however, we adopted the map reduced with the default 
parameters because the clumpy structure of the 
filament became more clearly visible (Fig.~\ref{figure:G1136}). A slight 
beam-smoothing was applied in the reduction process, i.e., the maps were 
smoothed with a Gaussian kernel of the size $3\farcs8$ (full width at half 
maximum; FWHM). The instrument beam HPBW used by CRUSH-2 was $19\farcs5$, and 
therefore the angular resolution of the final maps is $19\farcs9$ 
($\sim0.1$ pc at 1 kpc). The gridding was done with 
a cell size of $4\arcsec$. The resulting $1\sigma$ rms noise levels in the 
final co-added maps are $\sim40-90$ mJy~beam$^{-1}$ [Col.~(5) of 
Table~\ref{table:laboca}]. Assuming that the 870-$\mu$m dust opacity and the 
dust temperature are, respectively, 1.38 cm$^2$~g$^{-1}$ and 15 K 
(see Sect.~4), the above surface-brightness sensitivity levels translate into 
$1\sigma$ H$_2$ column-density detection thresholds of 
$N({\rm H_2})\simeq2.0-4.4\times10^{21}$ cm$^{-2}$. These correspond to visual 
extinction values of $A_{\rm V}=N({\rm H_2})/0.94\times10^{21}\simeq2.1-4.7$ 
mag (\cite{bohlin1978})\footnote{This $A_{\rm V}-N({\rm H_2})$ relationship 
is based on observations of diffuse interstellar medium, and therefore may 
not be exactly correct for dense molecular clouds.}. We note that 
employing the 'extended' option in the reduction process leads to maps with 
a higher noise level than the values of $\sim30-50$ mJy~beam$^{-1}$ 
resulting from the standard procedure (because large scale emission is tried 
to be preserved). Therefore, the noise in the map of G11.36, which was reduced 
in the standard way, is clearly lower than in the other cases.

\begin{table*}
\caption{Target fields mapped with LABOCA.}
\begin{minipage}{2\columnwidth}
\centering
\renewcommand{\footnoterule}{}
\label{table:laboca}
\begin{tabular}{c c c c c}
\hline\hline 
Field\tablefootmark{a} & Map size & No. of maps & On-source time & $1\sigma$ rms noise\tablefootmark{b}\\
      & [$\arcmin \times \arcmin$] & & [min] & [mJy~beam$^{-1}$] \\
\hline
G1.87-0.14 & $30.2\times 27.9$ & 3 & 50.7 & 80 -- 90 (90)\\
G2.11+0.00 & $31.3\times 26.7$ & 5 & 79.2 & 60\\
G11.36+0.80 & $30.1\times 27.7$ & 3\tablefootmark{c} & 48.9 & 40 -- 50 (40) \\
G13.22-0.06 & $33.1\times 32.6$ & 3 & 76.0 & 60 -- 90 (80)\\
\hline 
\end{tabular} 
\tablefoot{\tablefoottext{a}{The fields are named here after their approximate 
central Galactic coordinates ($l,\,b$).}\tablefoottext{b}{These rms noise 
values refer to the maps used in the analysis (see text). The value given in 
parenthesis is the $1\sigma$ noise adopted in the {\tt clumpfind} 
analysis.}\tablefoottext{c}{The observations were interrupted during the 
fourth mapping of this field, and therefore only the first three maps could 
be used.}}
\end{minipage} 
\end{table*}

\subsection{C$^{17}$O$(2-1)$ line observations}

From each target field, we selected seven to eight positions for 
single-pointing C$^{17}$O$(2-1)$ observations. These positions, which are 
listed in Table~\ref{table:targets}, were chosen from the \textit{Spitzer} 
8-$\mu$m images, and they correspond to (apparently) highly extincted parts 
along the filamentary structures near the map centres. The main purpose of 
making these line observations was to obtain the cloud radial velocity, which 
is needed to determine the cloud kinematic distance (Sect.~4.1).

The C$^{17}$O$(2-1)$ observations at 224\,714.199 MHz were carried out on 
18, 22, and 26 May 2011 with APEX using the Swedish Heterodyne Facility 
Instrument (SHeFI; \cite{belitsky2007}; \cite{vassilev2008a}) [the 
heterodyne-part of the project 087.F-9315(A,B)]. As a frontend we used the 
APEX-1 receiver of the SHeFI (\cite{vassilev2008b}). The backend was the 
Fast Fourier Transfrom Spectrometer (FFTS; \cite{klein2006}) with a 1 GHz 
bandwidth divided into 8\,192 channels. The resulting channel spacing is 
122 kHz or 0.16 km~s$^{-1}$. The telescope beam size (HPBW) at the observing 
frequency is $27\farcs8$. 

The observations were performed in the wobbler-switching mode with a 
$150\arcsec$ azimuthal throw (symmetric offsets) and a chopping rate of 0.5 Hz 
(2 s wobbler period). Total (on+off) integration time was 5.6 min per position.
The telescope pointing accuracy was checked by CO$(2-1)$ cross maps of the 
carbon star RAFGL1922, and was found to be below $\lesssim4\arcsec$. The 
focus was checked by measurements on Saturn. Calibration was done by means of 
the chopper-wheel technique, and the output intensity scale given by the system 
is $T_{\rm A}^*$, which represents the antenna temperature corrected for the 
atmospheric attenuation. The observed intensities were converted to the 
main-beam brightness temperature scale by $T_{\rm MB}=T_{\rm A}^*/\eta_{\rm MB}$, 
where $\eta_{\rm MB}=0.75$ is the main-beam efficiency at the frequency used. 
The single-sideband system temperature, in units of $T_{\rm MB}$, was in the 
range 387 -- 415 K. The absolute calibration uncertainty is estimated to be 
around 10\%.

The spectra were reduced using the Fortran 90 version of the CLASS programme 
from the GILDAS software 
package\footnote{Grenoble Image and Line Data Analysis Software is 
provided and actively developed by IRAM, and is available at 
{\tt http://www.iram.fr/IRAMFR/GILDAS}}. 
The individual spectra were averaged and the resulting spectra were 
Hanning-smoothed in order to improve the signal-to-noise ratio of the 
data. A first- or third-order polynomial was applied to correct the baseline 
in the final spectra. The resulting $1\sigma$ rms noise levels are 
$\sim77-91$ mK at the smoothed resolution (4\,095 channels). 

We note that the $^{17}$O nucleus has a nuclear spin of $I=5/2$, so it has an 
electric quadrupole moment ($-2.6\times10^{-26}$ cm$^2$). The latter couples 
to the electric-field gradient at the nucleus. This causes the rotational 
lines of C$^{17}$O to have a hyperfine structure. The C$^{17}$O$(2-1)$ line 
is split into nine hyperfine (hf) components, which cover a velocity range of 
about 2.36 km~s$^{-1}$. We fitted this hf structure using ``method hfs'' of 
CLASS90 to derive the LSR velocity (${\rm v}_{\rm LSR}$) of 
the emission, and FWHM linewidth ($\Delta {\rm v}$). The hf-line fitting can 
also be used to derive the line optical thickness, $\tau$. However, in all 
spectra the hf components are blended together, thus the optical 
thickness could not be reliably determined. The rest frequencies and 
relative weights of the hf components were taken from Ladd et al. (1998; 
Table 6 therein).

\begin{table}
\renewcommand{\footnoterule}{}
\caption{Target positions of the C$^{17}$O$(2-1)$ observations in the 
equatorial J2000.0 system.}
\begin{minipage}{1\columnwidth}
\centering
\label{table:targets}
\begin{tabular}{c c c}
\hline\hline 
Field/ & $\alpha_{2000.0}$ & $\delta_{2000.0}$ \\
position & [h:m:s] & [$\degr$:$\arcmin$:$\arcsec$]\\
\hline
{\bf G1.87-0.14} & \\
A \ldots & 17 50 29.6 & -27 26 02 \\
B \ldots & 17 50 31.7 & -27 25 24 \\
C \ldots & 17 50 35.3 & -27 25 09 \\
D \ldots & 17 50 37.2 & -27 24 08 \\
E \ldots & 17 50 37.4 & -27 24 36 \\
F \ldots & 17 50 38.0 & -27 23 02 \\
G \ldots & 17 50 38.0 & -27 23 41 \\
{\bf G2.11+0.00} & \\
A \ldots & 17 50 30.0 & -27 08 25 \\
B \ldots & 17 50 30.1 & -27 07 53 \\
C \ldots & 17 50 30.3 & -27 06 37 \\
D \ldots & 17 50 30.7 & -27 07 17 \\
E \ldots & 17 50 35.6 & -27 07 15 \\
F \ldots & 17 50 37.2 & -27 07 12 \\
G \ldots & 17 50 38.4 & -27 07 02 \\
H \ldots & 17 50 38.7 & -27 06 42 \\
{\bf G11.36+0.80} & \\
A \ldots & 18 07 35.0 & -18 43 51 \\
B \ldots & 18 07 35.7 & -18 42 34 \\
C \ldots & 18 07 35.8 & -18 43 23 \\
D \ldots & 18 07 36.4 & -18 44 04 \\
E \ldots & 18 07 36.8 & -18 41 17 \\
F \ldots & 18 07 39.6 & -18 42 14 \\
G \ldots & 18 07 40.5 & -18 43 16 \\
{\bf G13.22-0.06} & \\
A \ldots & 18 14 28.2 & -17 33 28 \\
B \ldots & 18 14 31.6 & -17 32 44 \\
C \ldots & 18 14 35.8 & -17 30 51 \\
D \ldots & 18 14 36.0 & -17 26 55 \\
E \ldots & 18 14 36.7 & -17 29 17 \\
F \ldots & 18 14 40.7 & -17 29 06 \\
G \ldots & 18 14 42.8 & -17 30 06 \\
\hline 
\end{tabular} 
\end{minipage} 
\end{table}

\section{Observational results}

\subsection{LABOCA 870-$\mu$m maps}

\begin{figure*}
\begin{center}
\includegraphics[scale=0.55]{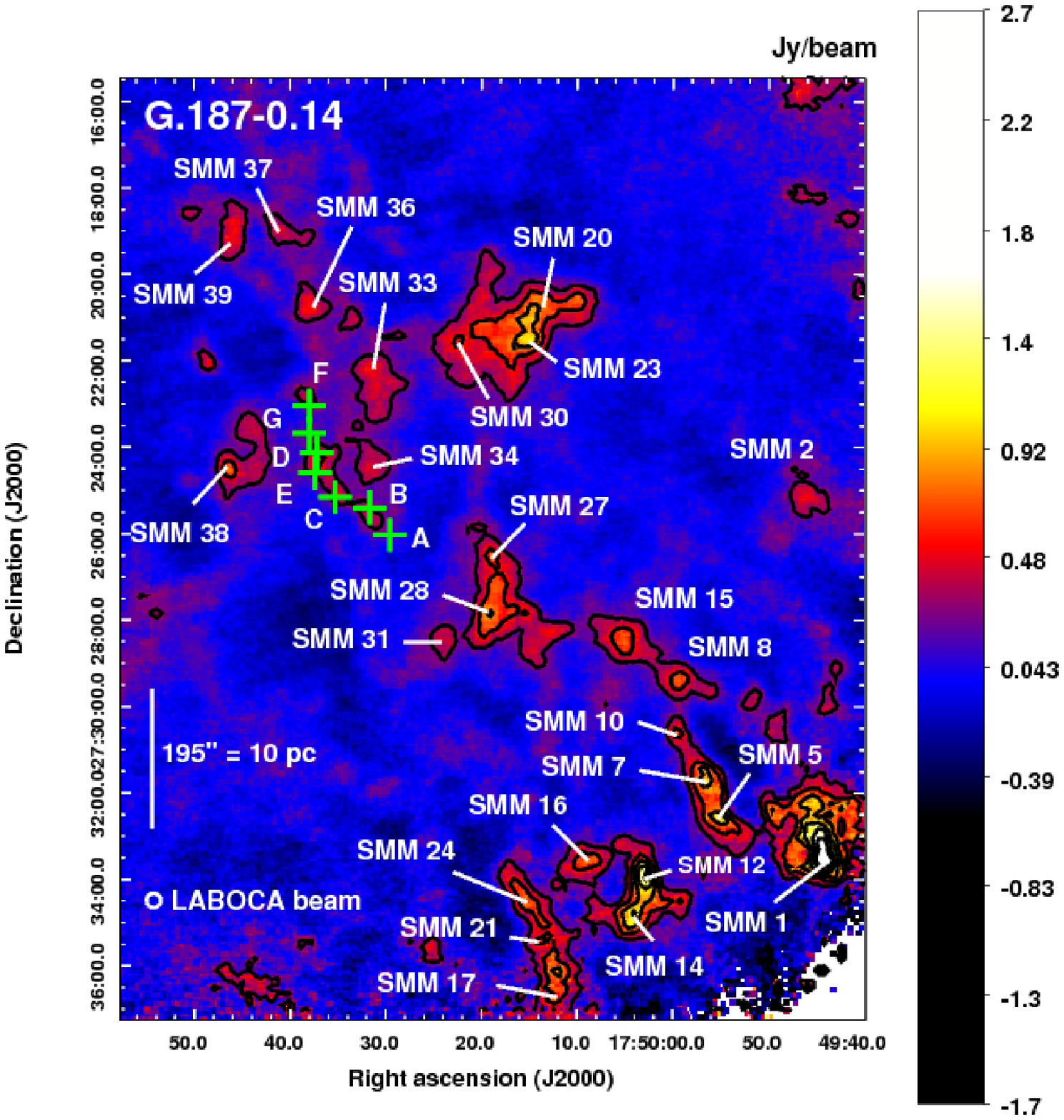}
\includegraphics[scale=0.53]{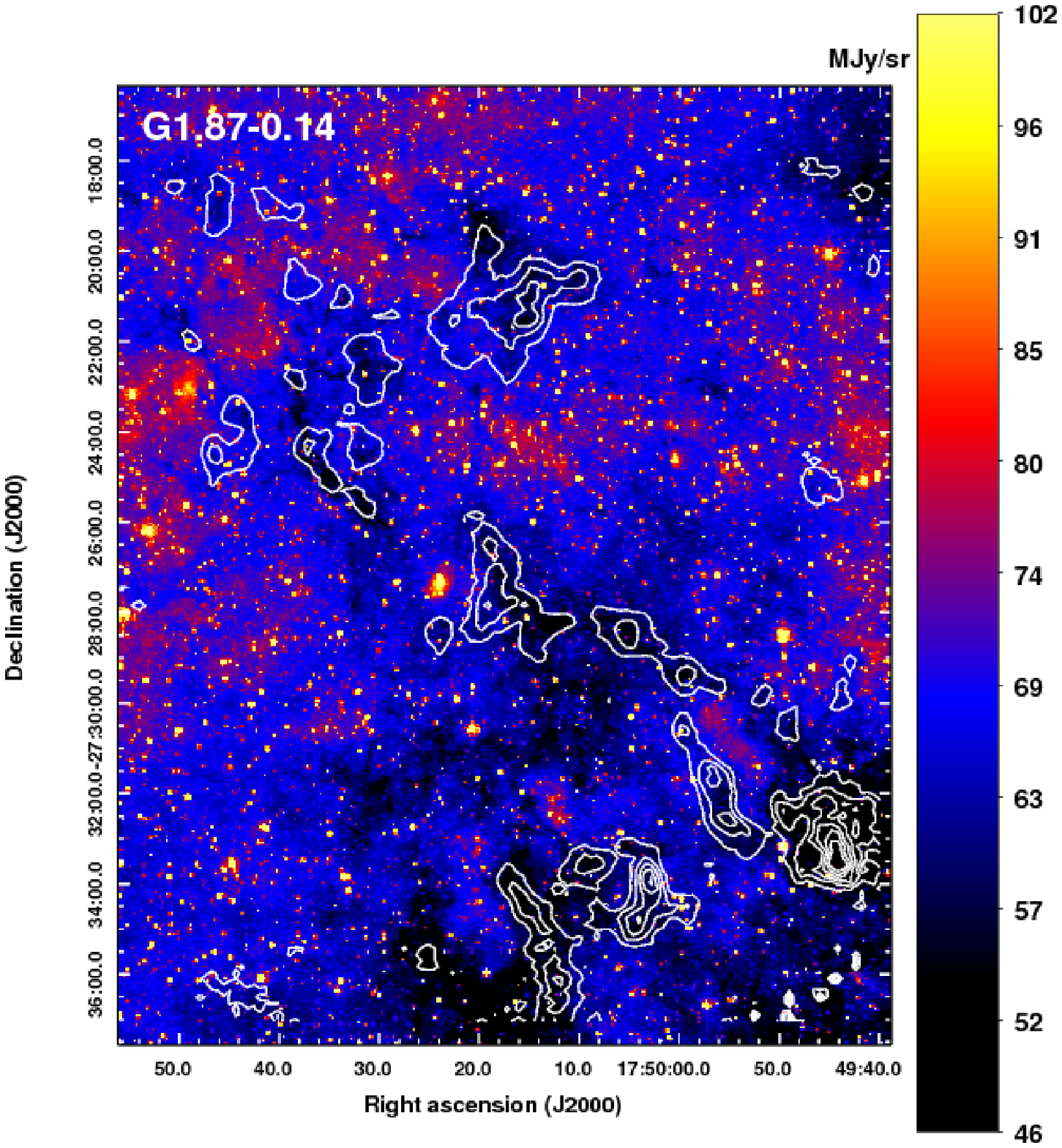}
\caption{\textbf{Left:} LABOCA 870-$\mu$m map of G1.87-0.14. The image is shown 
with linear scaling, and the colour bar indicates the surface-brightness scale 
in Jy~beam$^{-1}$. The overlaid contours go from 0.27 Jy~beam$^{-1}$ 
($3\sigma$) to 1.62 Jy~beam$^{-1}$, in steps of $3\sigma$. Selected clumps are 
labeled with their designation (as listed in Table~\ref{clumps}). The green 
plus signs indicate the positions of our C$^{17}$O$(2-1)$ observations (see 
Table~\ref{table:targets}). A scale bar indicating the 10 pc projected length 
is shown in the bottom left, with the assumption of a 10.57 kpc line-of-sight 
distance. The effective LABOCA beam of $19\farcs9$ is also shown in the lower 
left corner. \textbf{Right:} \textit{Spitzer} 8-$\mu$m image towards 
G1.87-0.14 overlaid with LABOCA contours from the left panel. The image is 
shown with linear scaling, where the scale limits are based on the IRAF 
z-scale algorithm of the DS9 programme. The colour bar shows the 
surface-brightness scale in MJy~sterad$^{-1}$. Note that not all IR-dark 
regions are seen in submm emission.}
\label{figure:G187}
\end{center}
\end{figure*}

\begin{figure*}
\begin{center}
\includegraphics[scale=0.481]{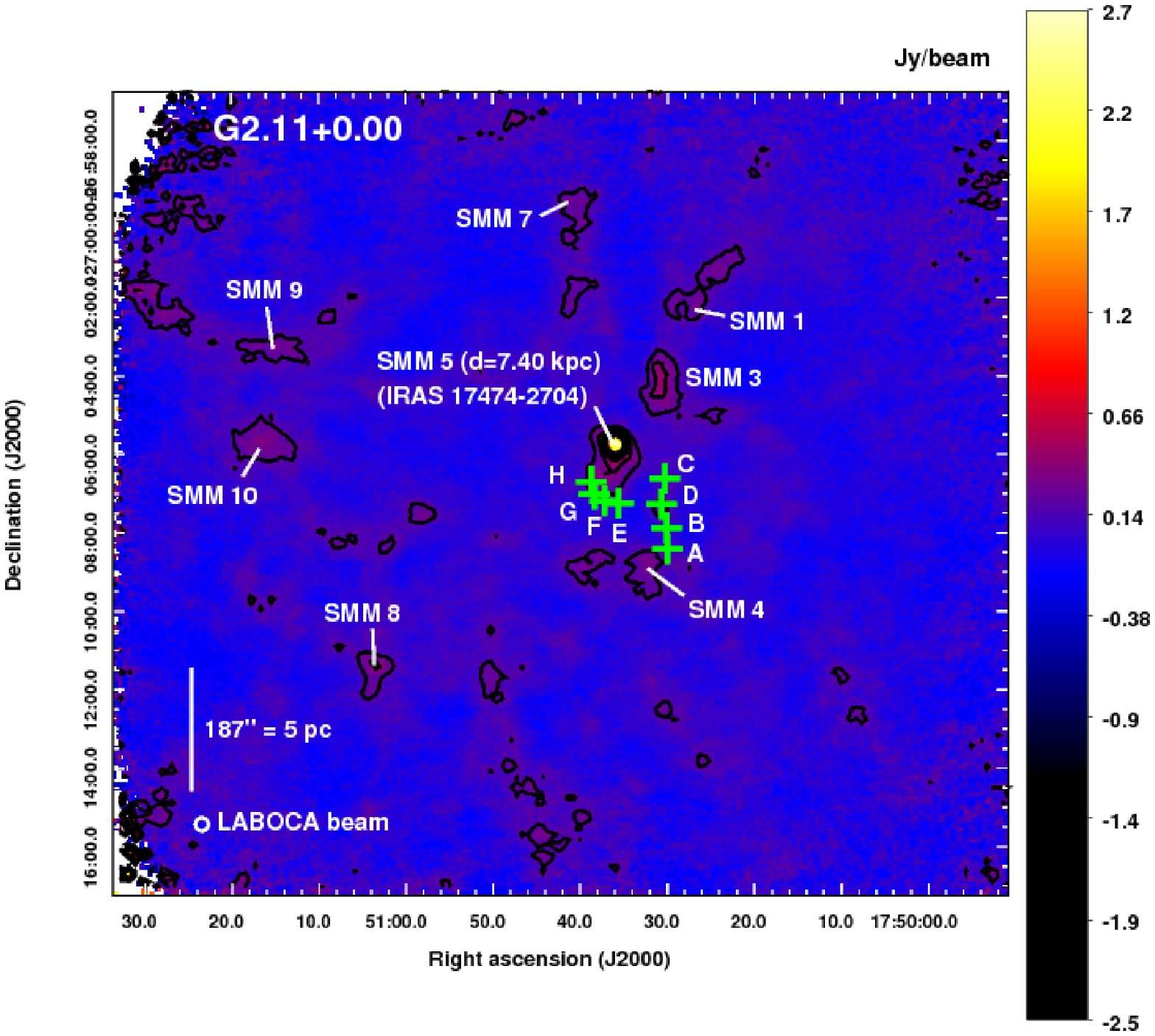}
\includegraphics[scale=0.497]{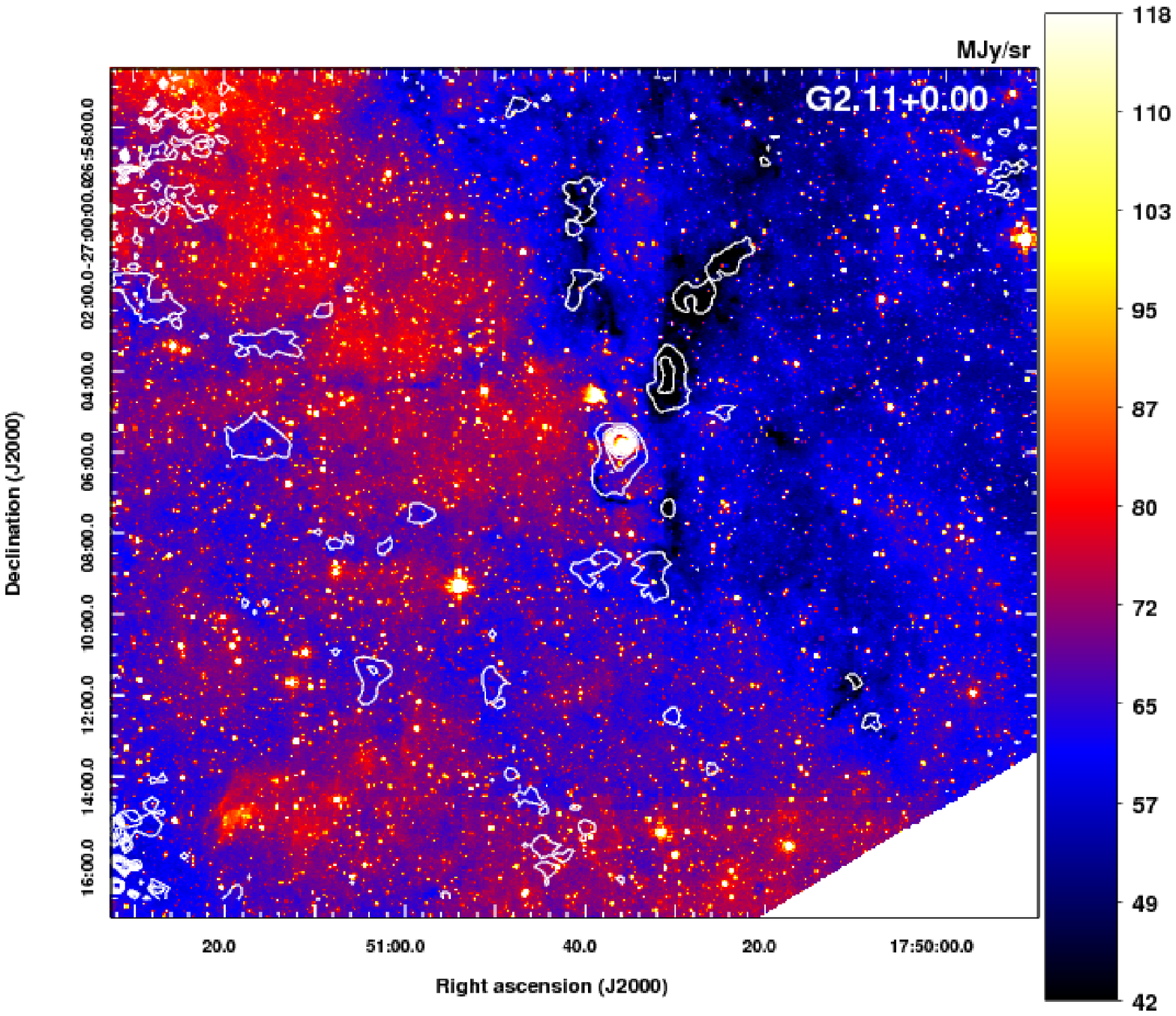}
\caption{Same as Fig.~\ref{figure:G187} but towards G2.11+0.00. The overlaid 
LABOCA contours go from 0.18 Jy~beam$^{-1}$ ($3\sigma$) to 1.08 Jy~beam$^{-1}$, 
in steps of $3\sigma$. A scale bar indicating the 5 pc projected length 
is shown in the bottom left, with the assumption of a 5.51 kpc line-of-sight 
distance. The clump SMM 5, which is associated with IRAS 17474-2704, lies at 
a distance of 7.40 kpc (see text).}
\label{figure:G211}
\end{center}
\end{figure*}

\begin{figure*}
\begin{center}
\includegraphics[scale=0.53]{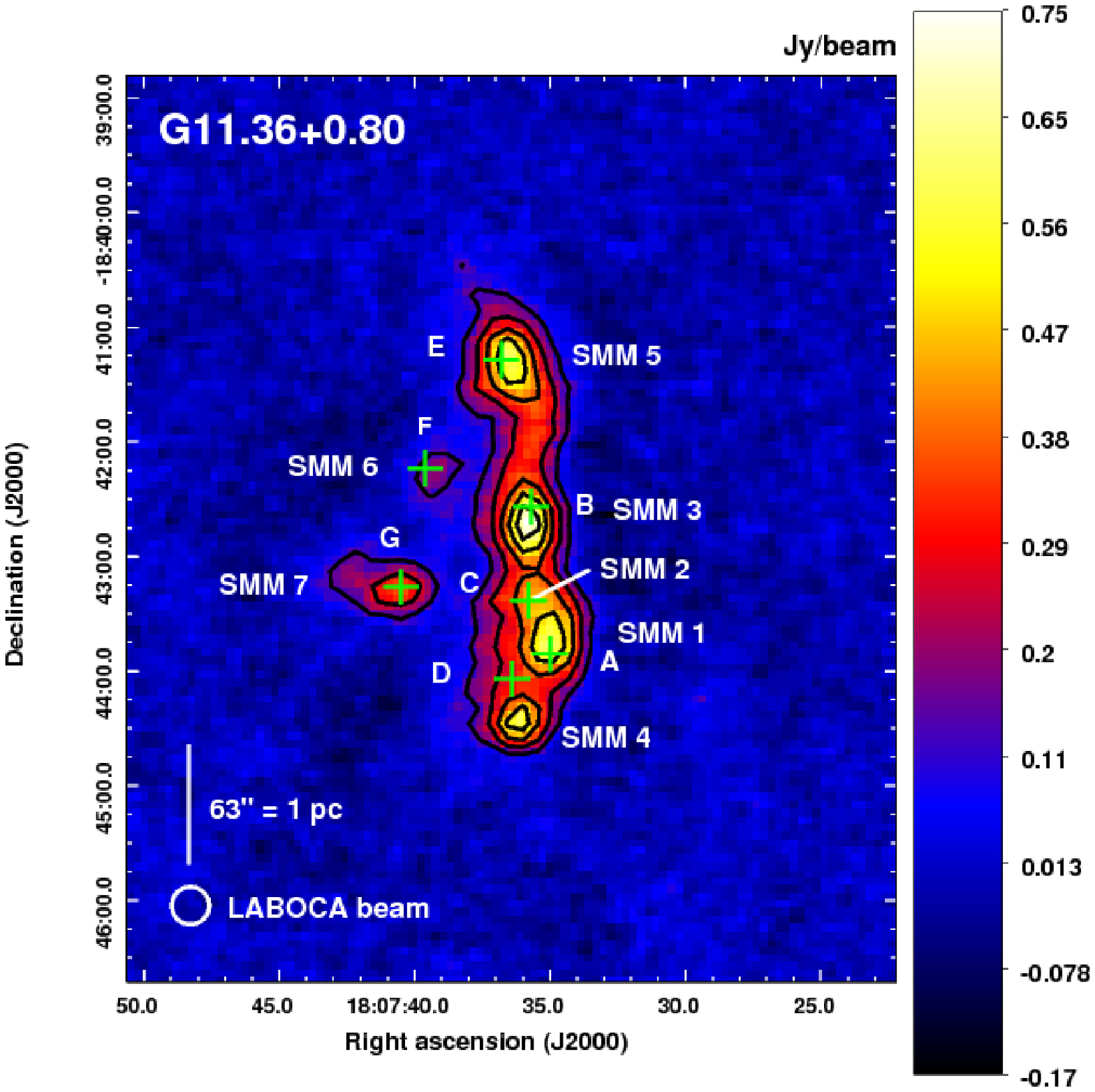}
\includegraphics[scale=0.6]{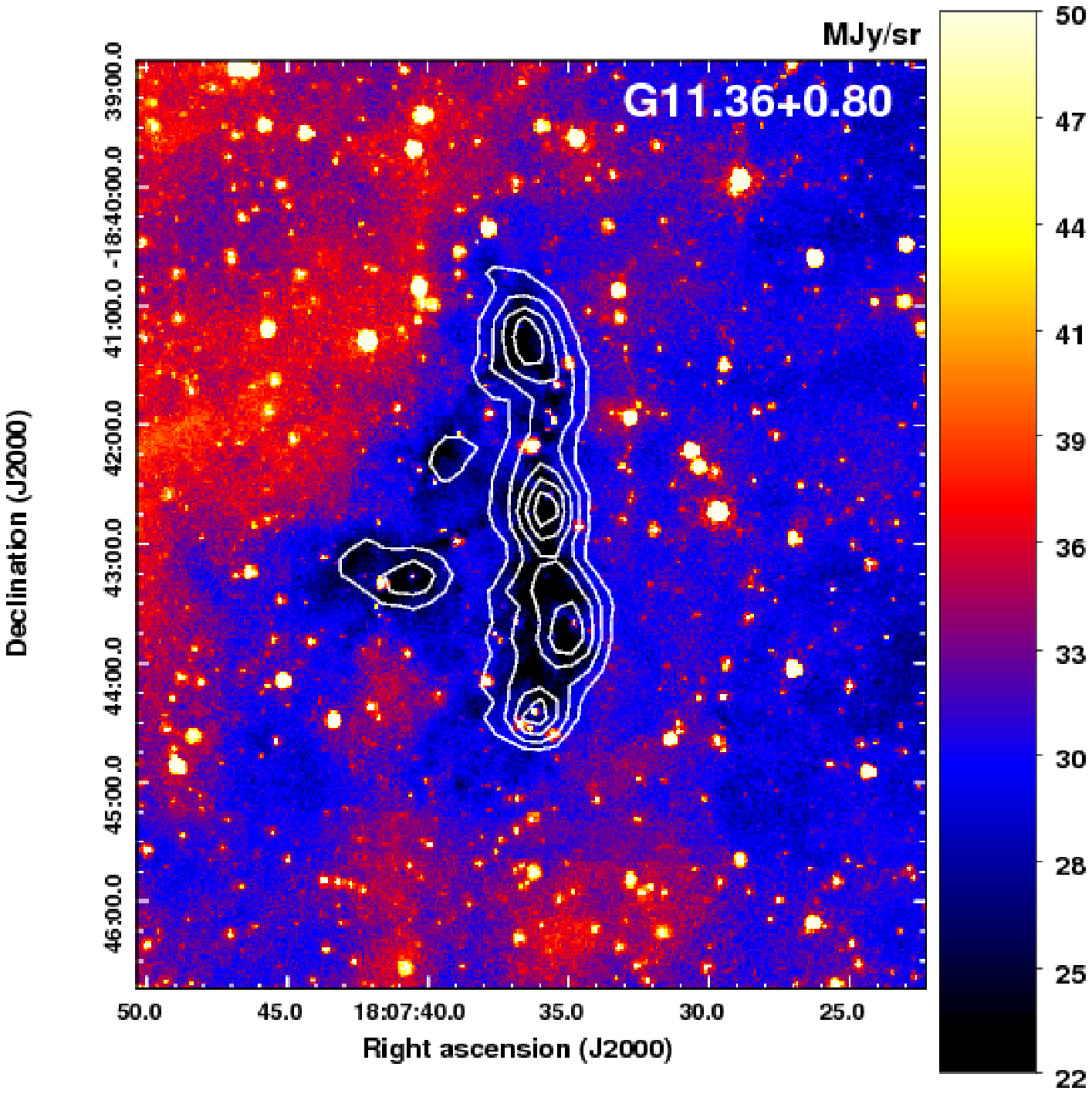}
\caption{Same as Fig.~\ref{figure:G187} but towards G11.36+0.80. The obtained 
LABOCA map is zoomed in towards the filamentary structure in the map centre as 
no submm dust emission was detected in other parts of the map. The overlaid 
LABOCA contours go from 0.12 Jy~beam$^{-1}$ ($3\sigma$) to 0.60 Jy~beam$^{-1}$, 
in steps of $3\sigma$. Note that most of the line-observation target positions 
are well matched with the submm peak positions. A scale bar indicating the 
1 pc projected length is shown in the bottom left, with the assumption of a 
3.27 kpc line-of-sight distance.}
\label{figure:G1136}
\end{center}
\end{figure*}

The obtained LABOCA maps are shown in 
Figs.~\ref{figure:G187}--\ref{figure:G1322}. In the right panel of each 
figure, we show the \textit{Spitzer} 8-$\mu$m image of the target field, 
overlaid with contours of the LABOCA submm dust emission. 

As can be seen from the maps, the fields contain filamentary structures and 
clumps of different projected shapes. It can also be seen, especially towards 
the G1.87 and G2.11 fields, that not all 8-$\mu$m dark features are seen in 
submm emission. These may be structures with too low a column density to be 
detected with the sensitivity limit of our data. On the other hand, as was 
pointed out by Wilcock et al. (2012), some of the 8-$\mu$m dark 
regions are not real dense clouds; they may just be dips in the MIR background 
that resemble the appearance of IRDCs. 

We note that in the 0.23 degr$^2$-sized G11.36 field, the filament in the map 
centre appeared to be the only submm-emitting object. Therefore, 
Fig.~\ref{figure:G1136} shows only the zoomed-in view towards the 
filament.

\subsection{Clump identification}

To systematically identify the submm clumps from the LABOCA maps, we employed 
the commonly used two-dimensional clumpfind algorithm, {\tt clfind2d}, 
developed by Williams et al. (1994). The algorithm requires two configuration 
parameters: \textit{i)} the intensity threshold, i.e., the lowest contour 
level, which determines the minimum emission to be included into the clump; 
and \textit{ii)} the contour level spacing, which determines the required 
``contrast'' between two clumps to be considered as different objects. We set 
both of these parameters to the classical value of $3\sigma$ 
($\sim120-270$ mJy~beam$^{-1}$), where the adopted $1\sigma$ sensitivity 
levels are given in Col.~(5) of Table~\ref{table:laboca} (the value in 
parenthesis when the noise level varies across the map). Only clumps with peak 
flux densities greater than $\sim 5\sigma$ were taken to be real. With these 
definitions, the number of clumps found by {\tt clfind2d} are 40, 10, 7, and 
34 in G1.87, G2.11, 11.36, and G13.22, respectively. This makes the total 
number of identified clumps in this survey to be 91. 

\begin{figure*}
\begin{center}
\includegraphics[scale=0.7]{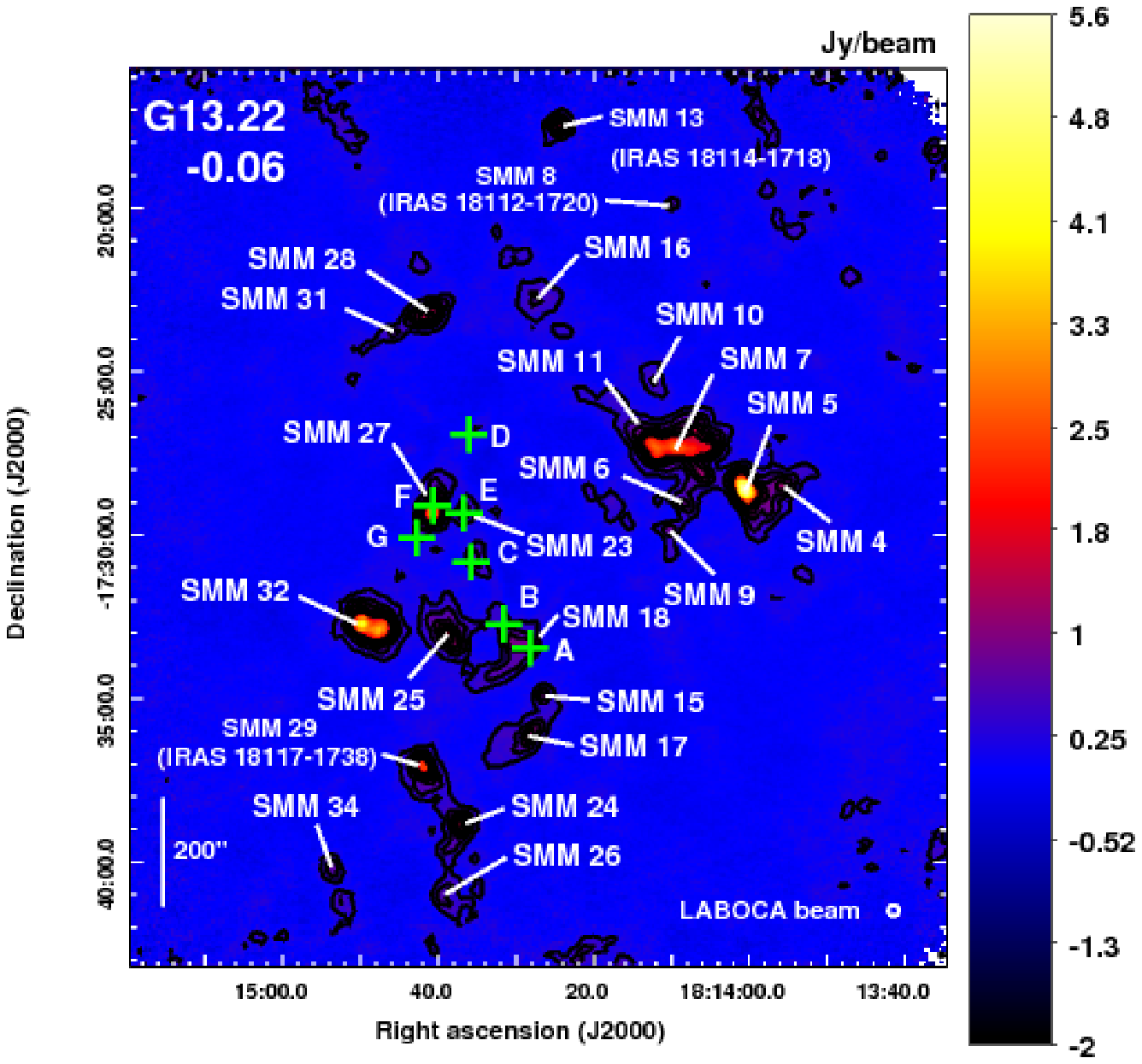}
\includegraphics[scale=0.5]{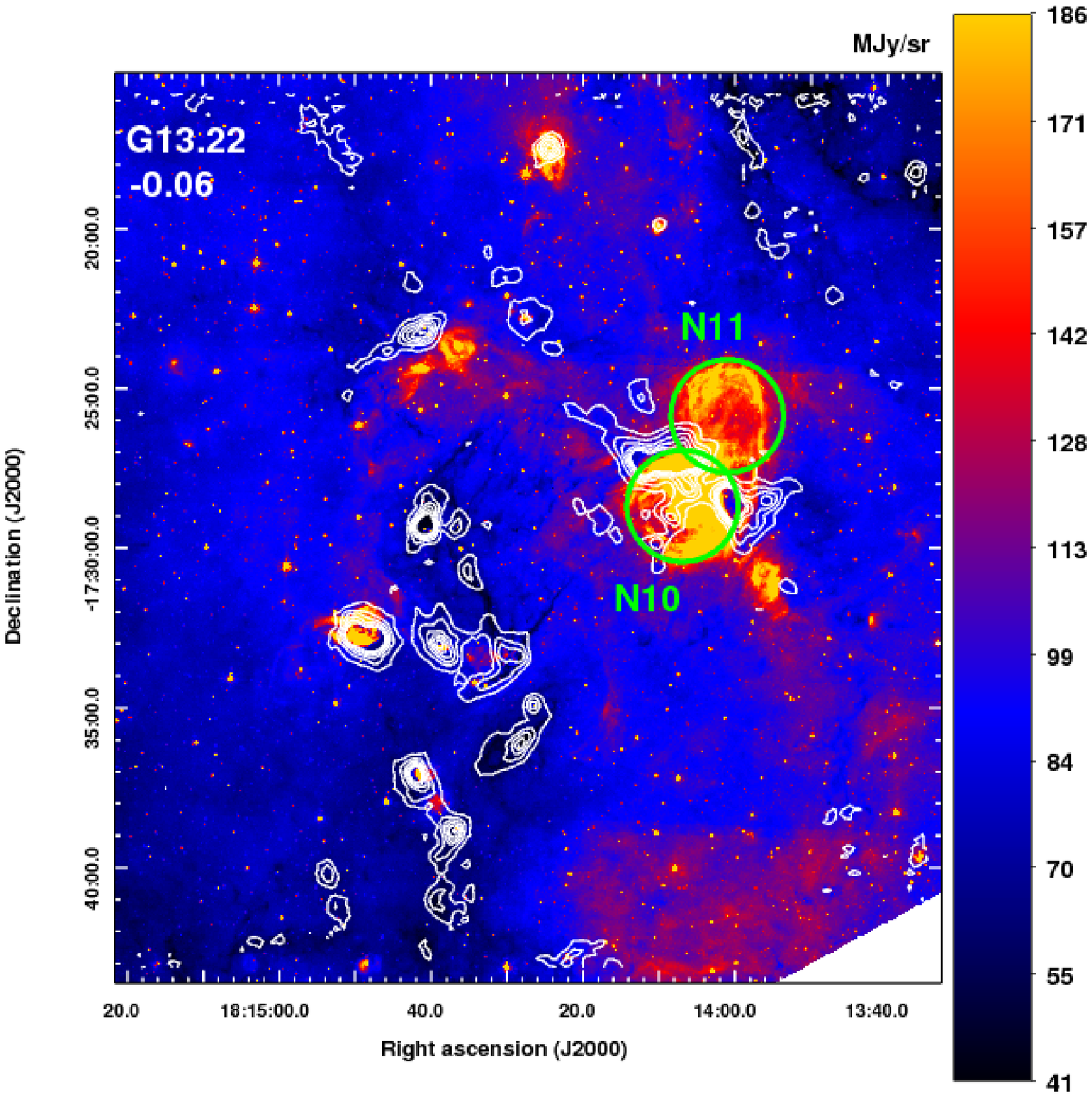}
\caption{Same as Fig.~\ref{figure:G187} but towards G13.22-0.06. The overlaid 
LABOCA contours go from 0.24 Jy~beam$^{-1}$ ($3\sigma$) to 1.44 Jy~beam$^{-1}$, 
in steps of $3\sigma$. The field contains several clumps at different kinematic 
distances, and therefore only the angular scale bar is shown. The green 
circles in the \textit{Spitzer} 8 $\mu$m image on the right panel indicate the 
positions and outer radii of the bubbles N10 and N11 from Churchwell et al. 
(2006; their Table~2).}
\label{figure:G1322}
\end{center}
\end{figure*}

In each field, the clumps are called SMM 1, SMM 2, etc., in order of 
increasing right ascension. The J2000.0 coordinates of the peak 870 $\mu$m 
emission, peak surface brightnesses, integrated flux densities (within 
$3\sigma$), and clump effective radii ($R_{\rm eff}=\sqrt{A/\pi}$, where $A$ is 
the projected area within the $3\sigma$ contour) are listed in Cols.~(2)--(6) 
of Table~\ref{clumps}. The quoted flux density uncertainties are based on the 
rms noise values and the 10\% absolute calibration error. The clump 
effective radii listed in Table~\ref{clumps} are not corrected for beam size. 
Note that the clumps SMM 32 and 40 in G1.87, SMM 2 in G2.11, 
SMM 6 in G11.36, and SMM 1, 2, and 8 in G13.22 are only barely 
resolved as their sizes are only slightly larger than the beam.
We also note that the clump SMM 32 in G13.22 (Fig.~\ref{figure:G1322}) 
could be clearly resolved by eye into two ``subclumps'', but they are treated 
as a single source by {\tt clfind2d} with our settings.

\addtocounter{table}{1}.

\subsection{C$^{17}$O$(2-1)$ spectra}

\begin{figure*}
\begin{center}
\includegraphics[width=3.1cm, angle=-90]{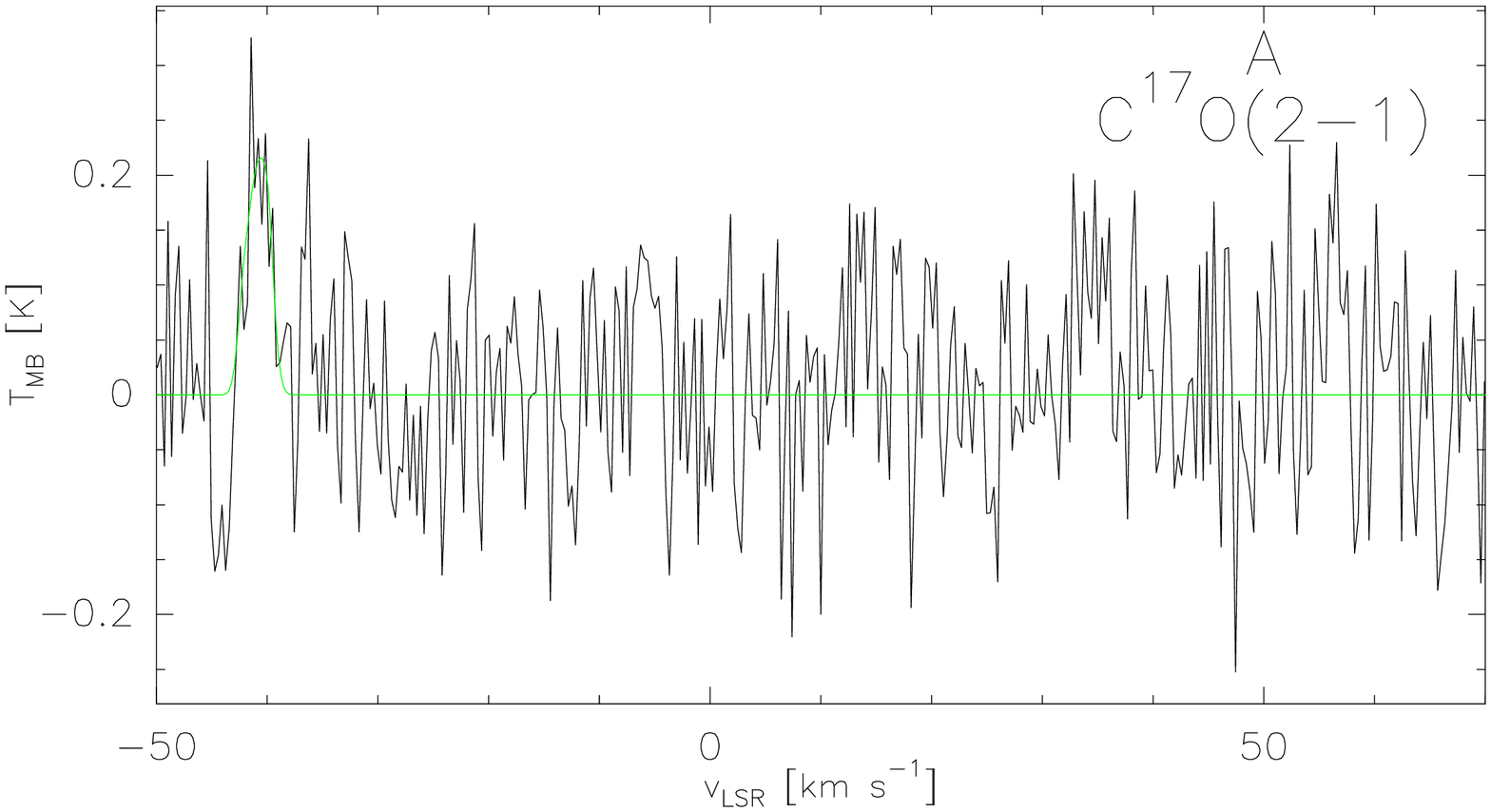}
\includegraphics[width=3.1cm, angle=-90]{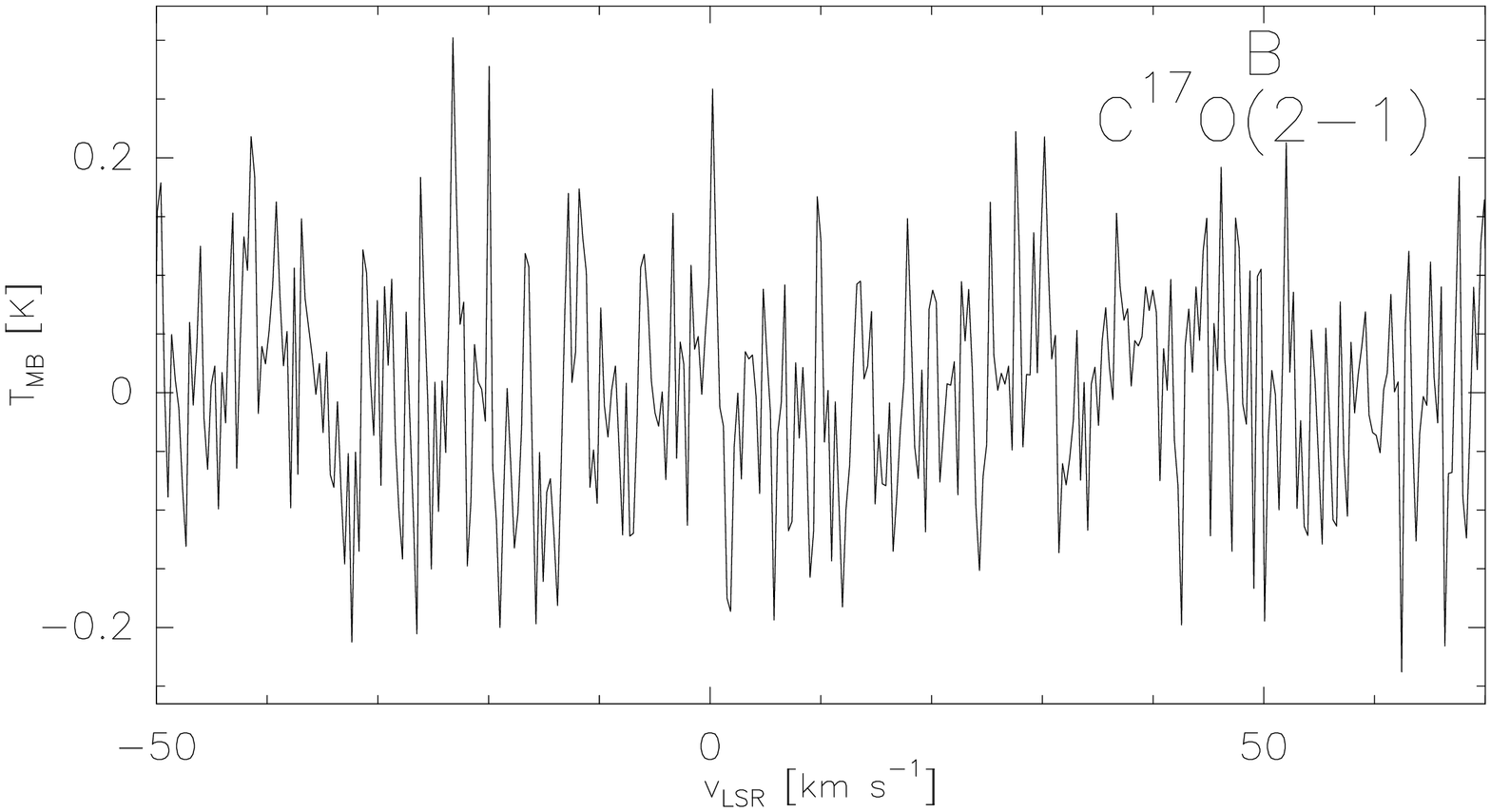}
\includegraphics[width=3.1cm, angle=-90]{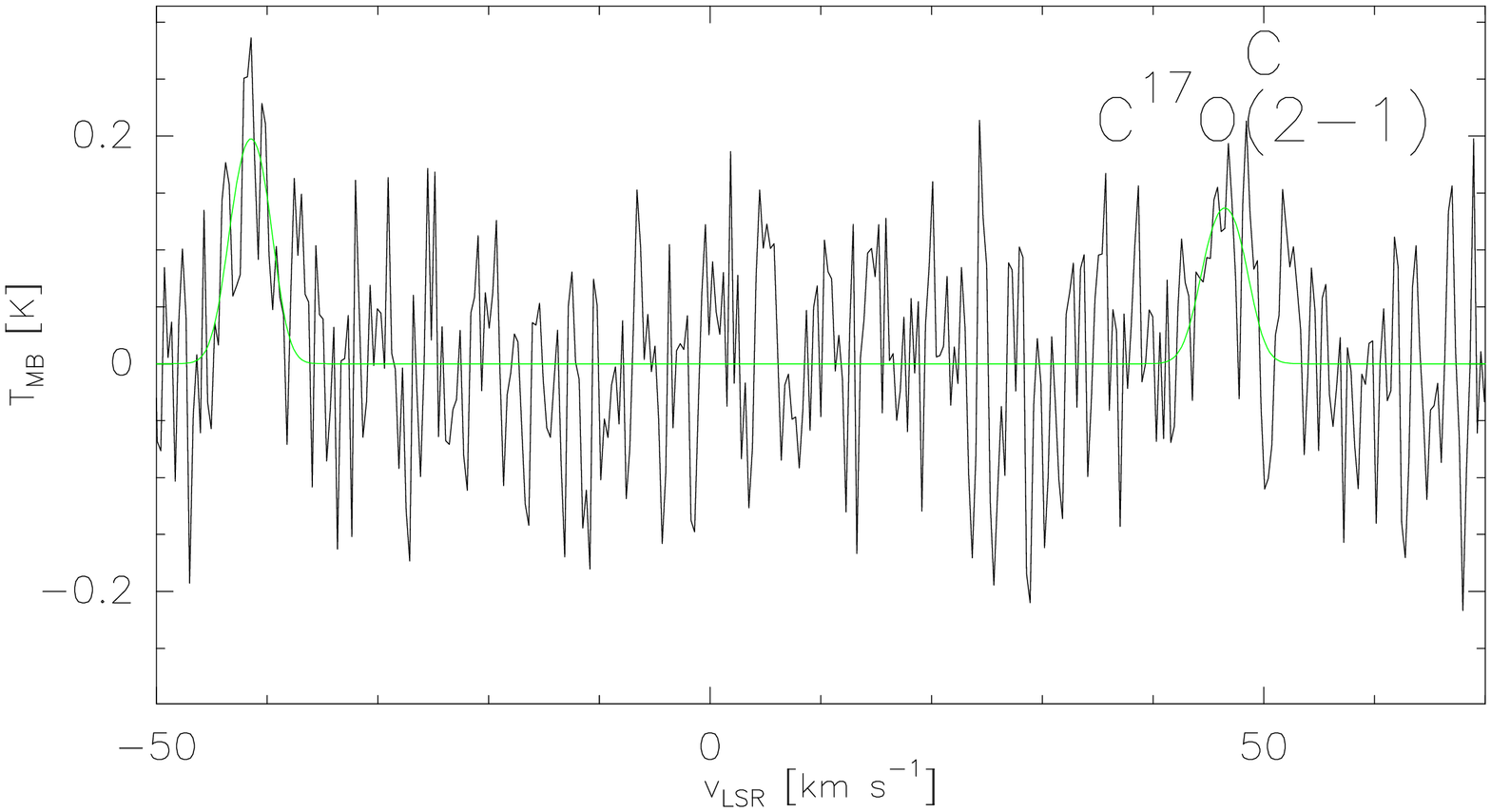}
\includegraphics[width=3.1cm, angle=-90]{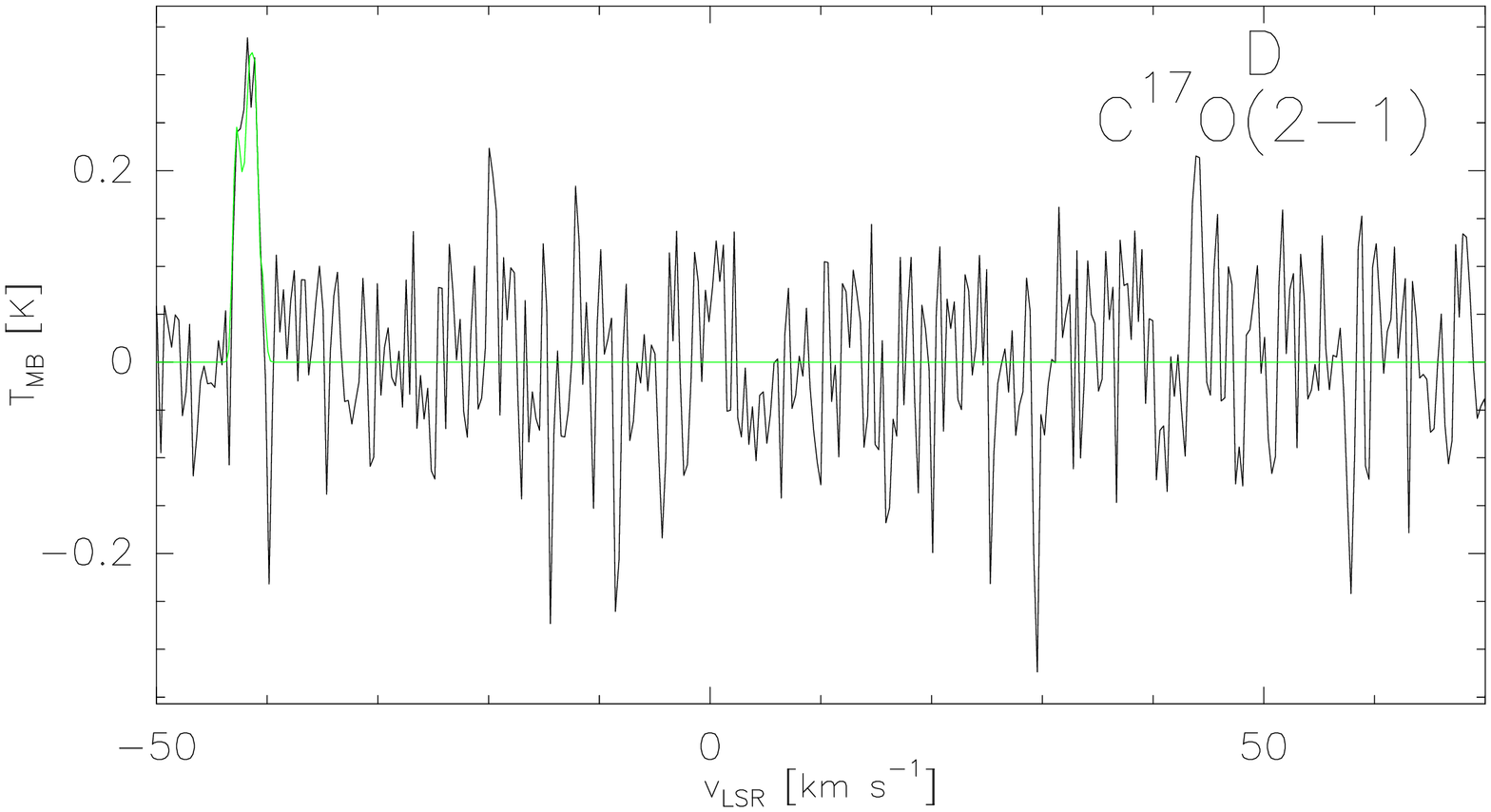}
\includegraphics[width=3.1cm, angle=-90]{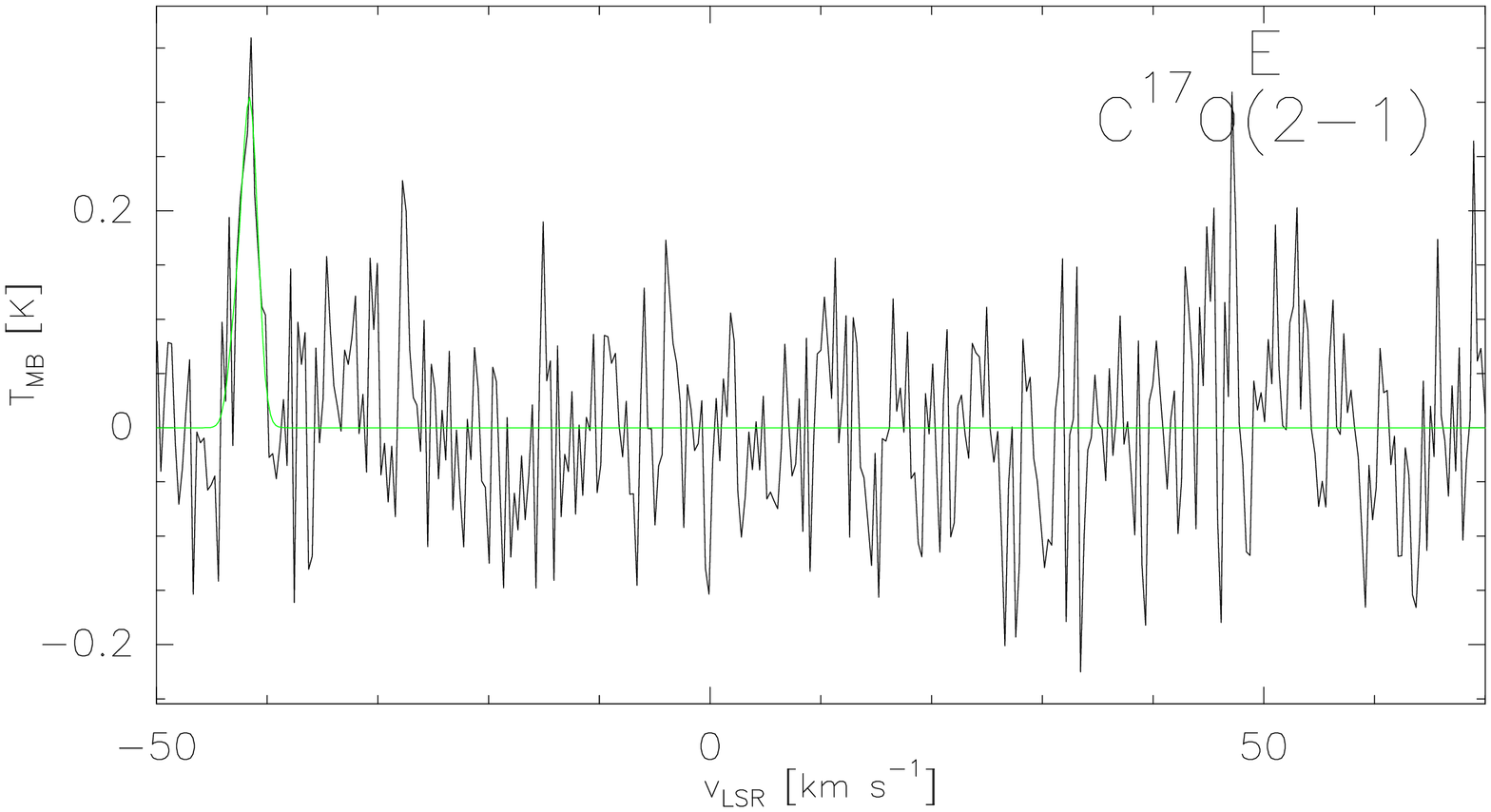}
\includegraphics[width=3.1cm, angle=-90]{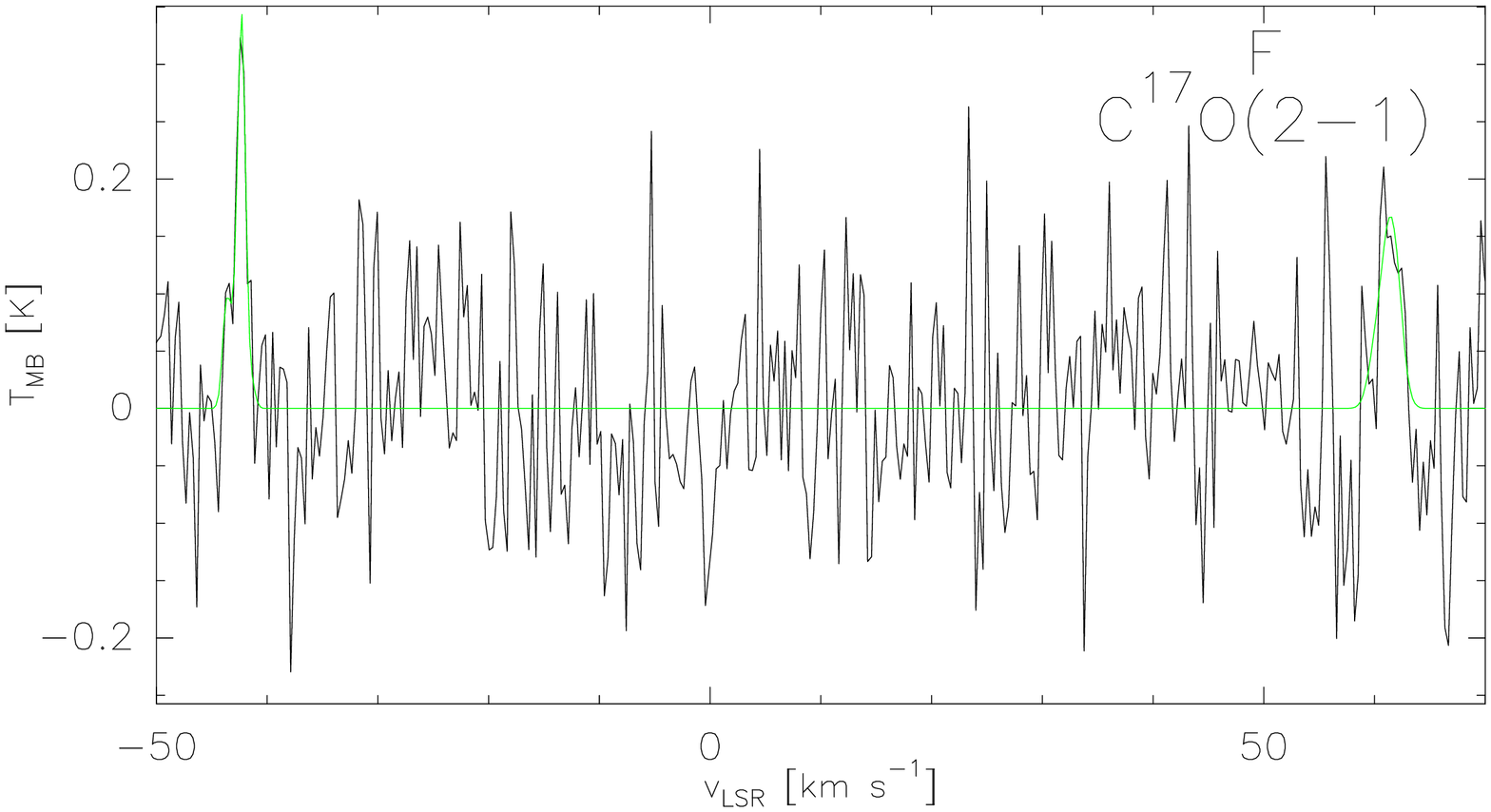}
\includegraphics[width=3.1cm, angle=-90]{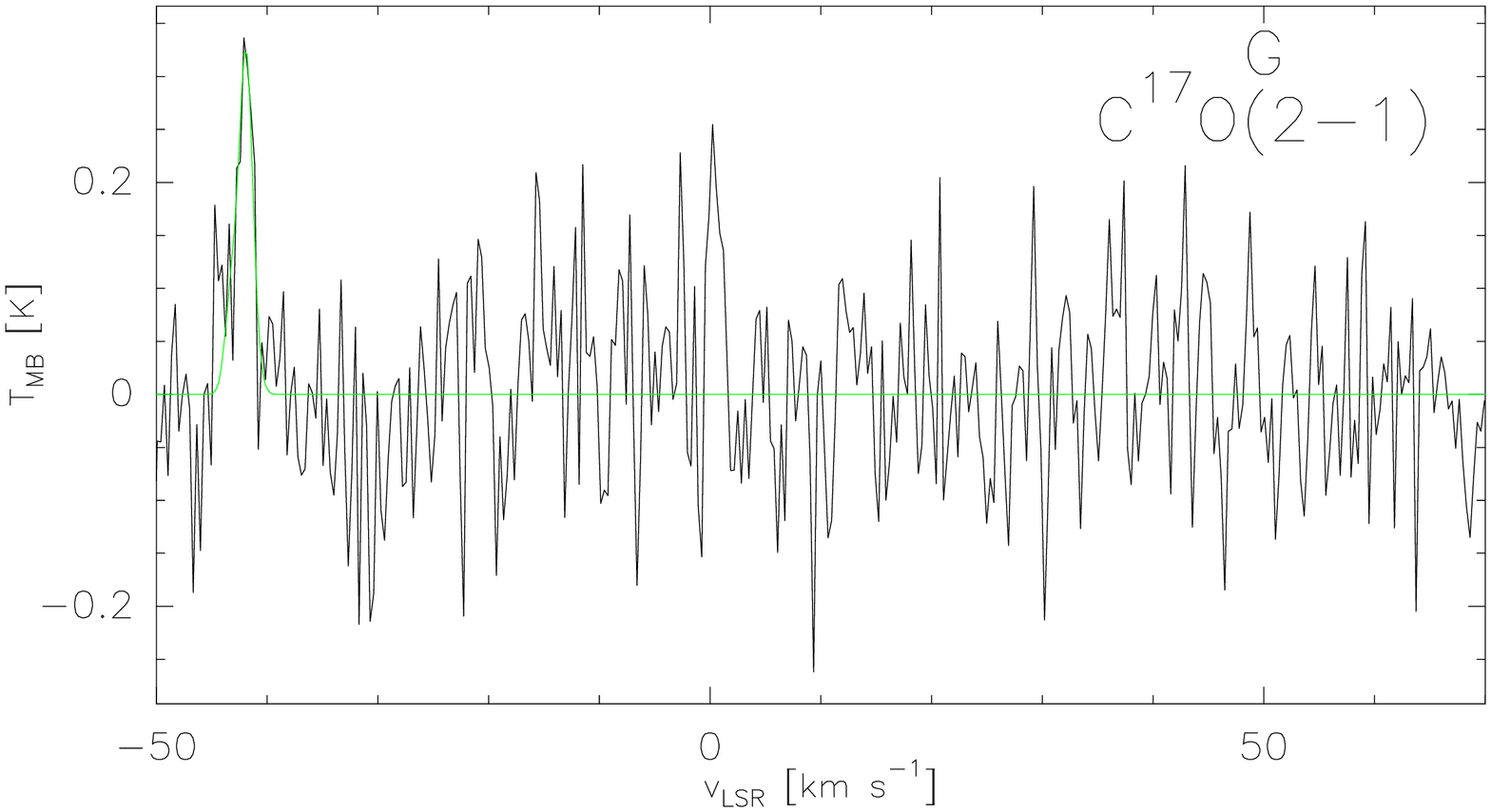}
\includegraphics[width=3.1cm, angle=-90]{blank.eps}
\includegraphics[width=3.1cm, angle=-90]{blank.eps}
\caption{Smoothed C$^{17}$O$(2-1)$ spectra towards selected positions in 
G1.87-0.14. Most lines are seen at $\sim -41$ km~s$^{-1}$. Hyperfine-structure 
fits to the lines are overlaid in green. No line emission is detected towards 
position B, whereas two velocity components are seen towards positions C and F.}
\label{figure:spectraG187}
\end{center}
\end{figure*}

\begin{figure*}
\begin{center}
\includegraphics[width=3.1cm, angle=-90]{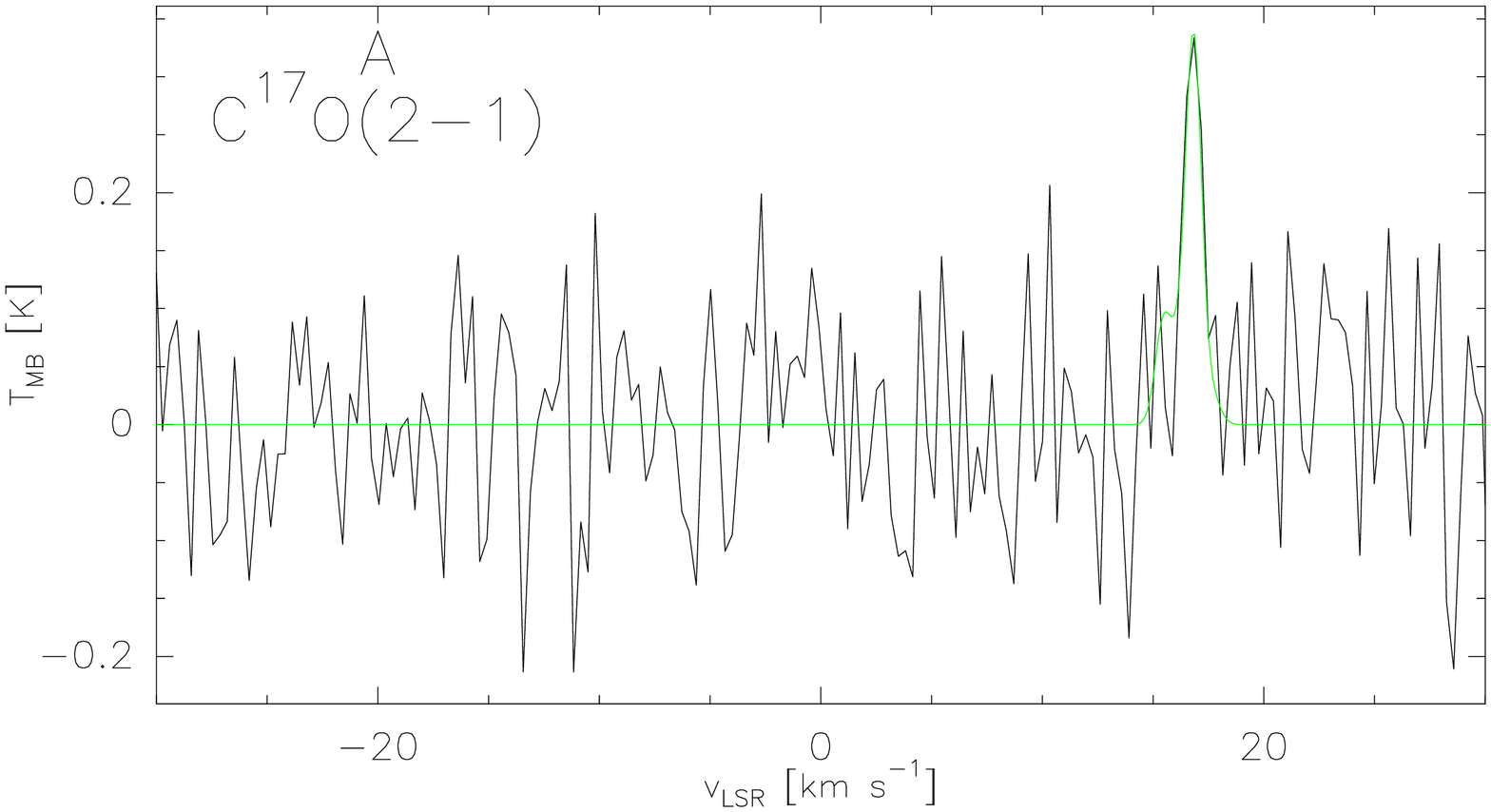}
\includegraphics[width=3.1cm, angle=-90]{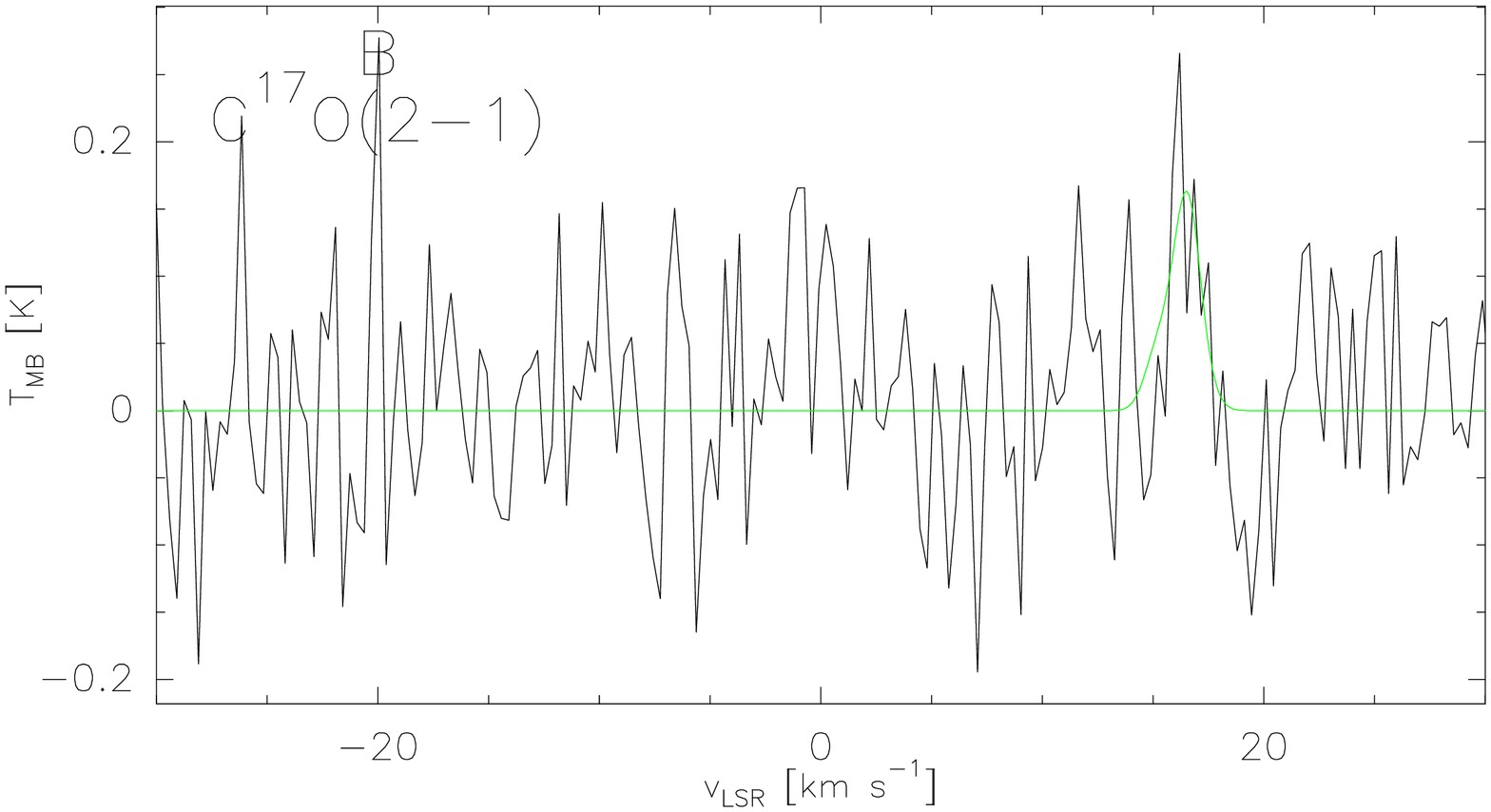}
\includegraphics[width=3.1cm, angle=-90]{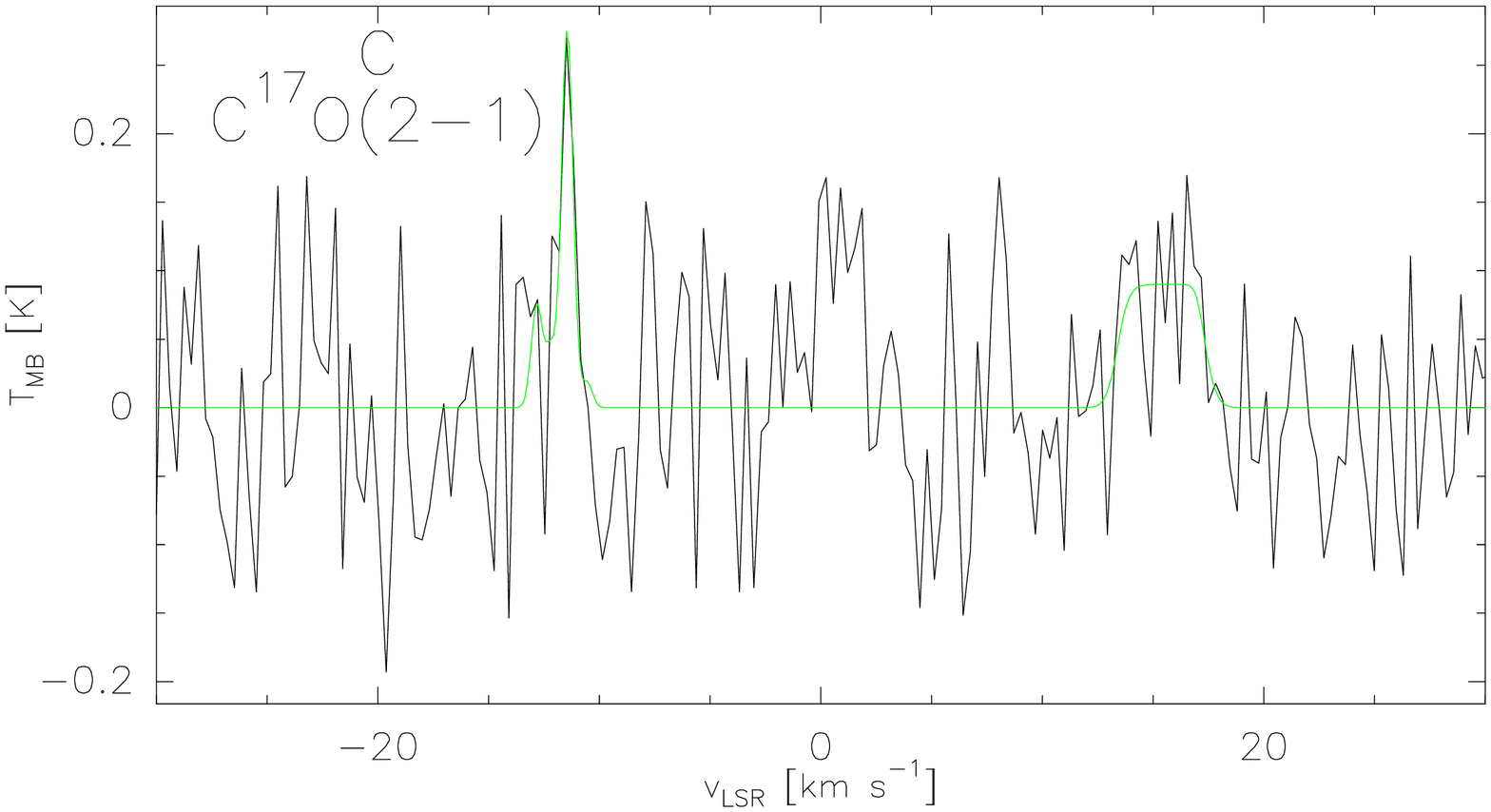}
\includegraphics[width=3.1cm, angle=-90]{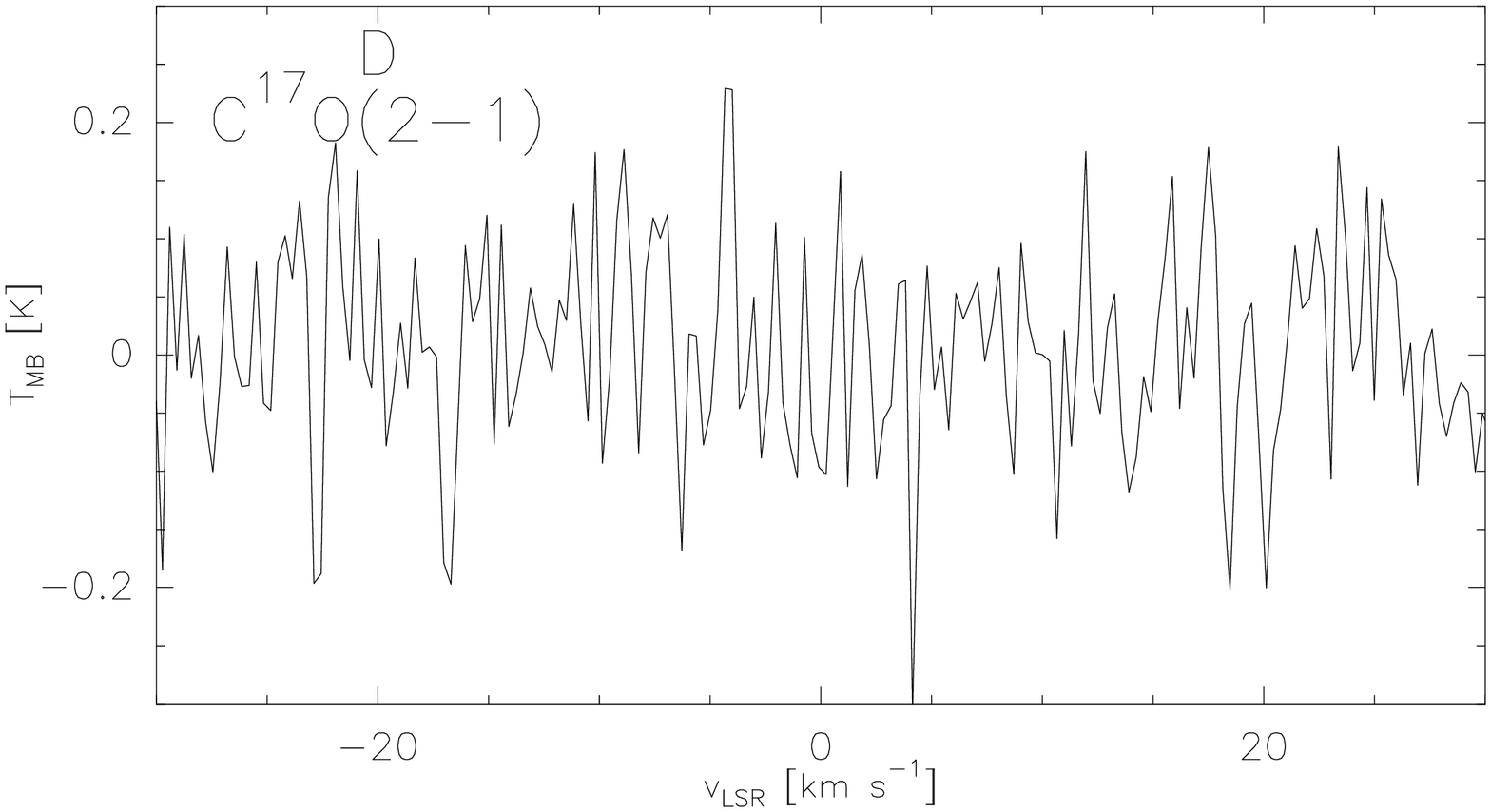}
\includegraphics[width=3.1cm, angle=-90]{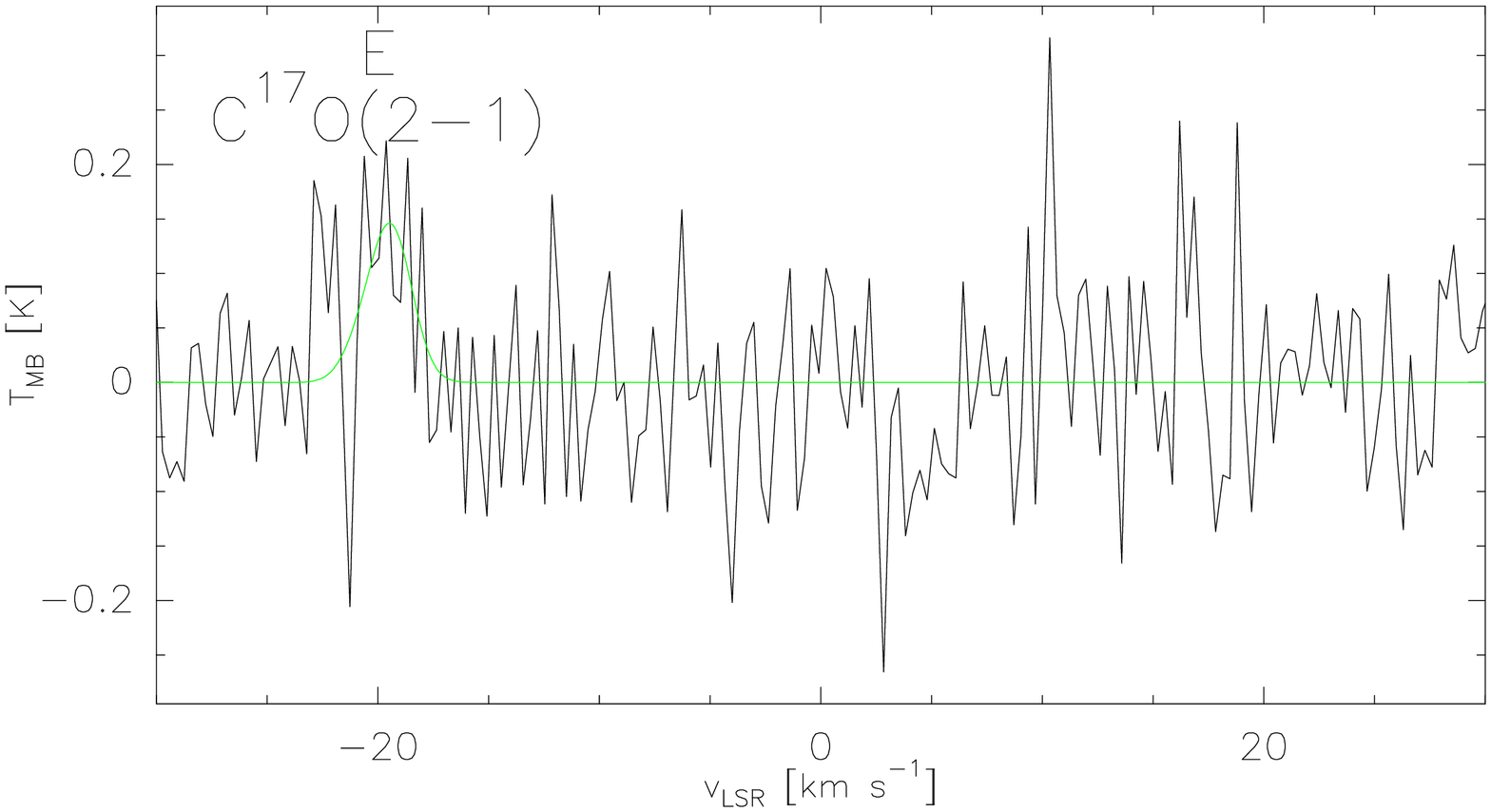}
\includegraphics[width=3.1cm, angle=-90]{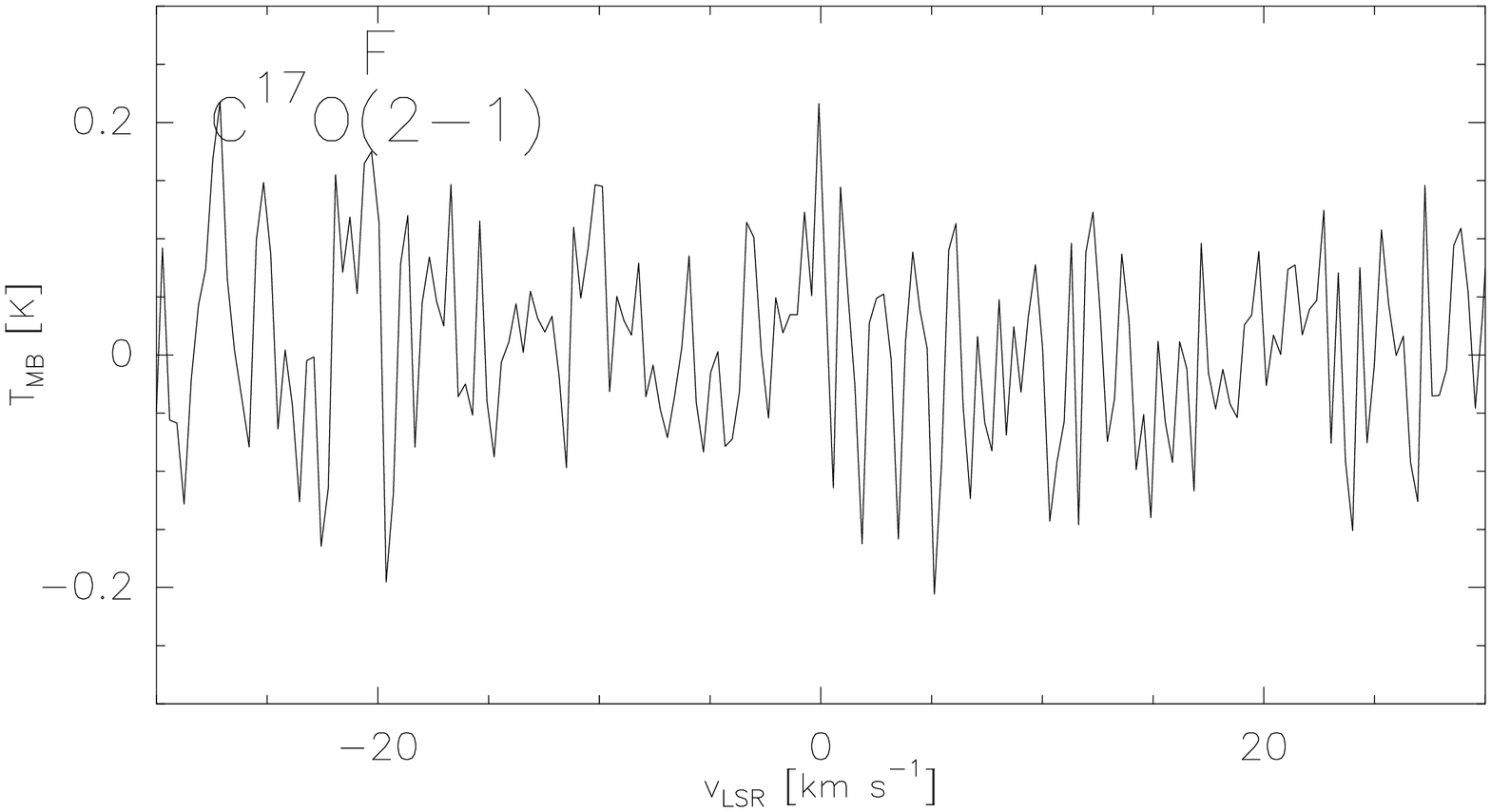}
\includegraphics[width=3.1cm, angle=-90]{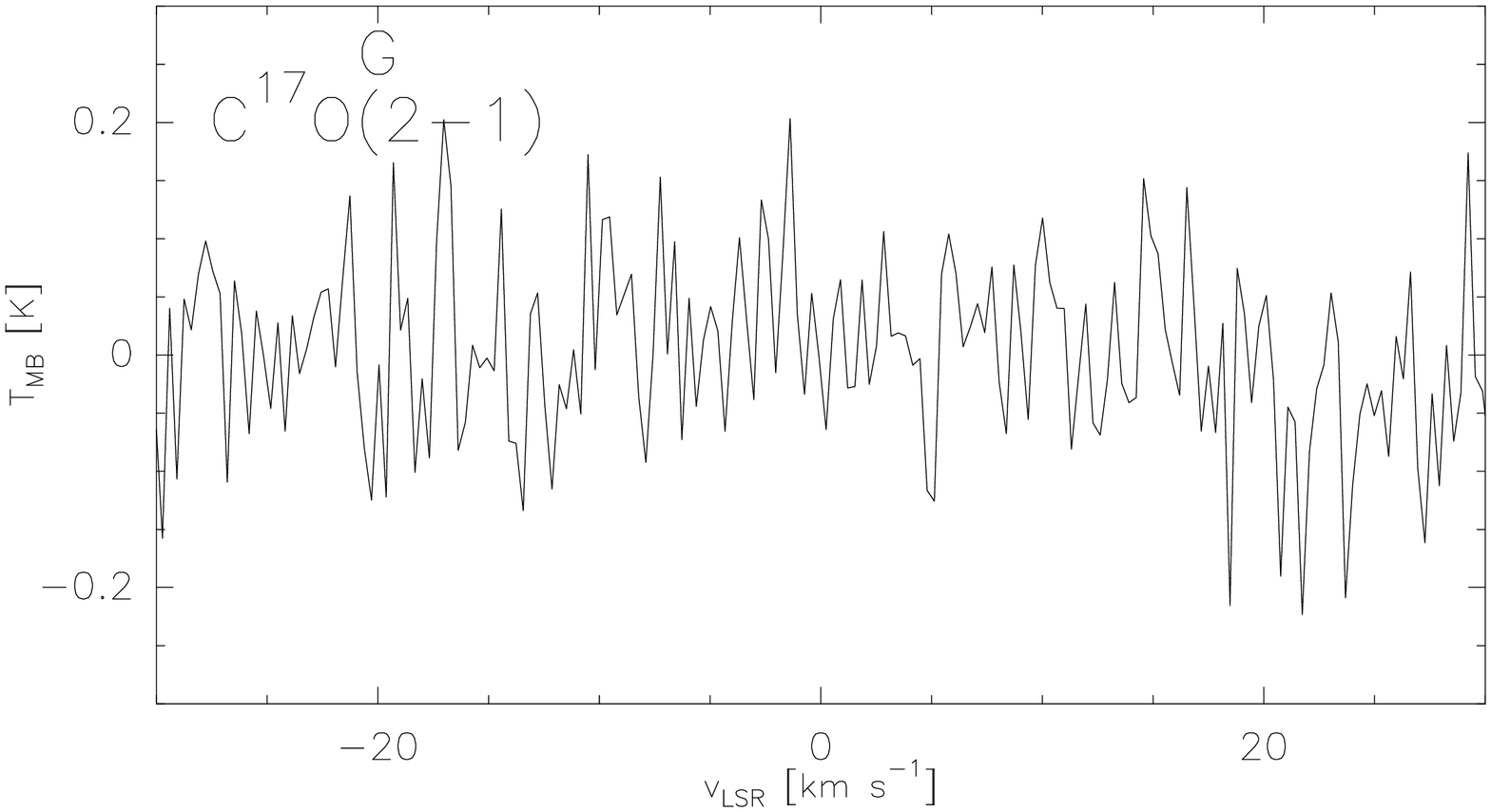}
\includegraphics[width=3.1cm, angle=-90]{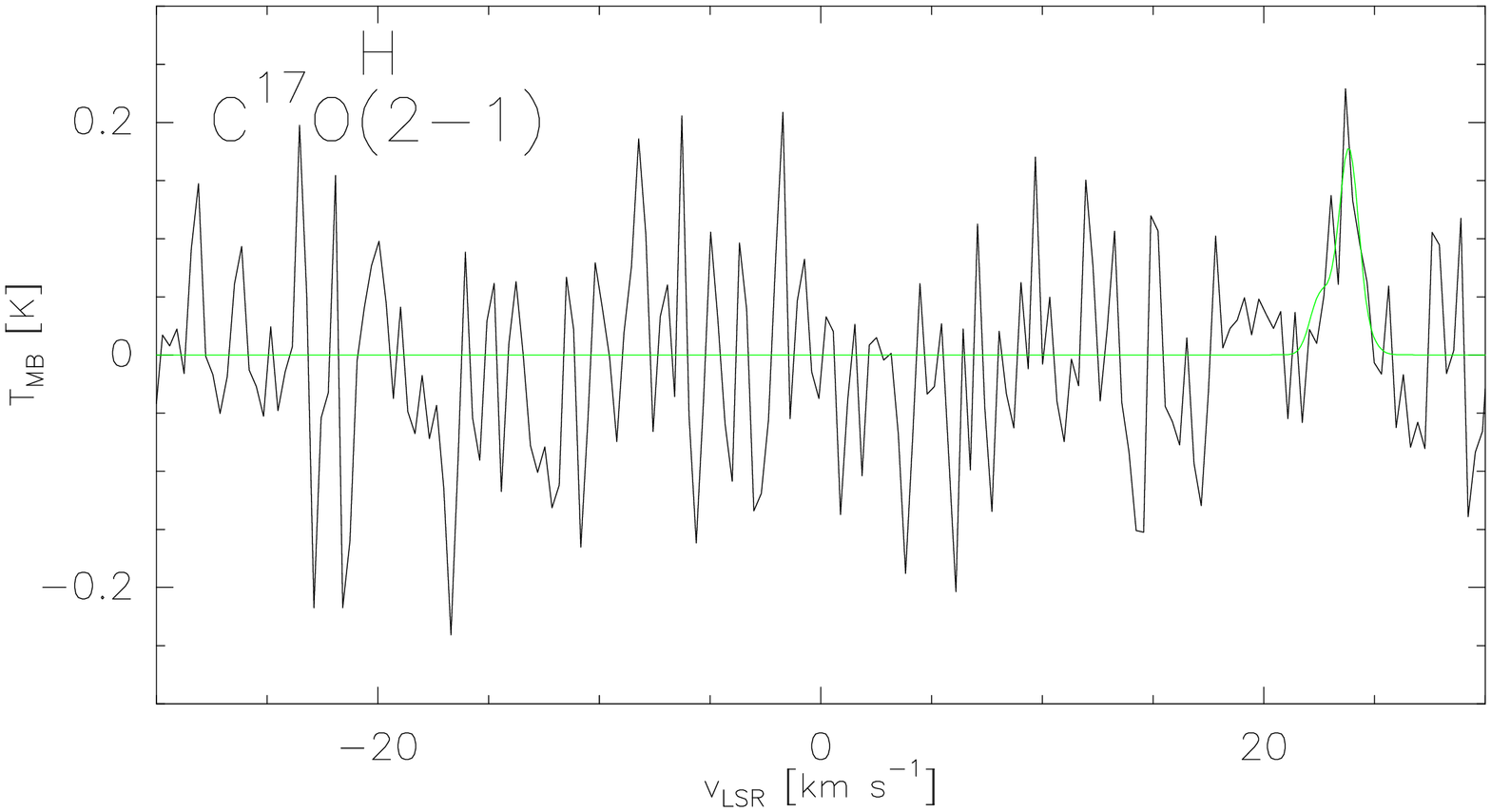}
\includegraphics[width=3.1cm, angle=-90]{blank.eps}
\caption{Same as Fig.~\ref{figure:spectraG187} but towards the selected 
positions in G2.11+0.00. Most detected lines are near $\sim16$ km~s$^{-1}$. 
No lines were detected towards positions D, F, and G. Two velocity components 
are seen towards position C.}
\label{figure:spectraG211}
\end{center}
\end{figure*}

\begin{figure*}
\begin{center}
\includegraphics[width=3.1cm, angle=-90]{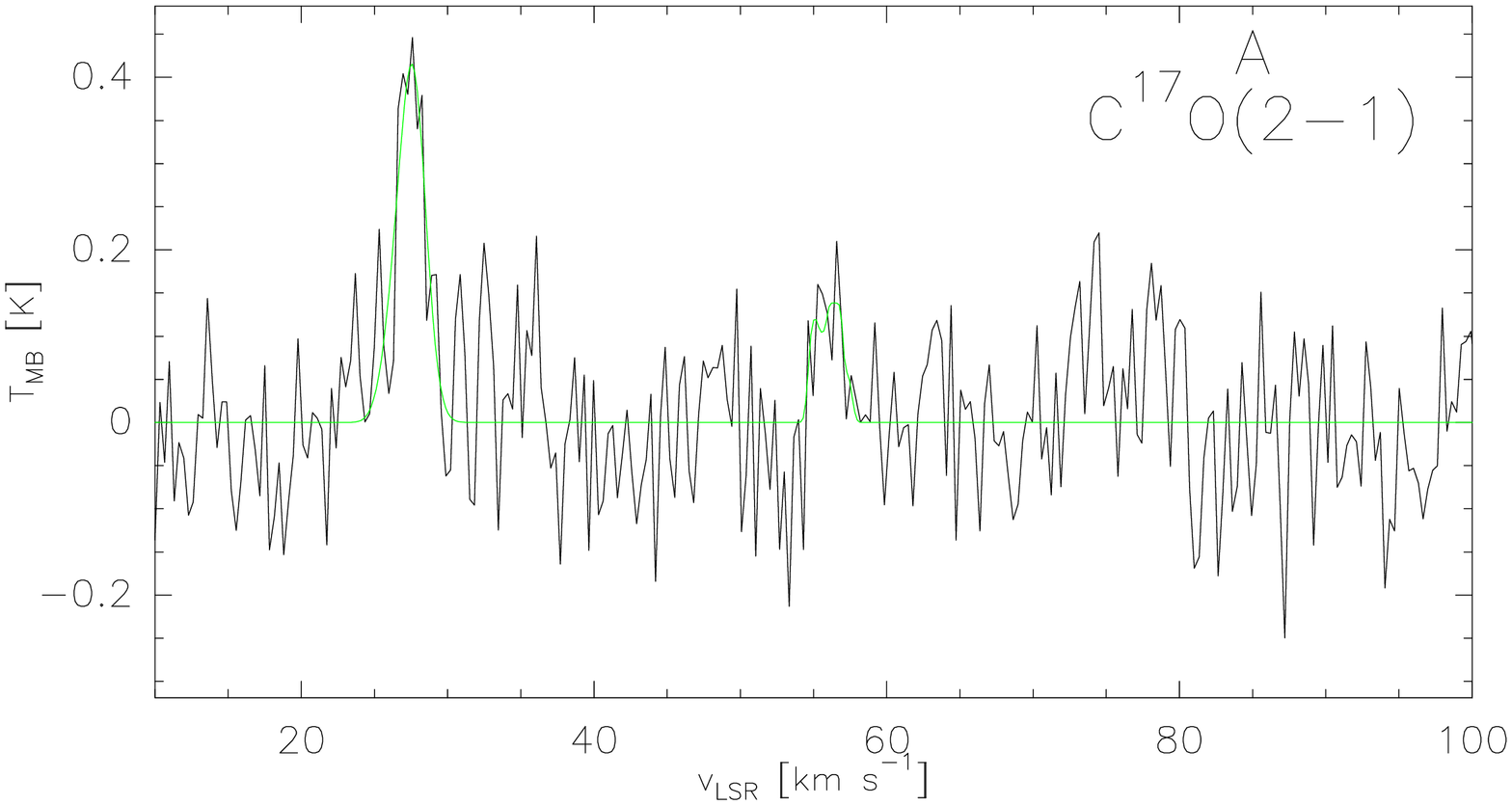}
\includegraphics[width=3.1cm, angle=-90]{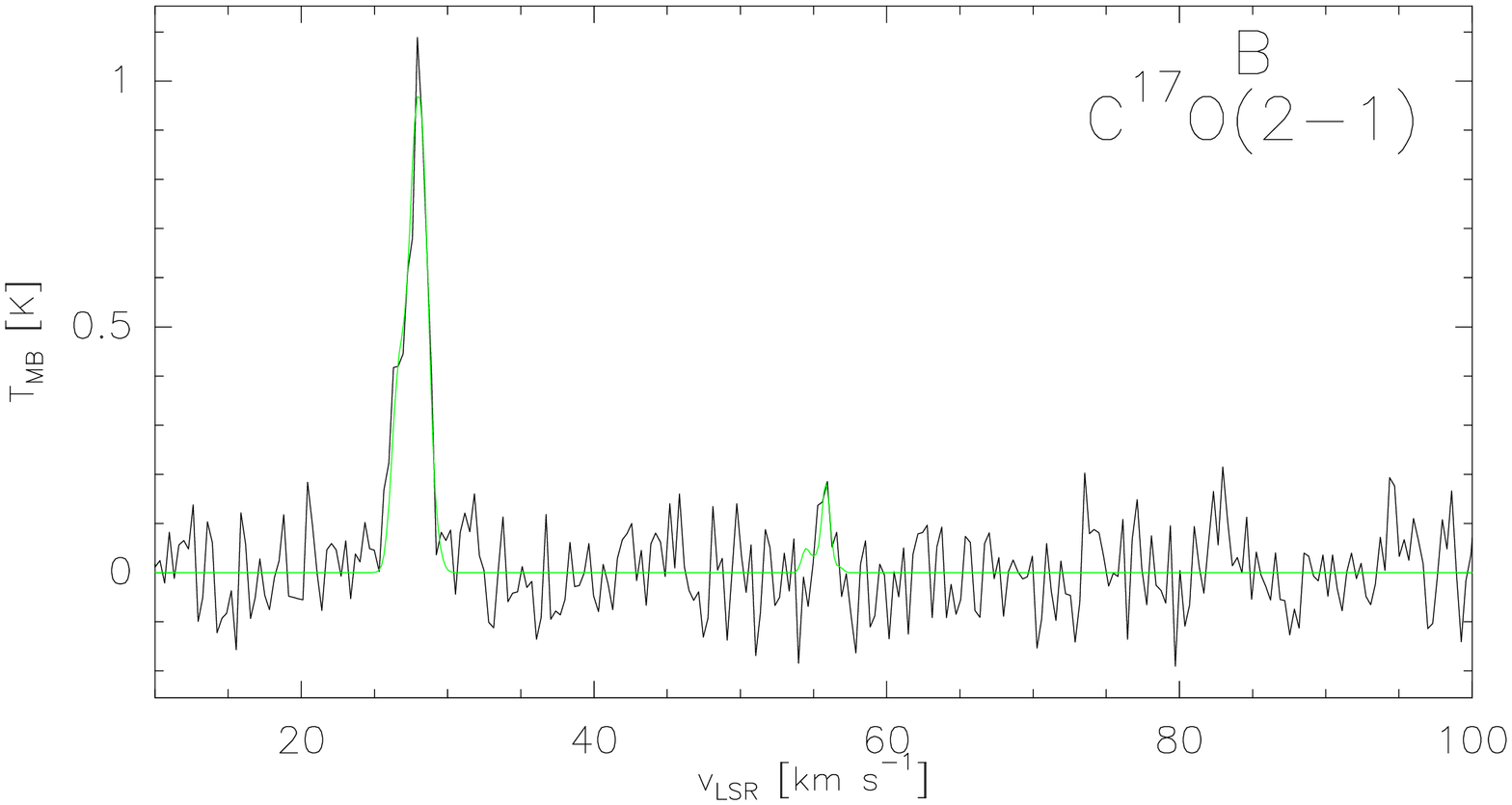}
\includegraphics[width=3.1cm, angle=-90]{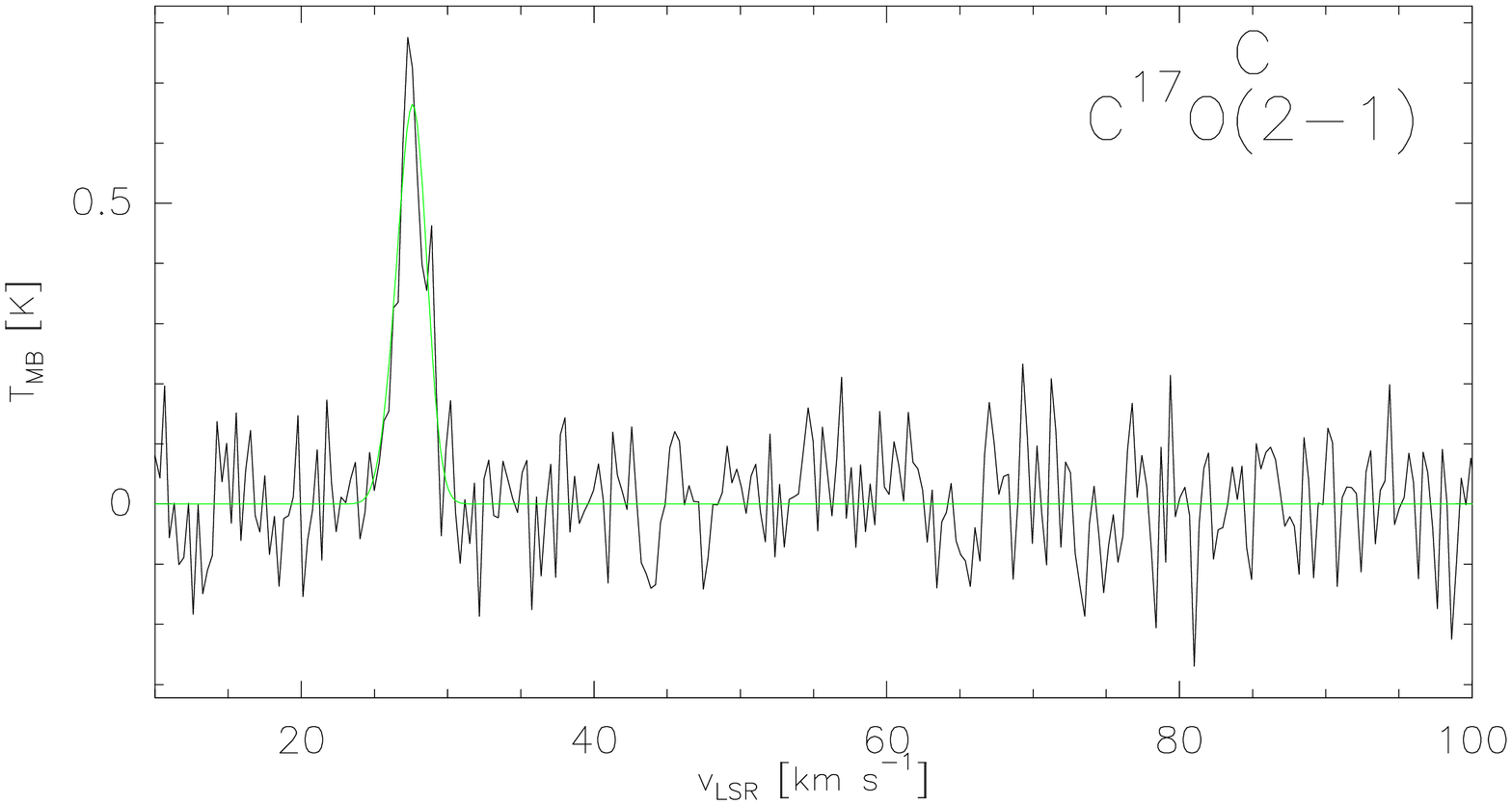}
\includegraphics[width=3.1cm, angle=-90]{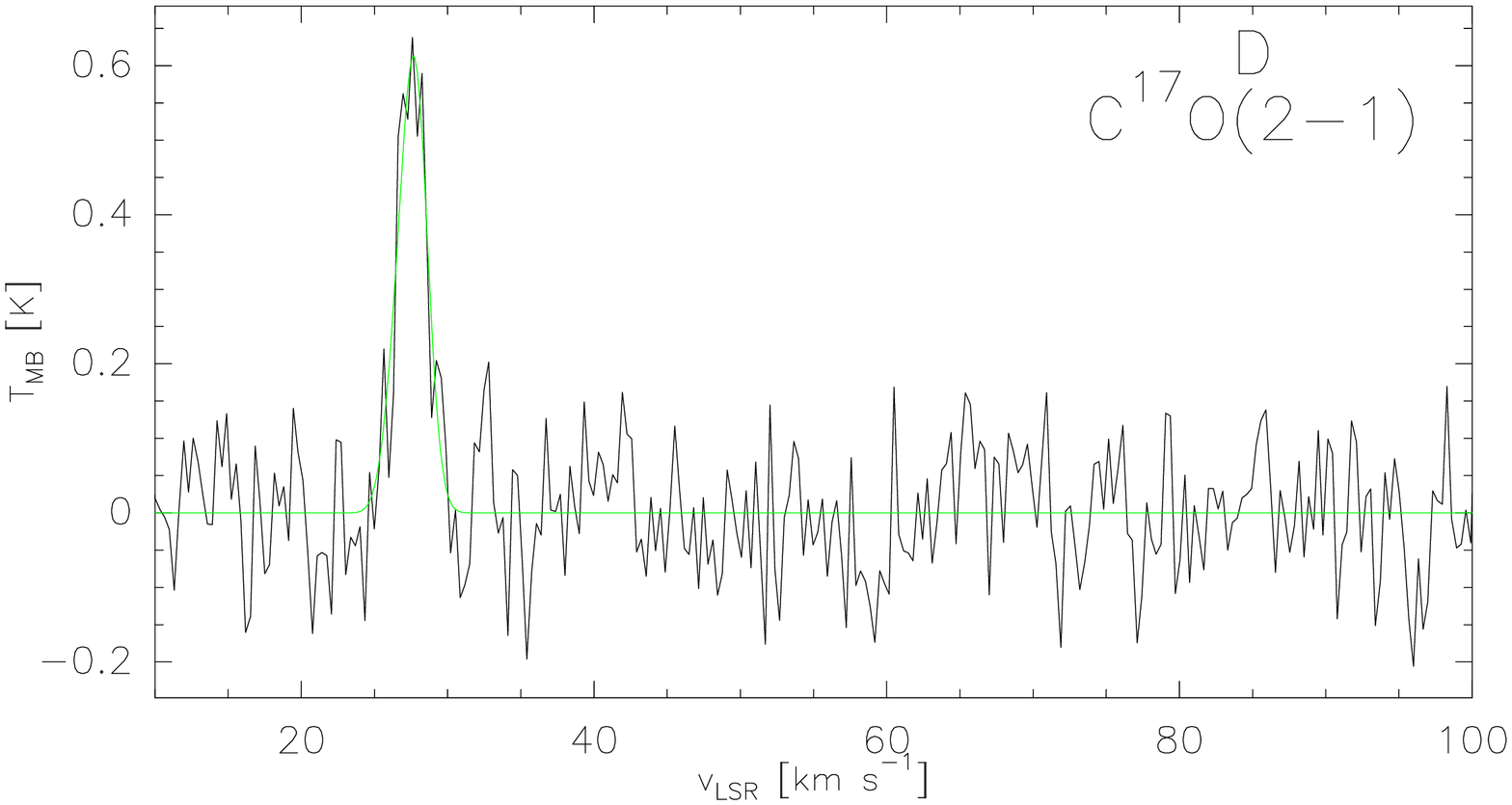}
\includegraphics[width=3.1cm, angle=-90]{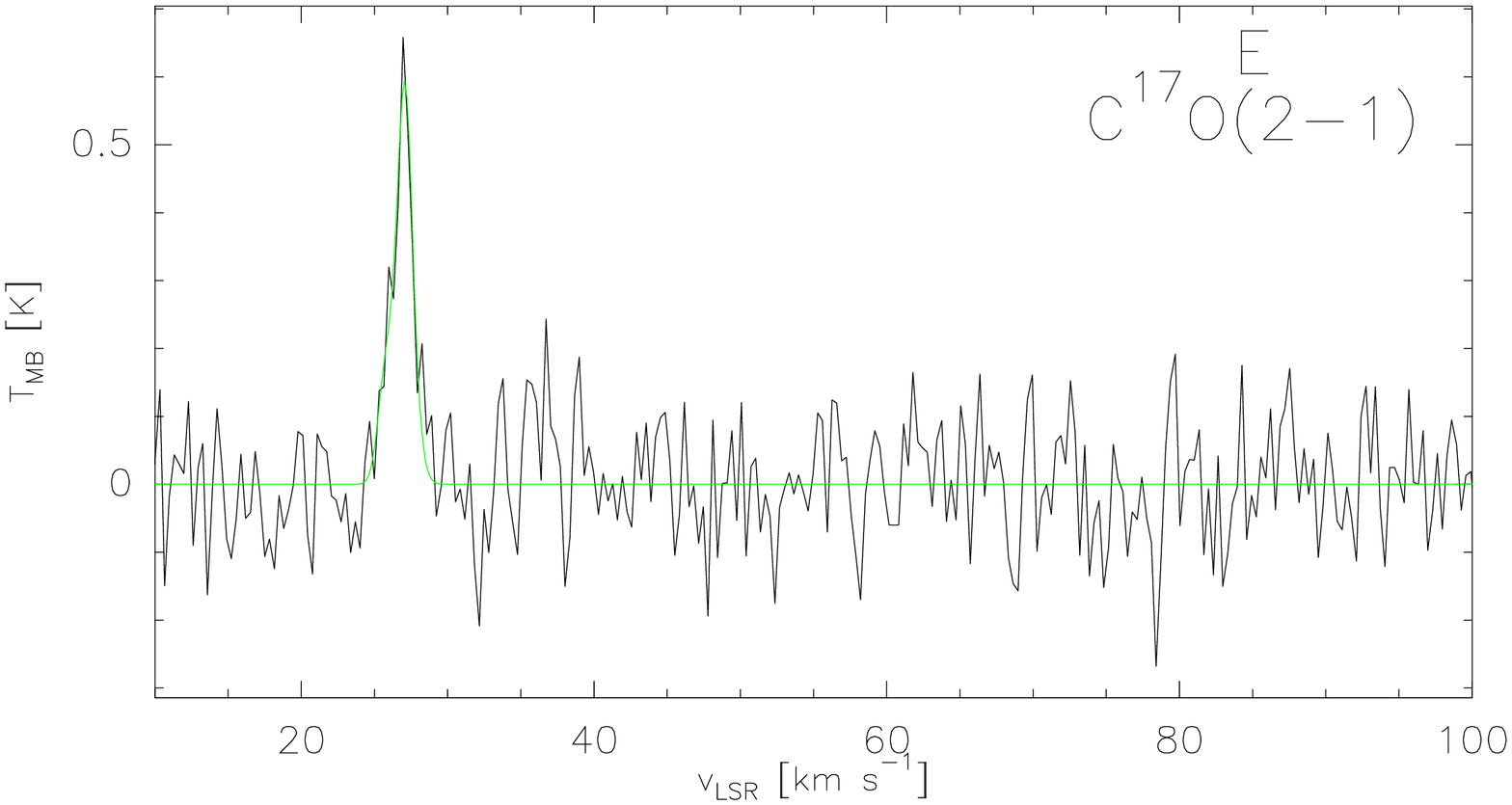}
\includegraphics[width=3.1cm, angle=-90]{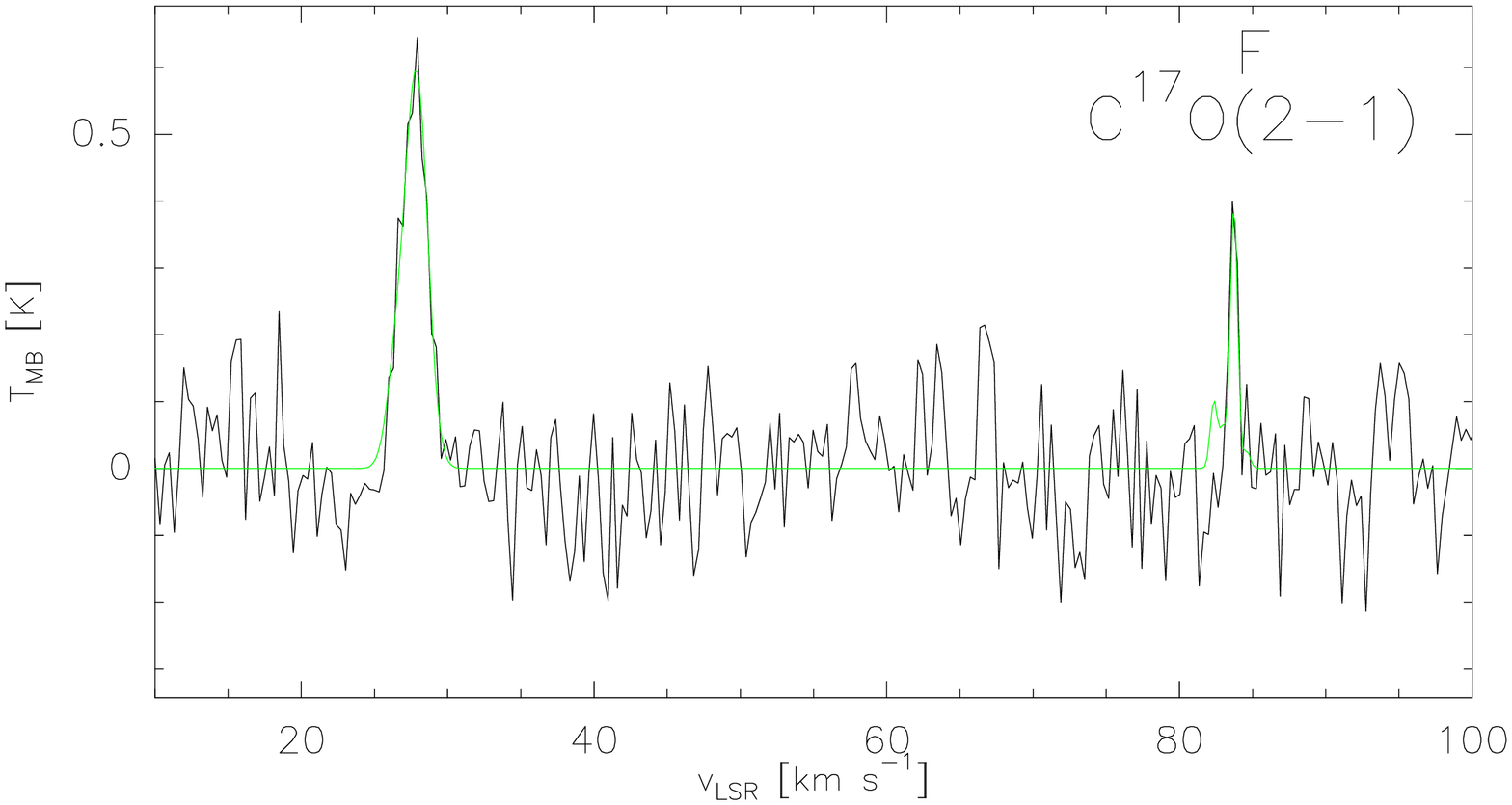}
\includegraphics[width=3.1cm, angle=-90]{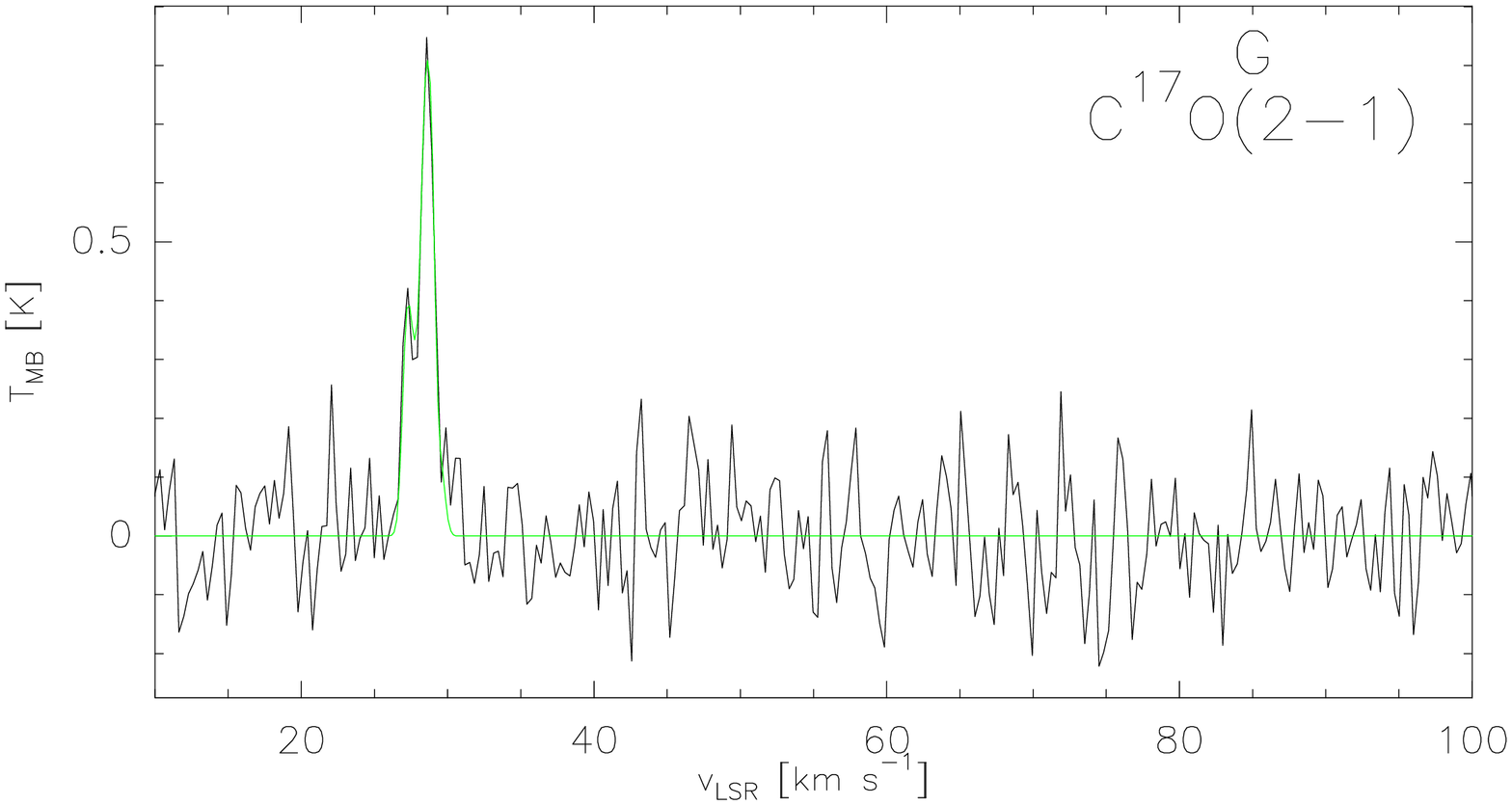}
\includegraphics[width=3.1cm, angle=-90]{blank.eps}
\includegraphics[width=3.1cm, angle=-90]{blank.eps}
\caption{Same as Fig.~\ref{figure:spectraG187} but towards the selected 
positions in G11.36+0.80. We note that clear line emission at $\sim28$ 
km~s$^{-1}$ is seen towards all positions. Two velocity components are 
detected towards positions A, B and F, but the secondary line is very weak in 
the former two case.}
\label{figure:spectraG1136}
\end{center}
\end{figure*}

\begin{figure*}
\begin{center}
\includegraphics[width=3.1cm, angle=-90]{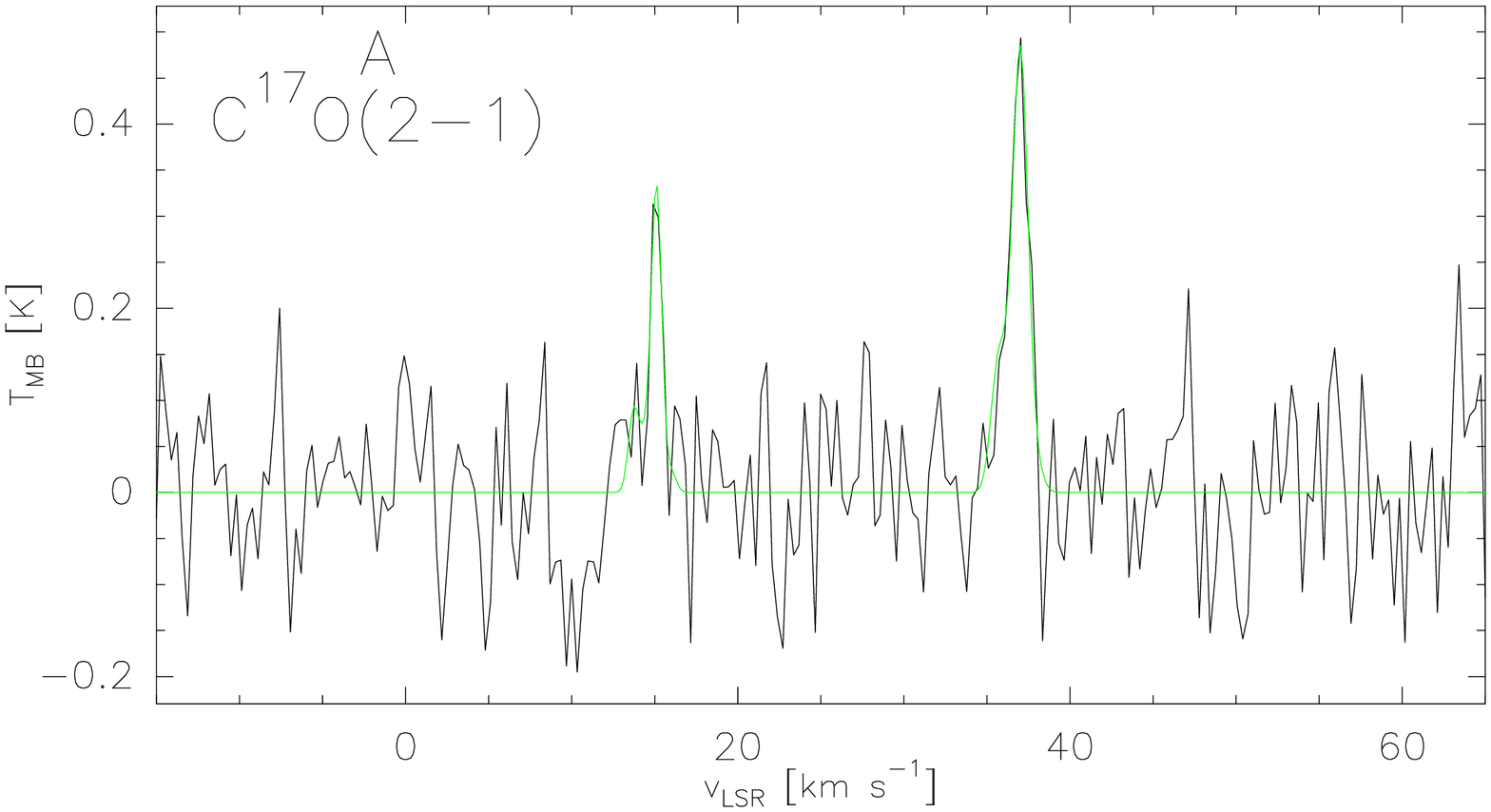}
\includegraphics[width=3.1cm, angle=-90]{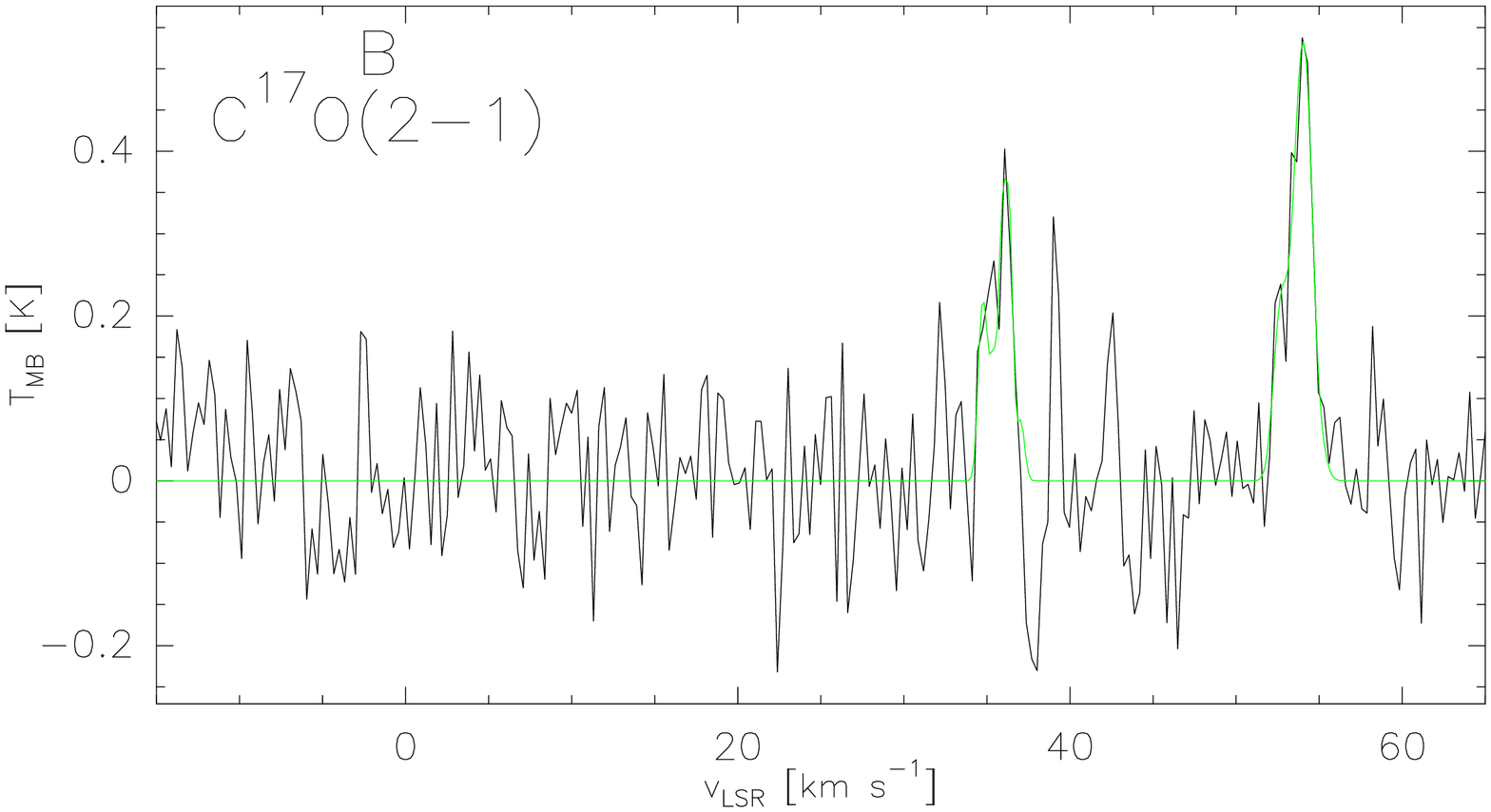}
\includegraphics[width=3.1cm, angle=-90]{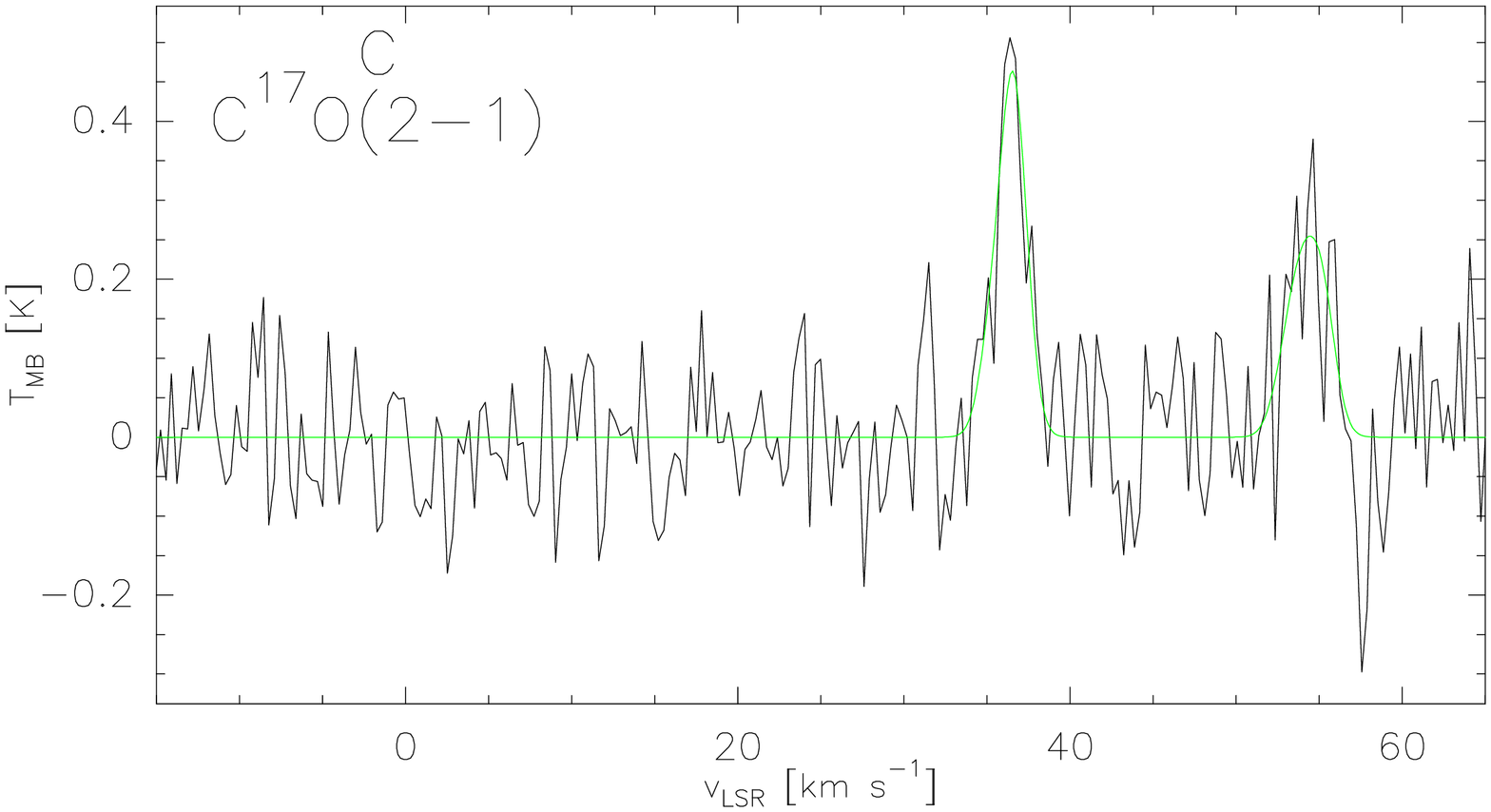}
\includegraphics[width=3.1cm, angle=-90]{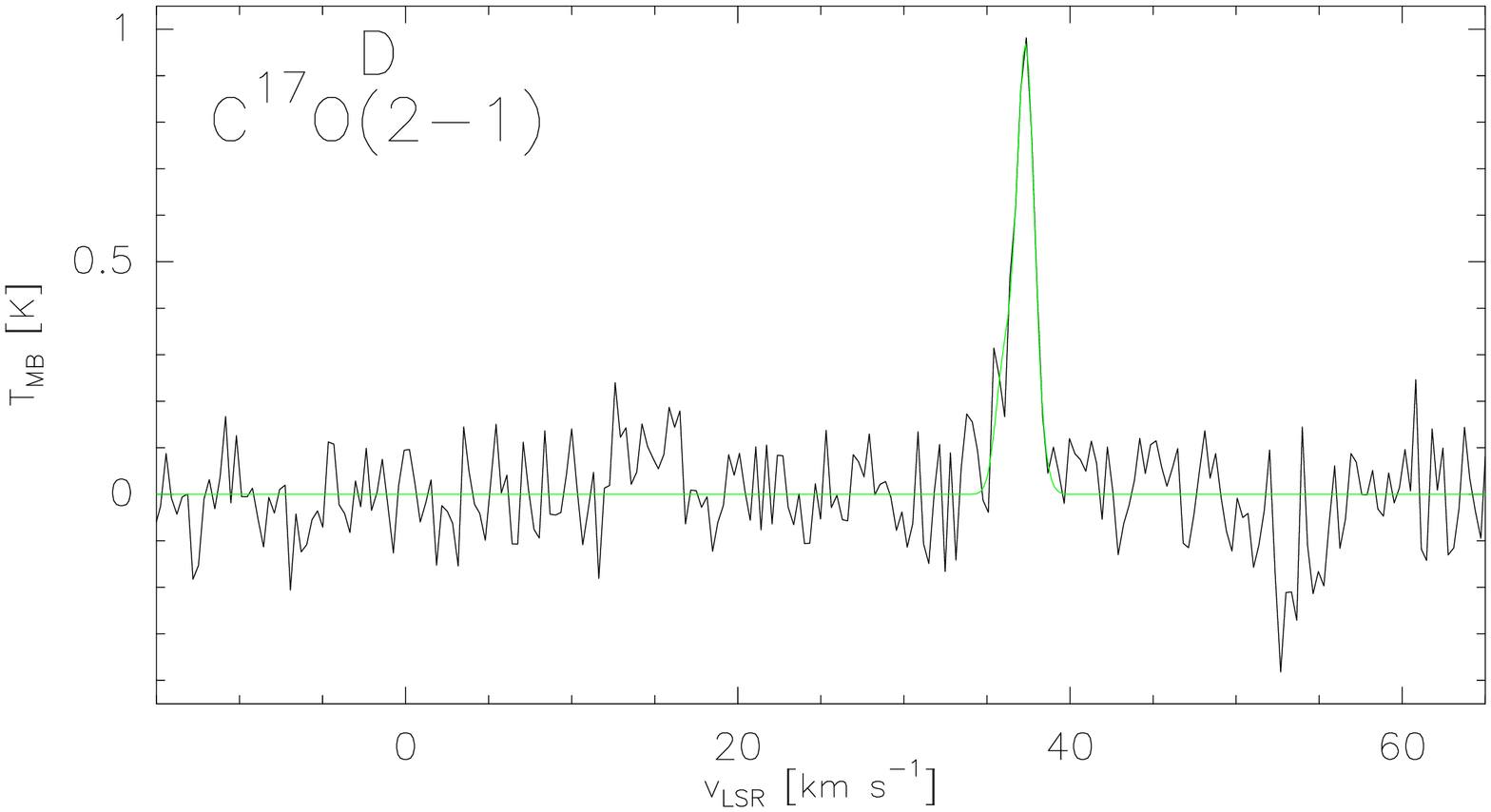}
\includegraphics[width=3.1cm, angle=-90]{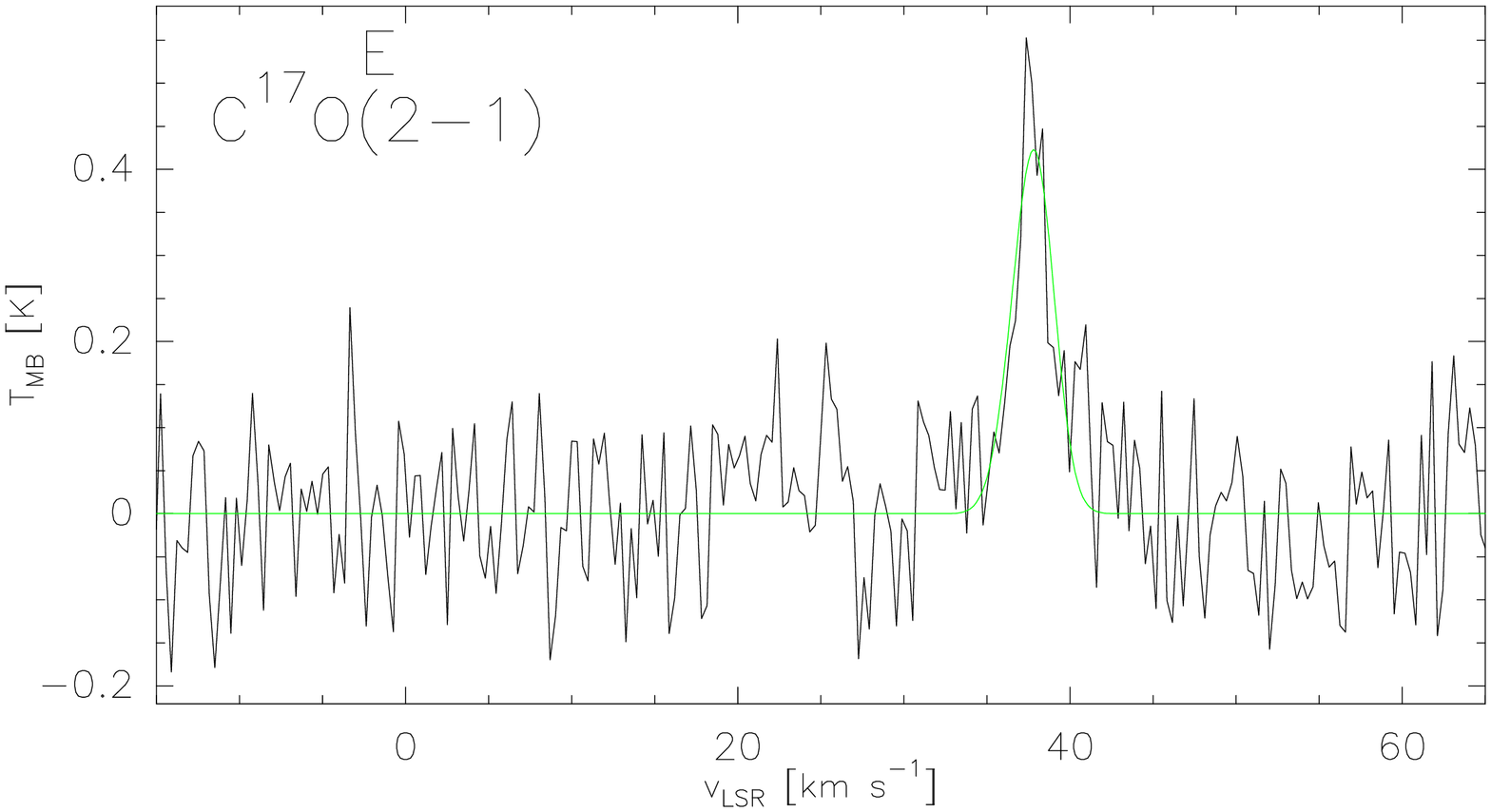}
\includegraphics[width=3.1cm, angle=-90]{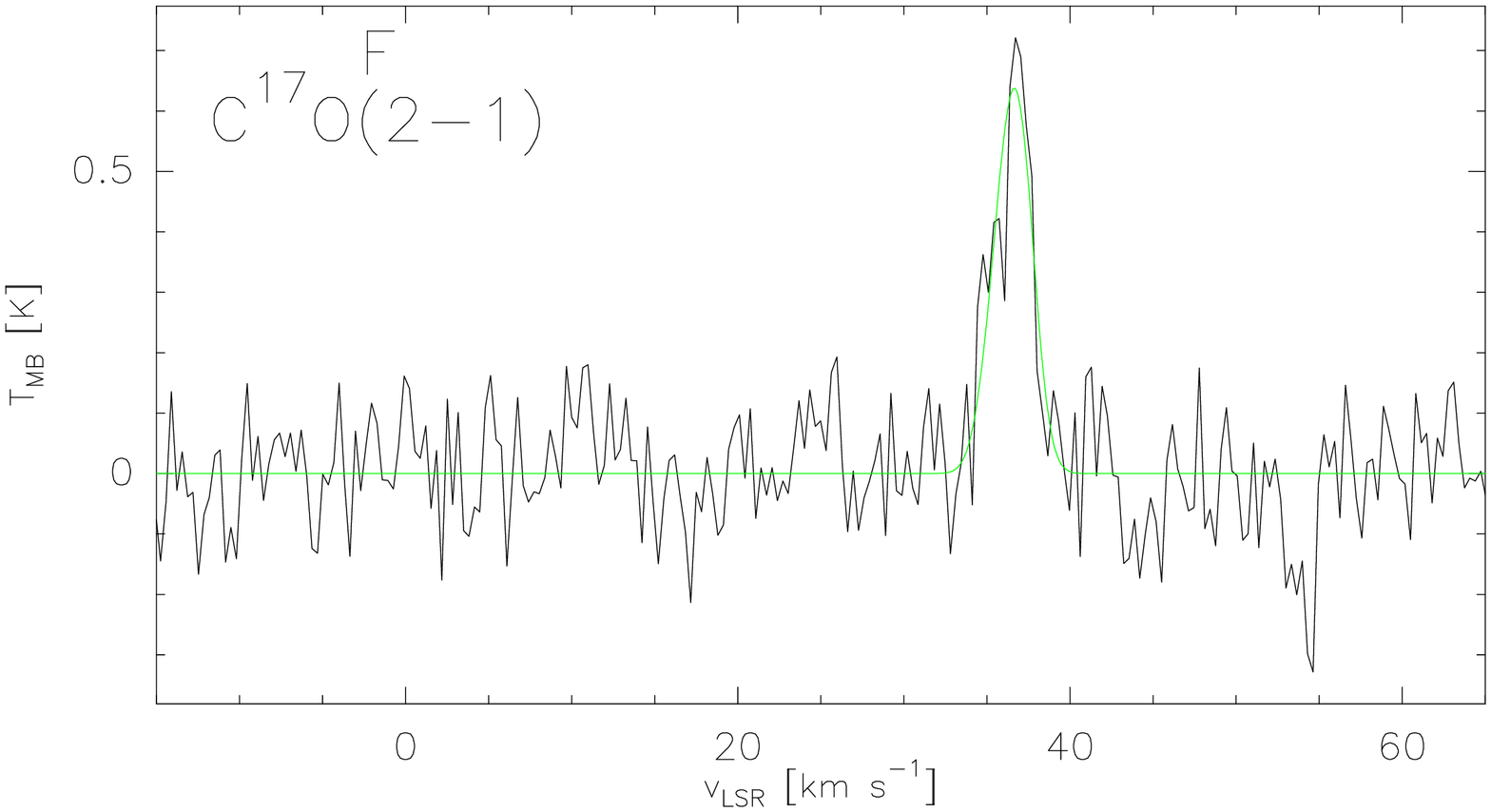}
\includegraphics[width=3.1cm, angle=-90]{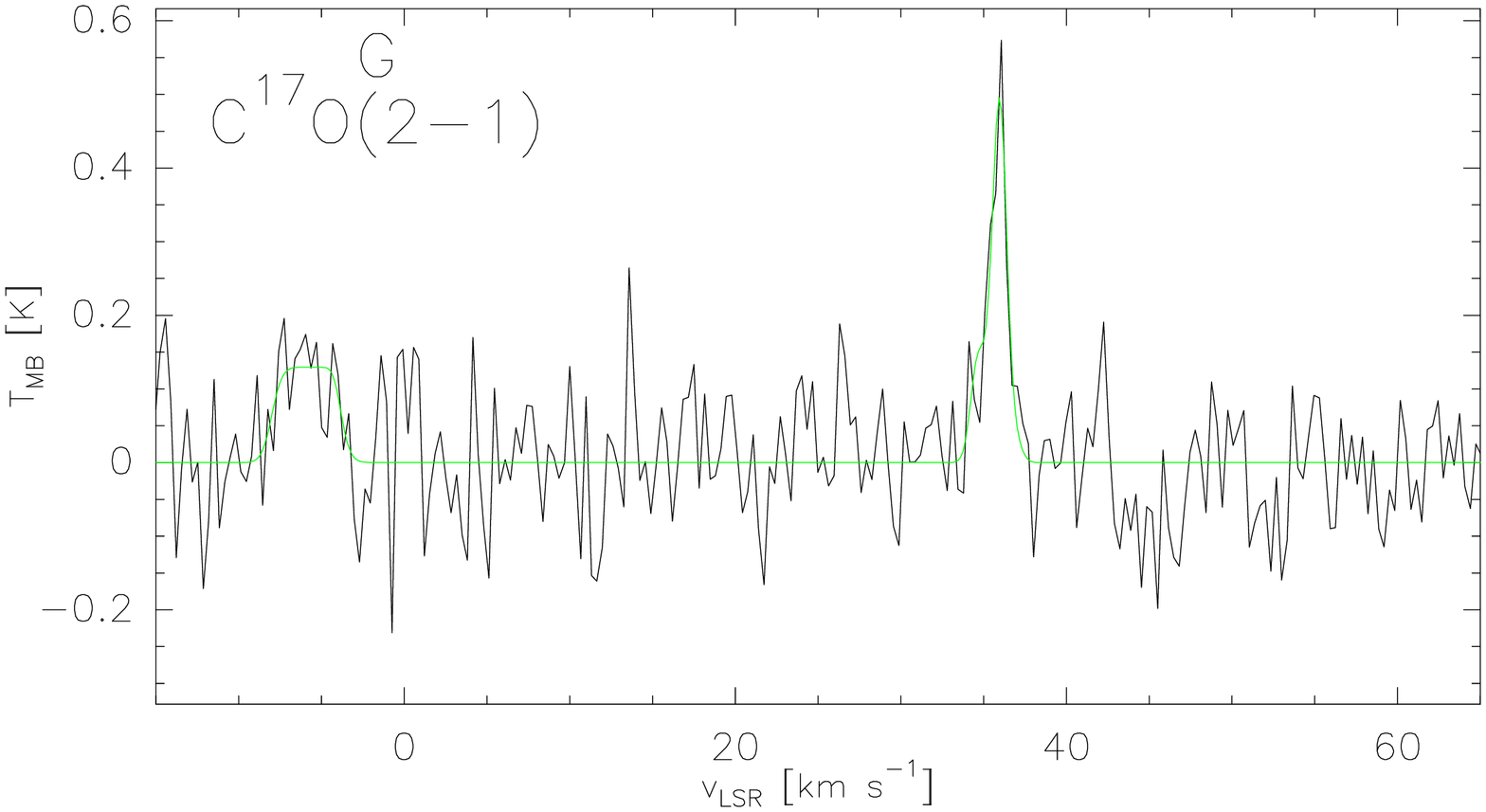}
\includegraphics[width=3.1cm, angle=-90]{blank.eps}
\includegraphics[width=3.1cm, angle=-90]{blank.eps}
\caption{Same as Fig.~\ref{figure:spectraG187} but towards the selected 
positions in G13.22-0.06. We note that clear line emission at $\sim37$ 
km~s$^{-1}$ is seen towards all positions. Two velocity components are 
detected towards positions A--C and G. }
\label{figure:spectraG1322}
\end{center}
\end{figure*}

As shown by the green plus signs in 
Figs.~\ref{figure:G187}--\ref{figure:G1322}, the selected C$^{17}$O$(2-1)$ 
observation target positions match very well the submm dust emission peaks 
only in G11.36. The target positions C and D in G1.87 are within the $3\sigma$ 
contour of SMM 35 (E is just outside). However, the target positions towards 
G2.11 do not have much, if any, associated dust emission (position D being an 
exception; it is coincident with SMM 2). Towards the G13.22 field the 
positions A, C, E, and F are rather a well matched with the peak positions of 
SMM 18, 22, 23, and 27, respectively. 
The obtained C$^{17}$O$(2-1)$ spectra towards all target positions are shown 
in Figs.~\ref{figure:spectraG187}--\ref{figure:spectraG1322}. Towards all 
fields, more than one velocity component is detected. This is unsurprising 
because we have observed along the Galactic midplane towards the inner Galaxy, 
where many molecular clouds along the line of sight can be expected. The 
fields G1.87, G2.11, and G13.22 show emission at both negative and positive 
LSR velocities. Inspecting the longitude-velocity maps of CO by Dame et al. 
(2001), this can be expected at the Galactic longitudes in question. We note 
that the critical density of the C$^{17}$O$(2-1)$ transition is $9.5\times10^3$ 
cm$^{-3}$ (assuming $T=15$ K and using the data from the LAMDA molecular 
database\footnote{{\tt http://www.strw.leidenuniv.nl/$\sim$moldata/}}; 
\cite{schoier2005}), and therefore the line emission originates in dense gas.

The C$^{17}$O$(2-1)$ line parameters are given in 
Table~\ref{table:lineparameters}. The LSR velocities and FWHM linewidths 
derived from hf-structure fits are given in Cols.~(2) and (3). The peak 
intensities derived through Gaussian fitting are listed in Col.~(6), and 
in Col.~(7) we give the integrated line intensities computed over the velocity 
range indicated in square brackets in the corresponding column. The 
uncertainties in the latter two parameters take into account the corresponding 
rms noise values and the 10\% calibration uncertainty. For non-detections, 
we provide the $3\sigma$ upper limit to the line intensity in Col.~(6). 
The other parameters listed in Table~\ref{table:lineparameters} are described 
later.

\begin{table*}
\caption{C$^{17}$O$(2-1)$ line parameters, column densities and fractional 
abundances, and CO depletion factors.}
{\scriptsize
\begin{minipage}{2\columnwidth}
\centering
\renewcommand{\footnoterule}{}
\label{table:lineparameters}
\begin{tabular}{c c c c c c c c c c}
\hline\hline 
Position\tablefootmark{a} & ${\rm v}_{\rm LSR}$ & $\Delta {\rm v}$ & $\sigma_{\rm NT}$ & $\frac{\sigma_{\rm NT}}{c_{\rm s}}$ & $T_{\rm MB}$ & $\int T_{\rm MB} {\rm dv}$\tablefootmark{b} & $N({\rm C^{17}O})$ & $x({\rm C^{17}O})$ & $f_{\rm D}({\rm CO})$\\
     & [km~s$^{-1}$] & [km~s$^{-1}$] & [km~s$^{-1}$] & & [K] & [K~km~s$^{-1}$] & [$10^{14}$ cm$^{-2}$] & [$10^{-8}$] &\\
\hline
{\bf G1.87-0.14} & \\
A \ldots & $-40.5\pm0.2$ & $1.36\pm0.58$ & $0.57\pm0.25$ & $2.5\pm1.1$ & $0.24\pm0.06$ & $0.63\pm0.13$ [-42.81, -37.90] & $3.4\pm0.7$ & $7.4\pm4.0$ & $5.0\pm2.7$ \\ 
B \ldots & \ldots & \ldots & \ldots & \ldots & $<0.26$ & \ldots & \ldots & \ldots & \ldots \\
C \ldots & $-41.2\pm0.4$ & $3.98\pm1.11$ & $1.69\pm0.47$ & $6.3\pm1.8$ & $0.20\pm0.06$ & $0.81\pm0.19$ [-45.15, -38.14] & $2.9\pm0.7$ & $3.1\pm1.0$ & $11.9\pm3.8$ \\
C(2nd) \ldots & $47.0\pm0.5$ & $2.23\pm0.61$ & $0.94\pm0.26$ & $3.5\pm1.0$ & $0.14\pm0.06$ & $0.70\pm0.18$ [42.58, 49.60] & $2.7\pm0.5$ & $2.9\pm0.8$ & \ldots \\
D \ldots & $-41.4\pm0.1$ & $0.55\pm0.44$ & $0.22\pm0.20$ & $0.8\pm0.7$ & $0.34\pm0.07$ & $0.65\pm0.11$ [-43.98, -40.01] & $2.8\pm0.5$ & $2.1\pm0.5$ & $17.6\pm4.2$ \\
E \ldots & $-41.6\pm0.2$ & $1.54\pm0.45$ & $0.65\pm0.19$ & $2.8\pm0.8$ & $0.30\pm0.07$ & $0.68\pm0.12$ [-44.22, -39.54] & $3.7\pm0.6$ & $2.1\pm0.5$ & $17.6\pm4.2$  \\
F \ldots & $-41.3\pm0.2$ & $0.81\pm0.76$ & $0.34\pm0.33$ & $1.5\pm1.4$ & $0.31\pm0.05$ & $0.43\pm0.10$ [-44.22, -40.94] & $2.3\pm0.5$ & $1.9\pm0.6$ & $19.4\pm6.1$ \\
F(2nd) \ldots & $61.5\pm0.3$ & $1.59\pm0.84$ & $0.67\pm0.36$ & $2.9\pm1.6$ & $0.18\pm0.05$ & $0.39\pm0.11$ [59.19, 63.17] & $2.1\pm0.6$ & $1.8\pm0.6$ & \ldots \\
G \ldots & $-41.9\pm0.1$ & $1.39\pm0.62$ & $0.59\pm0.26$ & $2.5\pm1.1$ & $0.28\pm0.08$ & $0.75\pm0.18$ [-45.39, -39.07] & $4.1\pm1.0$ & $4.6\pm1.7$ & $8.0\pm3.0$\\
{\bf G2.11+0.00} & \\
A \ldots & $16.8\pm0.1$ & $0.83\pm0.24$ & $0.35\pm0.10$ & $1.5\pm0.5$ & $0.35\pm0.04$ & $0.37\pm0.08$ [15.82, 18.16] & $2.0\pm0.5$ & $4.9\pm2.2$ & $5.5\pm2.5$\\
B \ldots & $16.5\pm0.3$ & $1.33\pm0.47$ & $0.56\pm0.20$ & $2.4\pm0.9$ & $0.19\pm0.07$ & $0.28\pm0.08$ [14.88, 18.16] & $1.5\pm0.5$ & $7.0\pm5.5$ & $3.8\pm3.0$\\
C \ldots & $15.7\pm0.3$ & $1.43\pm0.40$ & $0.60\pm0.17$ & $2.6\pm0.7$ & $0.10\pm0.06$ & $0.37\pm0.13$ [13.13, 17.92] & $2.0\pm0.7$ & $4.4\pm2.1$ & $6.1\pm2.9$\\
C(2nd) \ldots & $-11.5\pm0.1$ & $0.57\pm0.41$ & $0.23\pm0.18$ & $1.0\pm0.8$ & $0.25\pm0.04$ & $0.22\pm0.07$ [-12.43, -10.43; 95.1\%] & $1.2\pm0.4$ & $2.6\pm1.2$ & \ldots \\
D \ldots & \ldots & \ldots & \ldots & \ldots & $<0.26$ & \ldots & \ldots & \ldots & \ldots\\
E \ldots & $-19.4\pm0.3$ & $2.04\pm0.76$ & $0.86\pm0.32$ & $3.7\pm1.4$ & $0.16\pm0.07$ & $0.37\pm0.10$ [-20.91, -17.52] & $2.0\pm0.6$ & $6.2\pm3.4$ & $4.3\pm2.4$\\
F \ldots & \ldots & \ldots & \ldots & \ldots & $<0.27$ & \ldots & \ldots & \ldots & \ldots\\
G \ldots & \ldots & \ldots & \ldots & \ldots & $<0.23$ & \ldots & \ldots & \ldots & \ldots\\
H \ldots & $23.8\pm0.2$ & $0.85\pm0.42$ & $0.36\pm0.18$ & $1.5\pm0.8$ & $0.17\pm0.05$ & $0.25\pm0.09$ [22.02, 25.53] & $1.4\pm0.5$ & $1.3\pm0.5$ & $20.7\pm8.0$ \\
{\bf G11.36+0.80} & \\
A \ldots & $27.6\pm0.1$ & $1.93\pm0.34$ & $0.82\pm0.14$ & $3.5\pm0.6$ & $0.42\pm0.09$ & $1.05\pm0.12$ [24.39, 30.47] & $4.5\pm0.6$ & $3.1\pm0.5$ & $3.9\pm0.6$\\
A(2nd) \ldots & $56.4\pm0.2$ & $0.62\pm0.17$ & $0.25\pm0.08$ & $0.9\pm0.3$ & $0.16\pm0.05$ & $0.31\pm0.12$ [54.25, 58.31] & $1.7\pm0.6$ & $1.2\pm0.4$ & \ldots\\
B \ldots & $28.0\pm0.1$ & $1.06\pm0.12$ & $0.45\pm0.05$ & $1.9\pm0.2$ & $0.88\pm0.16$ & $2.04\pm0.23$ [25.09, 30.70] & $11.0\pm1.2$ & $7.3\pm1.1$ & $1.6\pm0.2$\\
B(2nd) \ldots & $55.8\pm0.2$ & $0.81\pm0.59$ & $0.34\pm0.26$ & $1.5\pm1.1$ & $0.18\pm0.04$ & $0.16\pm0.10$ [54.33, 57.37] & $0.9\pm0.5$ & $0.6\pm0.3$ & \ldots \\
C \ldots & $27.7\pm0.1$ & $2.01\pm0.21$ & $0.85\pm0.09$ & $3.7\pm0.4$ & $0.68\pm0.11$ & $1.73\pm0.21$ [23.73, 30.41] & $9.3\pm1.1$ & $4.8\pm0.8$ & $2.5\pm0.4$\\
D \ldots & $27.8\pm0.1$ & $1.95\pm0.21$ & $0.83\pm0.09$ & $3.6\pm0.4$ & $0.62\pm0.11$ & $1.55\pm0.19$ [24.62, 30.70] & $8.4\pm1.0$ & $5.6\pm0.9$ & $2.1\pm0.3$\\
E \ldots & $27.1\pm0.1$ & $1.18\pm0.25$ & $0.50\pm0.11$ & $2.2\pm0.5$ & $0.53\pm0.09$ & $1.11\pm0.15$ [24.39, 29.53] & $5.4\pm0.7$ & $3.8\pm0.7$ & $3.2\pm0.6$\\
F \ldots & $27.9\pm0.1$ & $1.50\pm0.45$ & $0.63\pm0.19$ & $2.7\pm0.8$ & $0.59\pm0.07$ & $1.32\pm0.17$ [25.56, 30.47] & $7.1\pm0.9$ & $12.5\pm3.2$ & $1.0\pm0.2$\\
F(2nd) \ldots & $83.7\pm0.1$ & $0.54\pm0.13$ & $0.22\pm0.06$ & $1.0\pm0.2$ & $0.43\pm0.06$ & $0.30\pm0.07$ [82.64, 84.98; 84.3\%] & $1.9\pm0.5$ & $3.3\pm1.1$ & \ldots \\
G \ldots & $28.6\pm0.1$ & $0.71\pm0.09$ & $0.29\pm0.04$ & $1.3\pm0.2$ & $0.64\pm0.15$ & $1.47\pm0.19$ [26.02, 31.17] & $6.0\pm0.8$ & $8.2\pm1.5$ & $1.5\pm0.3$\\
{\bf G13.22-0.06} & \\
A \ldots & $37.0\pm0.1$ & $1.06\pm0.17$ & $0.44\pm0.07$ & $1.7\pm0.3$ & $0.47\pm0.06$ & $0.69\pm0.11$ [34.45, 38.66] & $2.8\pm0.5$ & $1.4\pm0.3$ & $12.0\pm2.6$\\
A(2nd) \ldots & $15.1\pm0.1$ & $0.69\pm0.16$ & $0.29\pm0.07$ & $1.2\pm0.3$ & $0.35\pm0.05$ & $0.47\pm0.08$ [12.45, 16.90] & $1.9\pm0.3$ & $0.9\pm0.2$ & \ldots \\
B \ldots & $35.9\pm0.3$ & $1.11\pm1.08$ & $0.47\pm0.46$ & $2.0\pm2.0$ & $0.33\pm0.08$ & $0.56\pm0.10$ [33.74, 37.02] & $3.0\pm0.5$ & $1.5\pm0.4$ & $8.5\pm2.3$\\
B(2nd) \ldots & $54.0\pm0.3$ & $1.05\pm1.08$ & $0.44\pm0.46$ & $1.9\pm2.0$ & $0.48\pm0.09$ & $1.02\pm0.15$ [51.29, 56.91] & $5.5\pm0.8$ & $2.8\pm0.6$ & \ldots \\
C \ldots & $36.6\pm0.1$ & $1.71\pm0.30$ & $0.72\pm0.13$ & $3.1\pm0.6$ & $0.46\pm0.10$ & $1.07\pm0.16$ [33.51, 38.89] & $5.8\pm0.9$ & $3.4\pm0.8$ & $3.8\pm0.9$\\
C(2nd) \ldots & $54.9\pm0.2$ & $1.12\pm0.19$ & $0.47\pm0.08$ & $2.0\pm0.4$ & $0.27\pm0.09$ & $0.77\pm0.14$ [51.99, 56.67] & $4.2\pm0.8$ & $2.5\pm0.6$ & \ldots \\ 
D \ldots & $37.3\pm0.1$ & $1.17\pm0.11$ & $0.49\pm0.05$ & $2.1\pm0.2$ & $0.92\pm0.12$ & $1.69\pm0.20$ [34.68, 39.59] & $9.1\pm1.1$ & $12.3\pm4.1$ & $1.0\pm0.3$\\
E \ldots & $38.0\pm0.2$ & $2.58\pm0.66$ & $1.09\pm0.28$ & $4.1\pm1.1$ & $0.43\pm0.08$ & $1.41\pm0.22$ [34.91, 41.46] & $5.8\pm0.9$ & $3.9\pm0.8$ & $3.3\pm0.7$\\
F \ldots & $36.7\pm0.1$ & $1.47\pm0.29$ & $0.62\pm0.12$ & $2.3\pm0.5$ & $0.62\pm0.14$ & $1.88\pm0.23$ [33.51, 40.06] & $7.7\pm0.9$ & $1.3\pm0.2$ & $9.9\pm1.5$\\
G \ldots & $35.9\pm0.3$ & $0.87\pm1.08$ & $0.36\pm0.47$ & $1.6\pm2.0$ & $0.46\pm0.07$ & $0.75\pm0.12$ [33.98, 37.96] & $4.1\pm0.6$ & $4.0\pm1.2$ & $3.2\pm1.0$\\
G(2nd) \ldots & $-5.7\pm0.6$ & $2.06\pm1.31$ & $0.87\pm0.56$ & $3.8\pm2.4$ & $0.16\pm0.06$ & $0.57\pm0.15$ [-9.20, -3.11] & $3.1\pm0.8$ & $3.1\pm1.1$ & \ldots \\ 
\hline 
\end{tabular} 
\tablefoot{Columns~(2)--(10) of this table are as follows: (2) LSR velocity; 
(3) FWHM linewidth; (4) one-dimensional non-thermal velocity dispersion; (5) 
the ratio of the non-thermal velocity dispersion to the isothermal sound 
speed; (6) peak line intensity; (7) integrated intensity; (8) total C$^{17}$O 
column density; (9) fractional abundance of C$^{17}$O; (10) CO depletion 
factor (only towards the 'main' velocity components).\tablefoottext{a}{The 
additional velocity components are labeled with 
``(2nd)''.}\tablefoottext{b}{The velocity range used in the calculation is 
given in square brackets. The percentage value for G2.11-C(2nd) and 
G11.36-F(2nd) indicates the contribution of hf components' intensity lying 
within the detected line.}}
\end{minipage} }
\end{table*}

\subsection{Clump associations}

The \textit{Spitzer} 8- and 24-$\mu$m images were visually inspected to look 
for how the detected 870-$\mu$m clumps appear at these MIR wavelengths. The 
remarks concerning the 8/24 $\mu$m appearance of the clumps are given in the 
last column of Table~\ref{clumps}. From the fields G1.87, G2.11, G11.36, and 
G13.22, we found, respectively, 21, 5, 2, and 7 clumps that appear dark at 
both 8 and 24 $\mu$m (35 in total). The corresponding numbers of the clumps 
associated with either both 8 \textit{and} 24 $\mu$m emission, 
or only with 24 $\mu$m emission, are 16, 3, 5, and 27 (51 in total). 
The type of this MIR emission was found to be either point sources, group of 
point sources, extended, or diffuse-like. In addition, we found five clumps 
with an 8-$\mu$m point source near the submm peak position, but which appear 
dark at 24 $\mu$m. If the 8-$\mu$m source would be embedded within the clump, 
one would also expect to see emission at 24 $\mu$m (from warm dust). As noted 
in Table~\ref{clumps}, these 8-$\mu$m sources have 
\textit{Spitzer}-GLIMPSE $[3.6]-[4.5]$, $[4.5]-[5.8]$, and $[4.5]-[8.0]$ 
colours of $\simeq-0.03-0.80$, $0.16-0.78$, and $0.03-0.84$, 
respectively\footnote{From the GLIMPSE point source catalogue available at 
{\tt http://irsa.ipac.caltech.edu/}}. Because these colours are not ``red 
enough'' for sources to be protostellar in nature, they are likely to be chance 
projections of foreground stars where the emission is primarily photospheric 
(cf. \cite{gutermuth2008}; \cite{robitaille2008}). For sources near the 
Galactic plane and/or at large distances from the Sun, the foreground star 
population can be significant, and therefore five chance projections out of 91 
clumps (5.5\%) may not be surprising. We deal with these 
``foreground contaminated'' clumps as IR-dark, although in some studies such 
cases are excluded from the source sample [for example, this was the case in 
the study by Chambers et al. (2009), who called such sources ``blue cores'']. 
This makes the total number of IR-dark clumps in our survey to be 40.

We used the SIMBAD Astronomical 
Database\footnote{{\tt http://simbad.u-strasbg.fr/simbad/}} to search for 
possible source associations with our clumps. The resulting associations 
are given in Table~\ref{clumps}, where we list the sources within about one 
beam size ($\sim20\arcsec$) from the LABOCA peak position. 

In particular, 38 out of 91 clumps (42\%) were found to be associated with 
1.1-mm clumps from the Bolocam Galactic Plane Survey (BGPS; 
\cite{rosolowsky2010}; \cite{aguirre2011})\footnote{{\tt http://irsa.ipac.caltech.edu/data/BOLOCAM$_{-}$GPS/}}. The BGPS survey, with an effective FWHM 
resolution of $33\arcsec$, has detected and catalogued about 8\,400 clumps. 
Twenty-two clumps (24\%) are associated with SCUBA submm clumps from Di 
Francesco et al. (2008). The SCUBA maps of Di Francesco et al. (2008) are 
composed of a so-called Fundamental Map Data Set at 850 and 450 
$\mu$m (5\,061 objects), and an Extended Map Data Set at only 850 $\mu$m 
(6\,118 objects). Most of these associations (14 clumps) were found from the 
Extended Data Set (marked with ``JCMTSE''), and eight sources have counterparts 
in the Fundamental Data Set (marked with ``JCMTSF'').  Thirteen clumps, or 
14\% of the sources, were found to be associated with IRDCs identified by 
Peretto \& Fuller (2009)\footnote{The IRDC catalogue of Peretto \& Fuller 
(2009) is available at {\tt www.irdarkclouds.org/}}; these are marked with 
``SDC'' in Table~\ref{clumps}.

Some other associations worth mentioning here are as follows. Four clumps were 
found to be associated with \textit{IRAS} point sources: SMM 5 
in G2.11 with IRAS 17474-2704, and the clumps SMM 8, 13, and 29 in G13.22
with IRAS 18112-1720, 18114-1718, and 18117-1738, respectively. The clumps 
SMM 35 and 40 in G1.87, SMM 4 in G11.36, and SMM 5, 15, and 17 in 
G13.22 are associated with YSO candidates from Robitaille et al. (2008). 
Moreover, SMM 24 in G13.22 is associated with a candidate AGB star from 
Robitaille et al. (2008). Two MIR bubbles from the Churchwell et al. 
(2006) catalogue, namely N10 and N11, are associated with the concentration 
of several clumps in G13.22 (see Figs.~\ref{figure:G1322} and 
\ref{figure:bubbles}, and Sect.~5.1.2). 
Finally, we note that some of the clumps are associated with Class 
{\scriptsize II} methanol masers and UC H{\scriptsize II} regions, both of 
which are clear signposts of high-mass star formation. 
The diffuse/extended MIR emission seen towards some of the clumps 
is a typical characteristics of associated H{\scriptsize II} regions and 
photon-dominated regions (PDRs) surrounding them. 

In the cases of G2.11-SMM 9, G13.22-SMM 2, G13.22-SMM 9, and G13.22-SMM 11, 
the clump appears partly associated (in projection) with diffuse- and/or 
extended-like MIR emission, and partly IR dark. Especially in the N10/11 bubble 
environment, the clump classification into IR dark or IR bright was difficult 
because of the very bright and extended appearance of the region at 8 and 24 
$\mu$m. We note that classifying clumps in the above four cases is a 
subjective process, influenced by the adopted colour scale of the 8- and 
24-$\mu$m images. 

\section{Analysis and results}

\subsection{Kinematic distances}

The distance to the source is an important parameter when its physical 
properties, such as mass, are to be determined. Because our sources belong 
to the first quadrant ($0\degr \leq l < 90\degr$) in the inner Galaxy (i.e., 
interior to the solar circle), each radial velo\-city corresponds to two 
kinematic-distance values along the line of sight. However, in the first 
Galactic quadrant, the radial velocity of the source increases as a function 
of distance up to the tangent point. At this point, the source's velocity 
vector and the line of sight are aligned with each other, and the radial 
velocity has its maximum value. After passing the tangent point, the radial 
velocity starts to decrease as a function of distance, all the way down to 
negative values (\cite{romanduval2009}; Fig.~2 therein). In general, for 
sources associated with IRDCs, it is reasonable to assume, and often has 
been assumed, that they lie at the near distance, because in that case there 
is more background radiation against which to see the cloud in absorption.

To calculate the kinematic distances of G1.87, G2.11, and G11.36, we adopted 
the average C$^{17}$O$(2-1)$ radial velocities towards each field (using the 
velocities at which most lines were seen, and excluding the additional velocity 
components). The obtained average LSR velo\-cities for G1.87, G2.11, and 
G11.36 are, respectively, $-41.5$, 16.7 (from positions A and B), and 
27.8 km~s$^{-1}$. The clump SMM 5 (IRAS 17474-2704) in G2.11 is an exception. 
Although being near to our line observation positions in the plane of the sky, 
it has the peak Class {\scriptsize II} CH$_3$OH maser emission at 63 km~s$^{-1}$ 
(\cite{caswell1995}). The OH and H$_2$O masers towards this source also peak 
at comparable velocities (\cite{forster1989}). Therefore, for G2.11-SMM 5 
we adopted the velocity 63 km~s$^{-1}$. 

The average C$^{17}$O$(2-1)$ LSR velo\-city for G13.22 is 35.9 km~s$^{-1}$. 
However, from the HCO$^+$ and N$_2$H$^+$ survey of the BGPS 1.1-mm clumps by 
Schlingman et al. (2011), we obtained velo\-city information for the following 
clumps in G13.22: SMM 5, 10, 13, 17, 24, 25, 27, 28, 29, and 32. 
As shown in Fig.~\ref{figure:G1322}, the other clumps in G13.22 are seen in 
projection close to the above listed clumps. For example, the clumps near the 
bubbles N10/11, such as SMM 5 and 7, are likely 
physically connected, and SMM 31 is likely associated with SMM 28. 
Another example is SMM 26, which is likely a member of the filament connecting 
SMM 24 and 29. Therefore, for the field G13.22 we do not adopt our 
C$^{17}$O-derived radial velocity, but instead use the values from 
Schlingman et al. (2011). Sewilo et al. (2004), using the H$110\alpha$ and 
H$_2$CO observations, were able to distinguish between the near and far 
distances of G13.22-SMM 29 (IRAS 18117-1738); H$_2$CO absorption was seen 
between the source velocity and the velocity at the tangent point, placing 
the source at the far distance\footnote{We note that the H$110\alpha$ velocity 
of $40.6\pm3.8$ km~s$^{-1}$ is comparable to the HCO$^+$ velocity of 
44.64 km~s$^{-1}$ from Schlingman et al. (2011).}.

We employed the rotation curve of the Galaxy by Reid et al. (2009), 
which is based on direct measurements of trigonometric 
parallaxes and the proper motions of masers in high-mass star-forming regions. 
The best-fit rotation parameters of Reid et al. (2009) are 
($\Theta_0$, $R_0$)$=$(254 km~s$^{-1}$, 8.4 kpc), where $\Theta_0$ is the 
orbital velocity of the Sun around the Galactic Centre, and $R_0$ is the 
solar galactocentric distance. The resulting near and far kinematic distances, 
$d_{\rm near}$ and $d_{\rm far}$, and the Galactocentric distances, $R_{\rm GC}$, 
are given in Table~\ref{table:distances}. Unless otherwise stated, the near 
distance was adopted. We note that $R_{\rm GC}$ does not 
have a distance ambiguity. All the other clumps in G13.22 for which we adopted 
the distances derived using the data from Schlingman et al. 
(2011) are listed in Col.~(5) of Table~\ref{table:distances}. We note that our 
distances for the G13.22 clumps correspond to the values reported by 
Schlingman et al. (2011) who also used the Reid et al. (2009) rotation curve. 

As discussed above, the negative radial velocity of G1.87 places it at the far 
distance. In principle, it could also belong to the near 3-kpc arm at $d\sim5$ 
kpc (\cite{dame2008}; \cite{green2011}; \cite{tackenberg2012}), but the far 
solution is adopted in this work. Another noteworthy issue to raise is the 
fact that our line observations were only made towards the filamentary 
structures near the map centres, and therefore the derived radial velocities 
may not apply for all the clumps detected in the field. This may especially 
be the case towards G1.87, where many clumps are detected at low Galactic 
longitudes. Additional distance-uncertainty towards G1.87 is caused by the 
possible association of some of the clumps with the near 3-kpc arm.
On the other hand, the field G11.36, where only one velocity-coherent filament 
is detected, can be considered to have the most reliable distance estimate 
among our target fields. 

\begin{table*}
\renewcommand{\footnoterule}{}
\caption{Source distances\tablefootmark{a}.}
{\tiny
\begin{minipage}{2\columnwidth}
\centering
\label{table:distances}
\begin{tabular}{c c c c c}
\hline\hline 
Field/ & $d_{\rm near}$ & $d_{\rm far}$ & $R_{\rm GC}$ & Remark\tablefootmark{c}\\
clump\tablefootmark{b} & [kpc] & [kpc] & [kpc] & \\
\hline
{\bf G1.87-0.14} & ($6.22^{+0.38}_{-0.58}$)\tablefootmark{d} & $10.57^{+0.58}_{-0.38}$\tablefootmark{e} & 2.19 & $d_{\rm far}$ for the whole field\\
{\bf G2.11+0.00} & $5.51^{+0.63}_{-1.11}$ & $11.28^{+1.11}_{-0.63}$ & 2.90 & $d_{\rm near}$ for all clumps except SMM 5\\
SMM 5 & $7.40^{+0.10}_{-0.12}$ & $9.39^{+0.12}_{-0.10}$ & 1.04\\
{\bf G11.36+0.80} & $3.27^{+0.47}_{-0.56}$ & $13.20^{+0.56}_{-0.47}$ & 5.23 & \\
{\bf G13.22-0.06} & $3.54^{+0.39}_{-0.45}$ & $12.82^{+0.45}_{-0.39}$ & 5.02 & this $d_{\rm near}$ was not used for any of the clumps\\
SMM 5 & $4.24^{+0.31}_{-0.35}$ & $12.12^{+0.35}_{-0.31}$ & 4.38 & $d_{\rm near}$ also 
for SMM 2, 4, 6, 7, 9, 11, 12\\
SMM 10 & $4.31^{+0.30}_{-0.34}$ & $12.04^{+0.34}_{-0.30}$ & 4.32 & $d_{\rm near}$ also 
for SMM 1, 3, 8\\
SMM 13 & $1.95^{+0.61}_{-0.74}$ & $14.39^{+0.74}_{-0.61}$ & 6.52\\
SMM 17 & $3.56^{+0.39}_{-0.45}$ & $12.80^{+0.45}_{-0.39}$ & 5.00 & $d_{\rm near}$ also 
for SMM 15, 20\\
SMM 24 & ($3.98^{+0.34}_{-0.38}$) & $12.38^{+0.38}_{-0.34}$\tablefootmark{e} & 4.61 & $d_{\rm far}$ also for SMM 26, 33, 34 \\
SMM 25 & $4.47^{+0.28}_{-0.32}$ & $11.88^{+0.32}_{-0.28}$ & 4.17 & $d_{\rm near}$ also 
for SMM 18, 21 \\
SMM 27 & $3.56^{+0.39}_{-0.44}$ & $12.79^{+0.44}_{-0.39}$ & 5.00 & $d_{\rm near}$ also 
for SMM 22, 23 \\
SMM 28 & $4.46^{+0.28}_{-0.32}$ & $11.89^{+0.32}_{-0.28}$ & 4.19 & $d_{\rm near}$ also 
for SMM 14, 16, 19, 30, 31\\
SMM 29 & ($4.03^{+0.33}_{-0.38}$) & $12.33^{+0.38}_{-0.33}$\tablefootmark{e} & 4.57 & \\
SMM 32 & $4.41^{+0.29}_{-0.32}$ & $11.94^{+0.32}_{-0.29}$ & 4.23\\
\hline 
\end{tabular} 
\tablefoot{\tablefoottext{a}{The near and far kinematic distances 
were determined using the Reid et al. (2009) rotation curve of the Galaxy. 
The near distance was adopted unless otherwise stated.}\tablefoottext{b}{The 
fields G2.11 and G13.22 are known to contain sources at different distances 
than determined here from our C$^{17}$O radial velocities 
(see text).}\tablefoottext{c}{Other clumps for which the quoted distance was 
adopted.}\tablefoottext{d}{The negative LSR velocity of G1.87 places it at the far distance. However, it could also be a member of the near 3-kpc arm. The near distance derived here is comparable to the $\sim5$ kpc distance to the near 3-kpc arm.}\tablefoottext{e}{Far distance was adopted.}}
\end{minipage} }
\end{table*}

\subsection{Temperatures}

The dust and gas temperatures, $T_{\rm dust}$ and $T_{\rm kin}$, are also 
essential knowledge when studying the physics and chemistry of molecular 
clumps. To our knowledge, for only one clump in our sample, namely SMM 5 in 
G13.22, the gas kinetic temperature measurement has been published. 
Pillai et al. (2007) derived the NH$_3$ rotation temperature of 
$T_{\rm rot}=17.4\pm1.3$ K for SMM 5 (their source G13.18+0.06). Using the 
$T_{\rm rot}-T_{\rm kin}$ relationship from Tafalla et al. (2004), we obtain 
$T_{\rm kin}=20.5\pm1.7$ K\footnote{Pillai et al. (2007) only reported the 
$T_{\rm rot}({\rm NH_3})$ values. We note that the NH$_3$ radial velocity 
measured by Pillai et al. (2007) is in good agreement with the HCO$^+$ radial 
velocity from Schlingman et al. (2011).}. We note that 
at high densities of $n({\rm H_2})\gtrsim3\times10^4$ cm$^{-3}$, where 
collisional coupling between the gas and dust becomes efficient, the gas and 
dust temperatures are expected to be similar, $T_{\rm kin}\simeq T_{\rm dust}$ 
(e.g., \cite{galli2002}).

For those four clumps in our sample which are associated with 
\textit{IRAS} point sources, we estimated the dust temperature to be the same 
as the 60/100-$\mu$m colour temperature defined by Henning et al. (1990) as

\begin{equation}
T_{\rm dust}\simeq T_{\rm c}\left(\frac{60}{100}\right)=96\left[(3+\beta)\ln \left(\frac{60}{100}\right)-\ln \left(\frac{S_{60}}{S_{100}}\right) \right]^{-1} \,.
\end{equation}
In this formula, $\beta$ is the dust emissivity index, and $S_{\lambda}$ is the 
flux density at the wavelength $\lambda$. The value of $\beta$ was 
set to be 1.8 to be consistent with the Ossenkopf \& Henning (1994) dust 
model discussed in Sect.~4.3. For the \textit{IRAS} sources 17474-2704, 
18112-1720, 18114-1718, and 18117-1738, i.e., for the clumps G2.11-SMM 5, 
G13.22-SMM 8, G13.22-SMM 13, and G13.22-SMM 29, we derived the $T_{\rm dust}$ 
values of 30.0, 18.9, 35.5, and 18.9 K, respectively.

For the remaining 86 clumps the dust temperatures we assumed to be 
the following: 15 K for IR-dark clumps, 20 K for IR-bright clumps 
(8/24 $\mu$m emission), and 30 K for clumps associated with H{\scriptsize II} 
regions/radio sources. The choice of $T_{\rm dust}=15$ K or 20 K for most of our 
clumps is expected to be reasonable, because previous molecular-line 
observations of clumps within IRDCs have shown the typical gas kinetic 
temperature to lie in the range $T_{\rm kin}\approx10-20$ K (\cite{carey1998}; 
\cite{teyssier2002}; \cite{sridharan2005}; \cite{pillai2006}; 
\cite{sakai2008}, \cite{zhang2011}; \cite{devine2011}; \cite{ragan2011}). 
For comparison, in their study of a massive clump associated with an IRDC, 
Hennemann et al. (2009) derived the $T_{\rm dust}$ values of 22 K and 15 K for 
the clump's substructures ('cores') with and without 24-$\mu$m sources, 
respectively. Rathborne et al. (2010) also found that 24-$\mu$m bright 
clumps embedded in IRDCs are warmer than their 24-$\mu$m dark counterparts. 
The assumption of a slightly higher temperature of 30 K in clumps with 
embedded H{\scriptsize II} regions is supported by the 30-K temperature of 
the UC H{\scriptsize II} region G2.11-SMM 5 derived above, and also consistent 
with some other observational results (see, e.g., \cite{sreenilayam2011} and 
references therein).

Finally, we note that the BGPS 1.1-mm data available for 42\% of our clumps 
could, in principle, be used to estimate the dust colour temperature. However, 
there are three factors that would hamper this analysis: 
\textit{i)} the Bolocam 1.1-mm flux densities may be somewhat uncertain (due to 
spatial filtering and calibration issues), and should be multiplied by 
$1.5\pm0.15$ as recommended by Aguirre et al. (2011); \textit{ii)} one should 
make assumptions about the value of $\beta$ as above; and \textit{iii)} the 
LABOCA and Bolocam wavelengths are quite close to each other (0.87 mm vs. 1.1 
mm), and therefore the corresponding flux densities are comparable to each 
other. In addition, our LABOCA data should be smoothed to the $33\arcsec$ BGPS 
resolution for a proper comparison. For these reasons, it seems justified to 
assume that the $T_{\rm dust}$ va\-lues are similar to those observed in some 
other sources of similar type.

\subsection{Radii, masses, and H$_2$ column and number densities}

The linear clump effective radii (in pc) were derived from the angular radii 
and kinematic distances as 
$R_{\rm eff}[{\rm pc}]=R_{\rm eff}[{\rm rad}]\times d[{\rm pc}]$. The error in 
$R_{\rm eff}$ was computed from the average value between the $\pm$-distance 
errors quoted in Table~\ref{table:distances}.

The clump masses, $M$, over an effective area of radius $R_{\rm eff}$ were 
estimated from the integrated 870-$\mu$m flux densities and using the standard 
optically thin dust emission formulation [see, e.g., Eq.~(6) in 
\cite{miettinenharju2010}; hereafter MH10]. Following Hatchell et al. (2007), 
the peak surface brightness was used if it was higher than the integrated flux 
density (this was the case for G1.87-SMM 40, G2.11-SMM 2, G11.36-SMM 6, and 
G13.22-SMM 8; all but G2.11-SMM 2 are associated with previously known 
sources, and are therefore likely to be real). The distances and dust 
temperatures used were as explained above. The dust opacity per unit dust mass 
at 870 $\mu$m was taken to be $\kappa_{870}=1.38$ cm$^2$~g$^{-1}$. This value 
was extrapolated from the Ossenkopf \& Henning (1994) model describing 
graphite-silicate dust grains that have coagulated and accreted \textit{thin} 
ice mantles over a period of $10^5$ yr at a gas density
of $n_{\rm H}=n({\rm H})+2n({\rm H_2})\simeq 2n({\rm H_2})=10^5$ cm$^{-3}$.\footnote{We note that in the ATLASGAL study of Schuller et al. (2009), the value  
$\kappa_{870}=1.85$ cm$^2$~g$^{-1}$ was used.} In this model, the dust 
emissivity index is $\beta\simeq1.8$, as determined from the slope between 
350 and 1300 $\mu$m ($\kappa_{\lambda}\propto \lambda^{-\beta}$). For the average 
dust-to-gas mass ratio, 
$R_{\rm d} \equiv \langle M_{\rm dust}/M_{\rm gas} \rangle$, we adopted the 
canonical value $1/100$ (\cite{spitzer1978}). The uncertainty in mass was 
propagated from the uncertainties in flux density and source distance (the 
1.7-K temperature error was also employed in the case of G13.22-SMM 5).

The peak beam-averaged H$_2$ column densities, $N({\rm H_2})$, were computed 
from the peak surface brightnesses in a standard way [see, e.g., Eq.~(8) in 
MH10]. For this calculation we assumed a He/H abundance ratio 
of 0.1, which leads to the mean molecular weight per H$_2$ molecule of 
$\mu_{\rm H_2}=2.8$. Other parameters ($T_{\rm dust}$, $\kappa_{870}$, $R_{\rm d}$) 
were the same as used in the mass calculation. The error in $N({\rm H_2})$ 
is solely based on the uncertainty in the peak surface brightness (except in 
the case of G13.22-SMM 5, where $T_{\rm dust}$ error was also used). 
The volume-averaged H$_2$ number densities over $R_{\rm eff}$, 
$\langle n({\rm H_2}) \rangle$, were calculated using Eq.~(7) of MH10, and the 
corresponding errors were propagated from those of $M$ and $R_{\rm eff}$.

\begin{figure*}
\begin{center}
\includegraphics[scale=0.5]{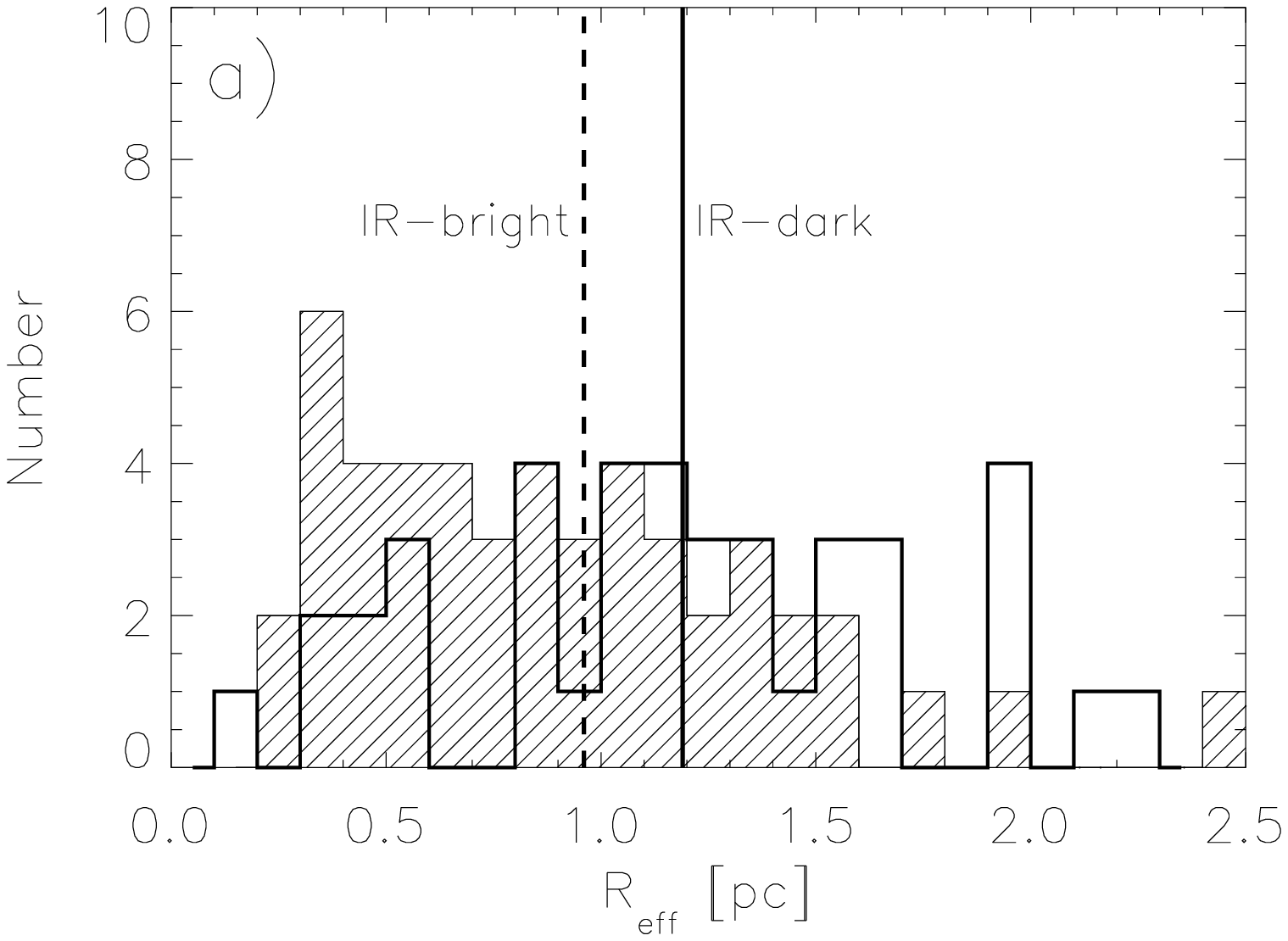}
\includegraphics[scale=0.5]{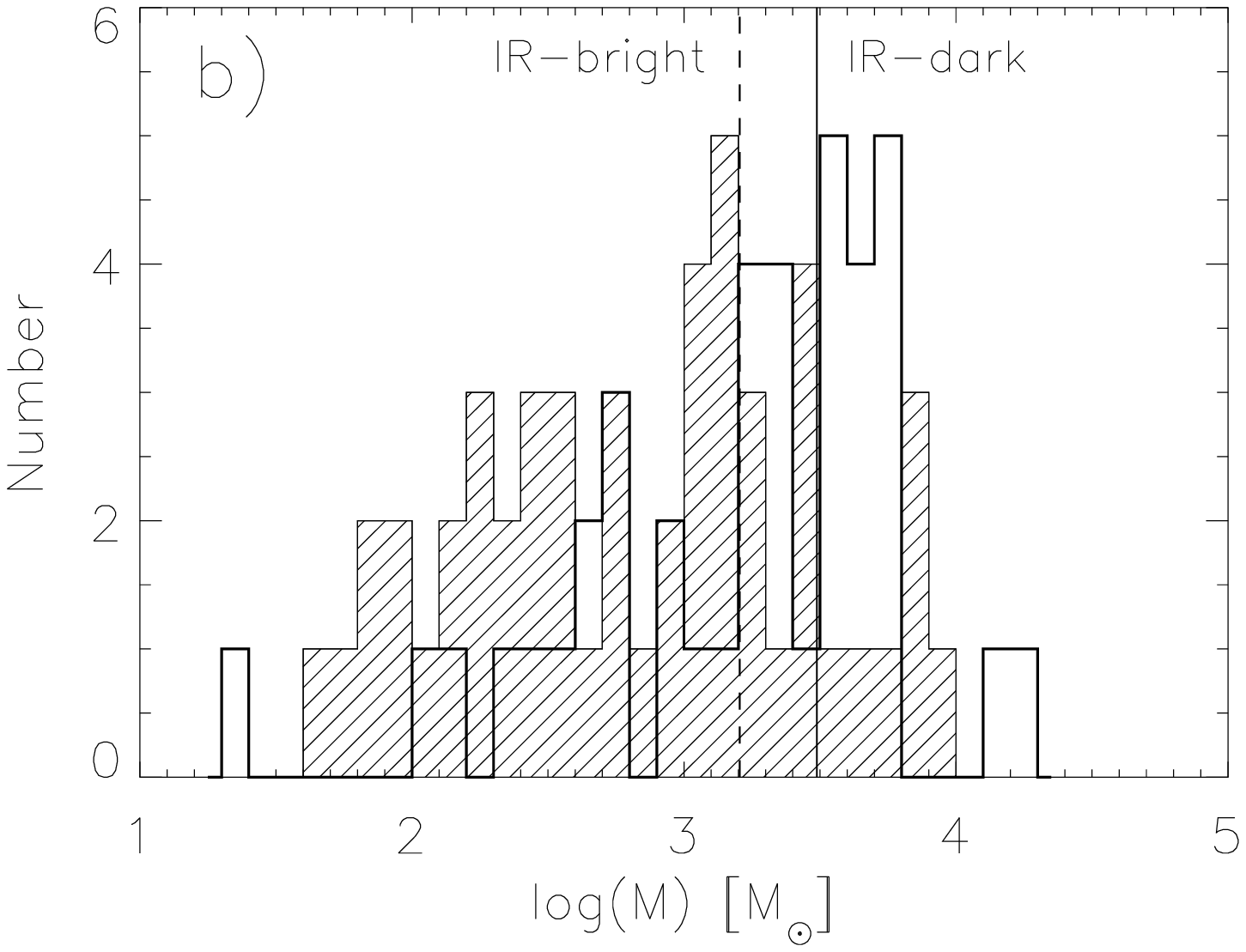}
\includegraphics[scale=0.5]{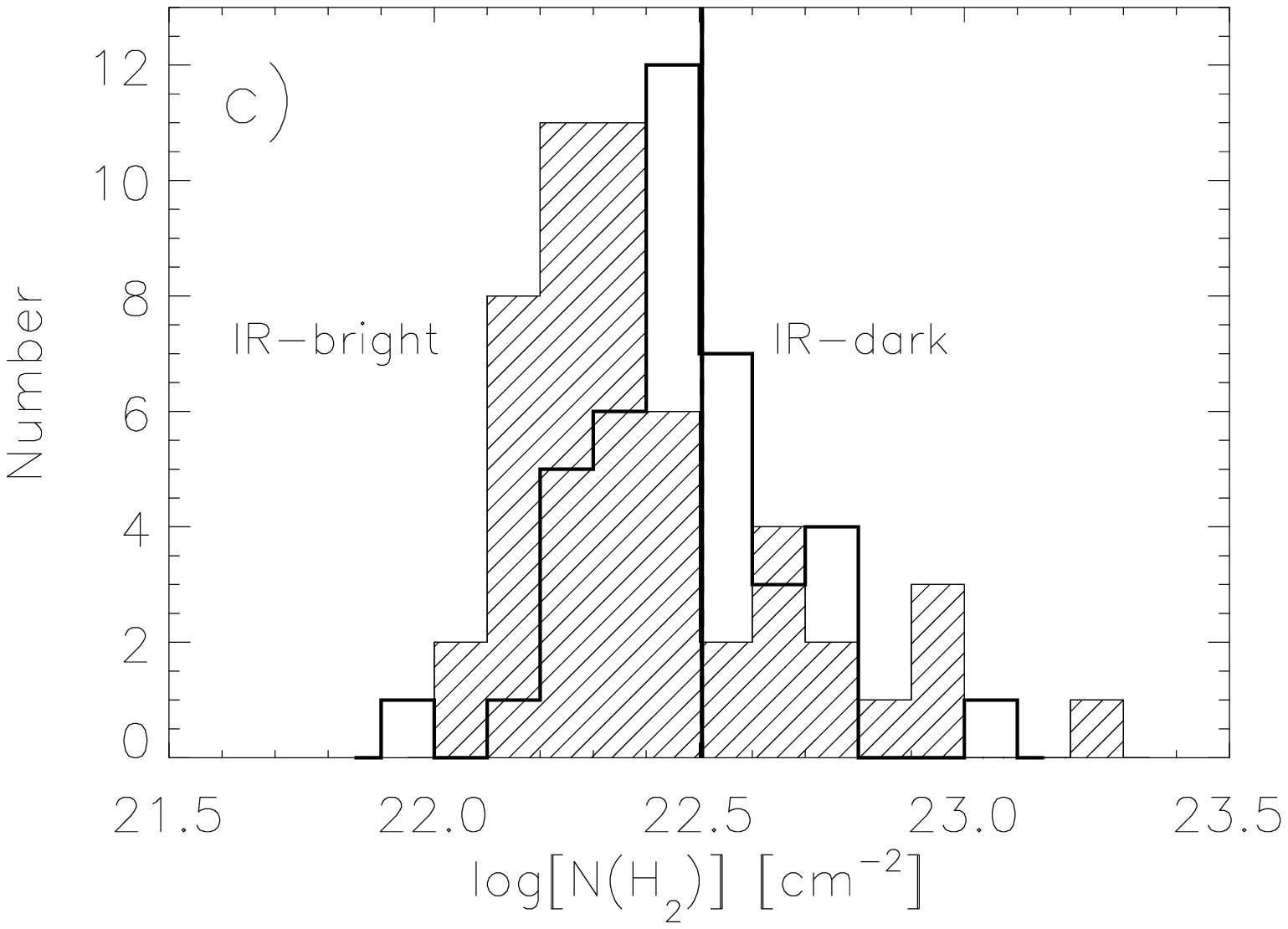}
\includegraphics[scale=0.5]{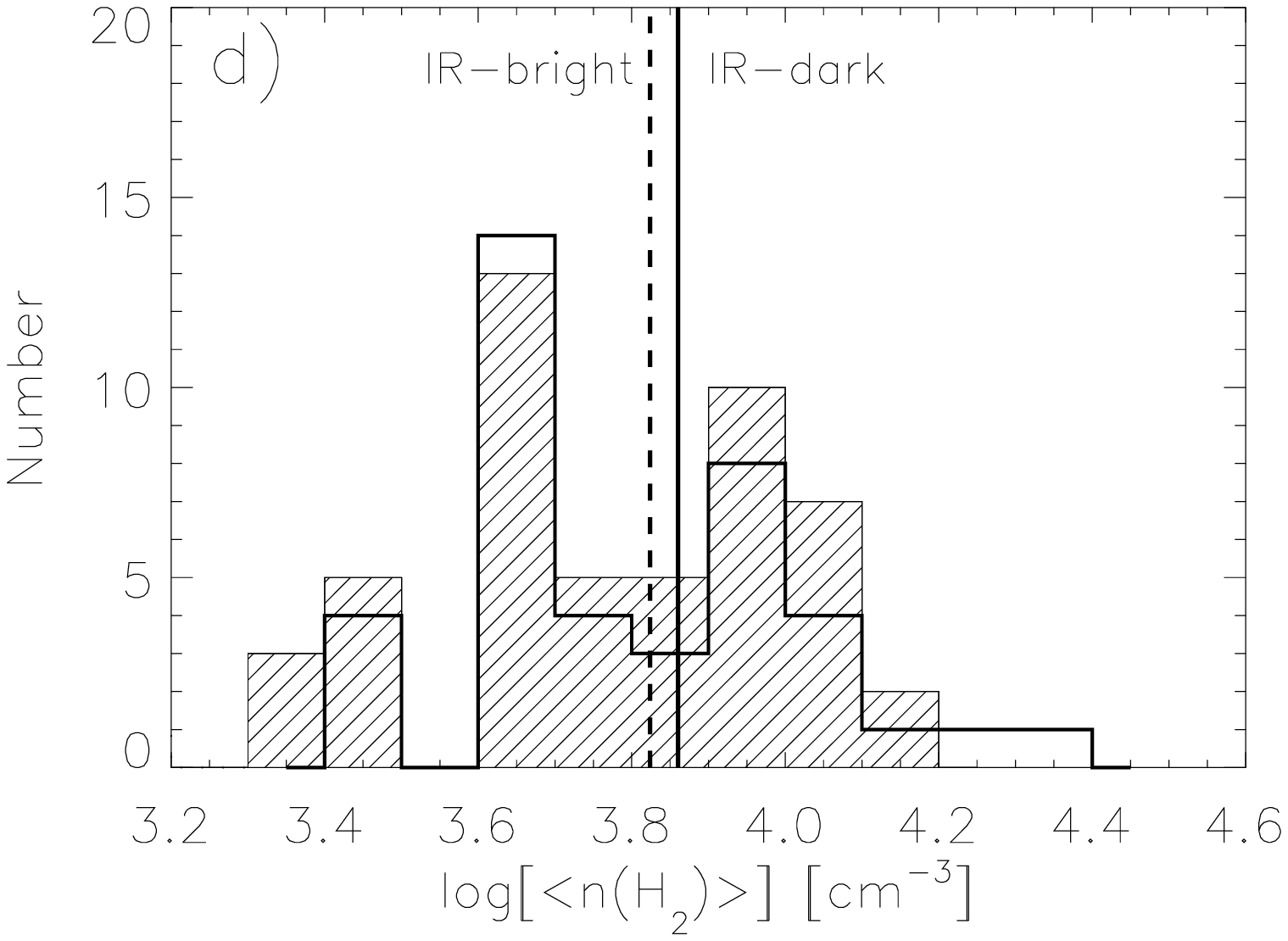}
\caption{Distributions of a) clump effective radii, b) masses, c) H$_2$ column 
densities, and d) H$_2$ number densities. Open histograms represent the IR-dark 
clumps, whereas shaded histograms represent the IR-bright clumps (those with 
8/24 $\mu$m emission). The solid and dashed vertical lines indicate the average 
values for IR-dark and IR-bright clumps, respectively. The average (median) 
radii for IR-dark and IR-bright clumps are 1.19 (1.14) and 0.96 (0.87) pc, 
respectively. The average (median) mass of IR-dark clumps is 3\,079 (2\,108) 
M$_{\sun}$, whereas that of IR-bright clumps is 1\,604 (804) M$_{\sun}$. The 
mean H$_2$ column density is $3.2\times10^{22}$ cm$^{-2}$ for both 
distributions (median values are $2.8\times10^{22}$ and $2.1\times10^{22}$ 
cm$^{-2}$ for IR-dark and IR-bright clumps, respectively). The average (median) 
H$_2$ number densities for IR-dark and IR-bright clumps are 7\,250 (6\,000) 
cm$^{-3}$ and 6\,667 (6\,000) cm$^{-3}$, respectively.}
\label{figure:histos}
\end{center}
\end{figure*}

The values of the physical parameters derived above are listed in 
Cols.~(6)--(9) of Table~\ref{clumps}. Their distributions are shown in 
Fig.~\ref{figure:histos}: panels a)--d) show the histograms of 
$R_{\rm eff}$, $M$, $N({\rm H_2})$, and $\langle n({\rm H_2}) \rangle$, 
respectively. The parameters' mean and median values are described in 
the figure caption.

\subsection{Non-thermal velocity dispersion}

The measured C$^{17}$O$(2-1)$ linewidths were used to calculate the non-thermal 
portion of the line-of-sight velocity dispersion, $\sigma_{\rm NT}$ 
(averaged over a $27\farcs8$ beam); see, e.g., Eq.~(1) in Miettinen (2012, 
hereafter M12). We also computed the ratio $\sigma_{\rm NT}/c_{\rm s}$, where 
$c_{\rm s}=\sqrt{k_{\rm B}T_{\rm kin}/\mu_{\rm p}m_{\rm H}}$ is the isothermal sound 
speed with $k_{\rm B}$ the Boltzmann constant, $\mu_{\rm p}=2.33$ the mean 
molecular weight per free particle for solar composition (He/H$=0.1$), and 
$m_{\rm H}$ the mass of the hydrogen atom. Some of our line observations are 
probing embedded YSOs, and $T_{\rm kin}=20$ K was assumed for these 
accordingly. For other positions, the value $T_{\rm kin}=15$ K was adopted. 
The values of $\sigma_{\rm NT}$ and $\sigma_{\rm NT}/c_{\rm s}$ are listed in 
Cols.~(4) and (5) of Table~\ref{table:lineparameters}. The uncertainties are 
based on the linewidth uncertainties. The vast majority of the positions 
show supersonic non-thermal motions ($\sigma_{\rm NT}>c_{\rm s}$), which 
are presumably due to turbulence. This is a general observational feature of 
the interstellar molecular clouds (e.g., \cite{larson1981}; \cite{mckee1992}; 
\cite{heyer2004}). 

\subsection{C$^{17}$O column densities, fractional abundances, and 
CO depletion factors}

The beam-averaged C$^{17}$O column densities, $N({\rm C^{17}O})$,  were derived 
following the standard LTE analysis outlined, e.g., in the paper by M12 
(Appendix~A.3 therein). In brief, we have assumed optically thin line 
emission, and computed the $N({\rm C^{17}O})$ values from the integrated line 
intensities. In the two cases where the detected line does not cover all the hf 
components (see Table~\ref{table:lineparameters}), the column densities were 
scaled by the inverse of the relative line strength within the detected line. 
The C$^{17}$O$(2-1)$ transition was assumed to be thermalised at the gas 
tempe\-rature of the target position as seen in previous studies (e.g., 
\cite{miettinen2011}). Therefore, the line excitation temperature was adopted 
to be $T_{\rm ex}=15$ K towards all the positions except those associated with 
MIR emission, where $T_{\rm ex}=20$ K was used. We note that if 
$T_{\rm ex}$ increases from 10 to 20 K, the column density decreases by a 
factor of 2.3 (or by a factor of 1.3 for $T_{\rm ex}$ of 15--20 K). The errors 
in $N({\rm C^{17}O})$ were propagated from those associated with the integrated 
intensity.

The fractional C$^{17}$O abundances were computed by
dividing the C$^{17}$O column density by the H$_2$ column density as 
$x({\rm C^{17}O})=N({\rm C^{17}O})/N({\rm H_2})$. For this purpose, the 
$N({\rm H_2})$ values were derived from the LABOCA maps smoothed to the 
$27\farcs8$ re\-solution of the line observations. The uncertainty in 
$x({\rm C^{17}O})$ was derived from both the errors in the C$^{17}$O and H$_2$ 
column densities.

The CO depletion factors, $f_{\rm D}$, were also derived following the 
analysis in M12 (see Appendix~A.4 and references therein). In summary, 
the Galactocentric distance of the source was first used to 
estimate the appropriate canonical (or undepleted) CO abundance. For example, 
the ``field'' $R_{\rm GC}$ values shown in Table~\ref{table:distances} lead to 
the CO abundances of $x({\rm CO})\simeq1.5-2.2\times10^{-4}$. We also employd 
the $R_{\rm GC}$-dependent $[{\rm ^{16}O}]/[{\rm ^{18}O}]$ 
ratio, which is $\simeq166-345$ for the ``field'' $R_{\rm GC}$ values, and 
adopted the $[{\rm ^{18}O}]/[{\rm ^{17}O}]$ ratio of 
3.52. This way, we estimated the canonical C$^{17}$O abundance and calculated 
the depletion factor as 
$f_{\rm D}=x({\rm C^{17}O})_{\rm can}/x({\rm C^{17}O})_{\rm obs}$.
The $f_{\rm D}$ uncertainty was propagated from that in the observed fractional 
abundance. The results of the calculations presented in this section are shown 
in Table~\ref{table:lineparameters} (the last three columns). We note that we 
only report the values of $f_{\rm D}$ for the ``main'' velocity components, 
because the line profiles in these cases are more reliable than those of the 
additional velocity components. We also stress that the formal 
$f_{\rm D}$ errors probably underestimate the true uncertainties by a factor of 
$\gtrsim2-3$ because of the uncertainties in the assumptions used 
($T_{\rm ex}$, oxygen-isotopic ratios, etc.). Therefore, some of the $f_{\rm D}$ 
values might be significantly lower than reported here. In this regard, the 
apparent variation of $f_{\rm D}$ between different sources might not be robust.

\subsection{Analysis of the filamentary IRDC G11.36+0.80}

The G11.36 cloud represents the best example of a filamentary IRDC in our 
survey. In addition, as demonstrated in Fig.~\ref{figure:G1136}, our C$^{17}$O 
line observations probe the clumps along the filament. For these reasons, we 
will analyse the filament's properties in more detail here. 

The north-south oriented G11.36 filament has a total projected length of 
about $4.05\arcmin$ or 3.85 pc, and its average radius is about 0.4 pc. 
By excluding the clumps SMM 6 and 7, which 
lie on the side of the filament, we can estimate the mass of the filament to 
be $\sim762$ M$_{\sun}$ as the sum of the clump masses within it. This makes 
the filament's mass per unit length, or line mass, to be $M_{\rm line}\sim198$ 
M$_{\sun}$~pc$^{-1}$. The average projected separation between the five clumps 
in the filament is about 0.9 pc. 

As shown in Col.~(5) of Table~\ref{table:lineparameters}, the filament appears 
to be dominated by supersonic non-thermal motions 
($\sigma_{\rm NT}/c_{\rm s}\simeq1.5-3.7$). Therefore, to examine the dynamical 
state of the fi\-lament, we calculate its virial mass per unit length as 
$M_{\rm line}^{\rm vir}=2 \langle \sigma^2\rangle /G$, where 
$\langle \sigma^2\rangle$ is the square of the total (thermal+non-thermal) 
velocity dispersion, i.e., the square of the effective sound speed 
($c_{\rm eff}$), and $G$ is the gravitational constant (\cite{fiege2000a}). 
Assuming that the gas kinetic temperature is $T_{\rm kin}=15$ K, and using 
the average non-thermal velocity dispersion of 0.69 km~s$^{-1}$, we obtain 
$c_{\rm eff}=0.73$ km~s$^{-1}$. Thus, we derive the values 
$M_{\rm line}^{\rm vir}\simeq246$ M$_{\sun}$~pc$^{-1}$ and 
$M_{\rm line}/M_{\rm line}^{\rm vir}\simeq0.8$. This implies that G11.36 as a 
whole is close to virial equili\-brium.

The magnetohydrodynamic ``sausage''-type instability theory predicts that 
growing perturbations can fragment a self-gravitating fluid cylinder into 
successive condensations with almost periodic separations (e.g., 
\cite{chandrasekhar1953}; \cite{nagasawa1987}; see also \cite{jackson2010}). 
This separation distance corresponds to the wavelength of the fastest growing 
mode, $\lambda_{\rm max}=2\pi/k_{\rm max}$ ($k_{\rm max}$ being the wavenumber), 
which, in turn, is twice the wavelength of the axisymmetric perturbations to 
which the cylinder is unstable. Jackson et al. (2010) and M12 applied 
the sausage-type instability predictions about clump separations to the 
filamentary IRDCs ``Nessie'' Nebula and G304.74+01.32, respectively. 
Both studies found that the observed clump separations agree 
with theoretical predictions when non-thermal (turbulent) motions are taken 
into account. In this case, the fastest growing mode in an isothermal, 
infinitely long gas cylinder appears at 

\begin{equation}
\lambda_{\rm max}\simeq22H_{\rm eff}=22\times \frac{c_{\rm eff}}{\sqrt{4\pi G \rho_0}}\simeq 6.2\times \frac{c_{\rm eff}}{\sqrt{G \rho_0}} \,,
\end{equation}
where $H_{\rm eff}$ is the effective radial scale height with $\rho_0$ the 
central gas-mass density along the cylinder's axis. If we compute $\rho_0$ 
assuming that the central number density is $10^5$ cm$^{-3}$, which seems 
reasonable because even the volume-average densities are $\sim10^4$ cm$^{-3}$ 
(Table~\ref{clumps}), we derive the values $H_{\rm eff}\simeq0.04$ pc and 
$\lambda_{\rm max}\simeq0.9$ pc. This is in excellent agreement with the 
observed average clump separation of $\sim0.9$ pc. The cloud may therefore lie 
close to the plane of the sky. Moreover, the 
wavelength of the fastest growing perturbation appears to be approximately 
equal to the filament's diameter of $\sim0.8$ pc, in accordance with theory 
(\cite{nakamura1993}).

In this theoretical framework, the clump masses should be less 
than $M\sim \lambda_{\rm max}M_{\rm line}$ (see \cite{jackson2010}). 
Using either the observed line mass or the $M_{\rm line}^{\rm vir}$ value, 
the predicted maximum mass is about 180 or 220 M$_{\sun}$, respectively. 
The estimated clump masses in the filament are in the range $\sim101-192$ 
M$_{\sun}$, which conforms to the theoretical expectation. 
The fragmentation timescale for a filament of radius $R_{\rm fil}$
is expected to be comparable to its radial signal crossing time,
$\tau_{\rm cross}=R_{\rm fil}/\langle \sigma^2\rangle^{1/2}$ [see Eq.~(26) in 
\cite{fiege2000b}]. For G11.36, this is estimated to be $\sim5.4\times10^5$ yr.

\subsection{Virial analysis of the clumps in G11.36+0.80}

\begin{figure}[!h]
\centering
\resizebox{0.8\hsize}{!}{\includegraphics[angle=0]{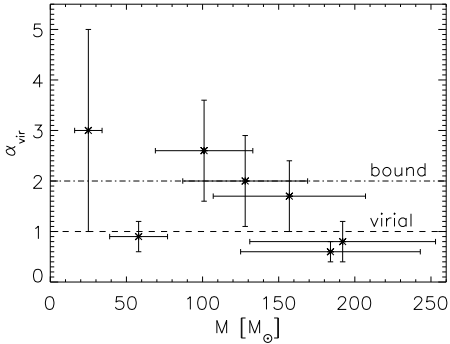}}
\caption{Virial parameter vs. mass for the clumps in G11.36. The dashed line 
indicates the virial-equilibrium limit of $\alpha_{\rm vir}=1$, and the 
dash-dotted line shows the limit of gravitational boundedness or 
$\alpha_{\rm vir}=2$.}
\label{figure:virial}
\end{figure}

We employed the C$^{17}$O$(2-1)$ linewidths and Eq.~(3) in M12 
to calculate the virial masses, $M_{\rm vir}$, of the clumps in G11.36. It was 
assumed that the clumps have a density profile of the form 
$n(r)\propto r^{-1.6}$ found by, e.g., Beuther et al. (2002) for high-mass 
star-forming clumps. The corresponding virial parameters were derived 
following the definition of Bertoldi \& McKee (1992), i.e., 
$\alpha_{\rm vir}=M_{\rm vir}/M$. If $\alpha_{\rm vir}=1$, the clump is in virial 
equilibrium, whereas the clumps with $\alpha_{\rm vir}\leq 2$ are taken to 
gravitationally bound. The derived values of $M_{\rm vir}$ and $\alpha_{\rm vir}$ 
are given in Table~\ref{table:virial}, and $\alpha_{\rm vir}$ is also plotted as 
a function of the clump mass in Fig.~\ref{figure:virial}. The error in 
$M_{\rm vir}$ was propagated from those associated with $\Delta {\rm v}$ and 
$R_{\rm eff}$, whereas the error of $\alpha_{\rm vir}$ includes the uncertainties 
in both the mass values $M_{\rm vir}$ and $M$. Although the uncertainties are 
large, it seems that five out of seven clumps in G11.36 are gravitationally 
bound. The corresponding ratio is 4/5 when considering only the clumps in 
the filament We note that the submm peak of SMM 4 was missed by the line 
observations, position D being closest to it. Therefore, the virial parameter 
of SMM 4 should be interpreted with some caution.  

\begin{table}
\renewcommand{\footnoterule}{}
\caption{Virial masses and parameters for the clumps in G11.36.}
\begin{minipage}{1\columnwidth}
\centering
\label{table:virial}
\begin{tabular}{c c c}
\hline\hline 
 Source\tablefootmark{a} & $M_{\rm vir}$ & $\alpha_{\rm vir}$ \\
 & [M$_{\sun}$] & \\
\hline
SMM 1(A) & $253\pm90$ & $2.0\pm0.9$ \\
SMM 2(C) & $259\pm66$ & $1.7\pm0.7$ \\
SMM 3(B) & $115\pm26$ & $0.6\pm0.2$ \\
SMM 4(D) & $258\pm64$ & $2.6\pm1.0$\\
SMM 5(E) & $150\pm55$ & $0.8\pm0.4$ \\
SMM 6(F) & $74\pm41$ & $3.0\pm2.0$ \\
SMM 7(G) & $49\pm10$ & $0.9\pm0.3$ \\
\hline 
\end{tabular} 
\tablefoot{\tablefoottext{a}{The letter in parenthesis indicates the line 
observation position as shown in Fig.~\ref{figure:G1136}. Note that position 
D is not very well coincident with the submm peak of SMM 4.}}
\end{minipage} 
\end{table}

\section{Discussion}

\subsection{On the nature of the detected clumps}

The sizes and masses of the clumps detected in this study indicate 
that they are the precursors/formation sites of stellar clusters and groups, 
rather than those of single stars or low-order multiples. 
Among our sample, 40 clumps are IR dark, and 51 clumps are IR bright, 
leading to the relative fractions of 44\% and 56\%, respectively. 
Some of the detected IR-dark clumps may represent the 
high-mass starless ``cores'' (HMSCs). However, it is fully possible that at 
least some of them do contain embedded YSOs, but which are not bright enough 
to be detected with the present sensitivity. The clumps associated with 
point-like MIR emission at 8 and 24 $\mu$m, or only at 24 $\mu$m, are likely 
to host YSOs. The only exceptions may be the cases where the clump is partly 
IR dark, and partly associated with extended MIR emission, possibly from 
the nearby sources as in the case of the N10/11 bubble environment. 
For example, star-formation activity within a clump, such as mass 
accretion, heats the surrounding dust causing it to emit the 24-$\mu$m IR 
radiation. Towards eight clumps, a group of 8-$\mu$m point 
sources is resolved. This indicates that the clumps were fragmented into 
smaller units, and the formation of a stellar group/cluster is taking place. 
For comparison, from their total sample of 190 clumps associated with IRDCs, 
Chambers et al. (2009) found that 98 (52\%) contain 24-$\mu$m sources, a 
percentage similar to ours. Parsons et al. (2009) found that from their 
(sub)sample of 69 clumps within IRDCs, 48 (70\%) were associated with embedded 
24-$\mu$m source(s). More recently, Tackenberg et al. (2012) found that only 
$\sim23\%$ of the clumps in their ATLASGAL-survey study showed no signs of 
IR emission.

\begin{figure}[!h]
\centering
\resizebox{0.8\hsize}{!}{\includegraphics[angle=0]{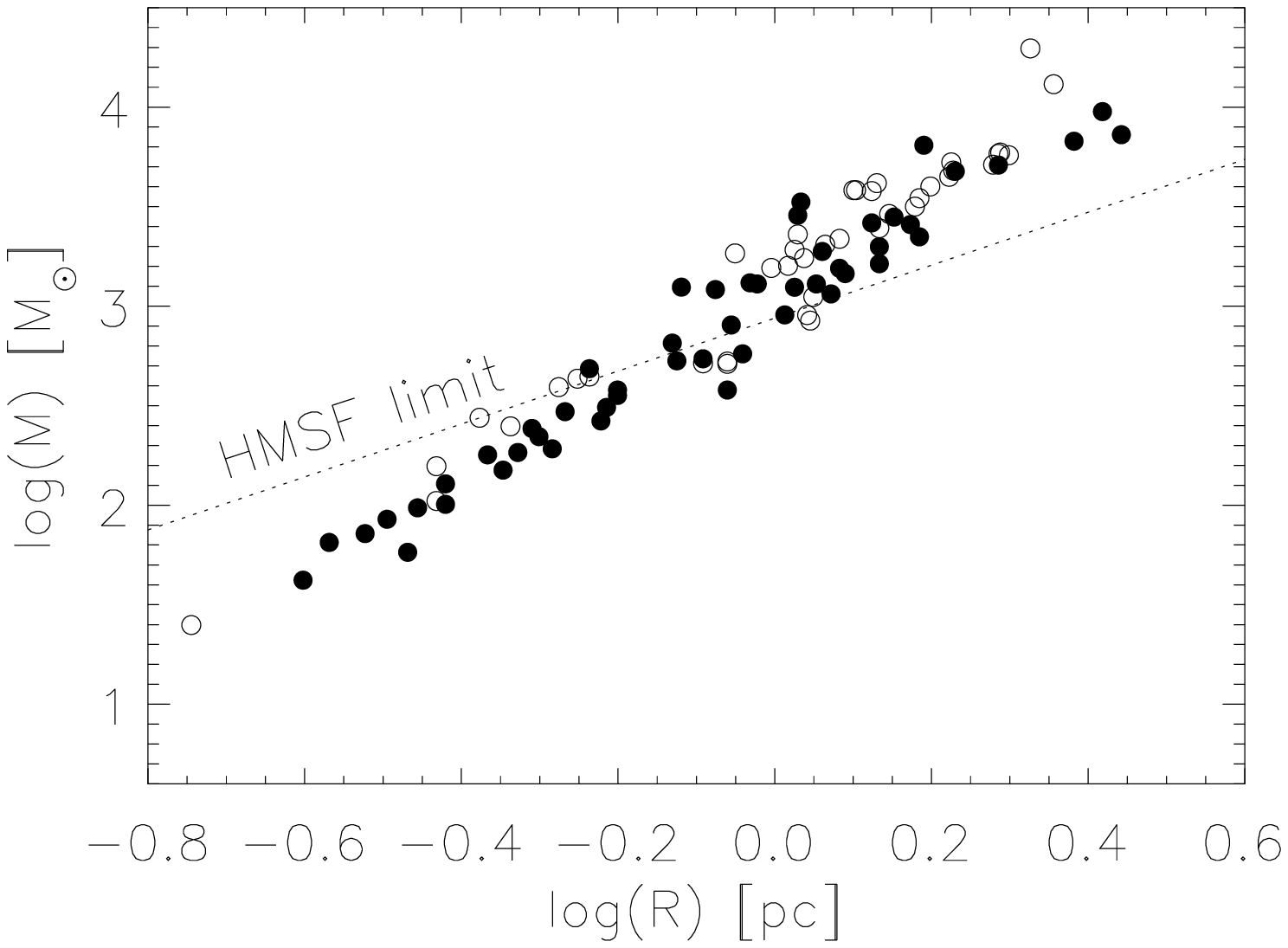}}
\caption{Relation between mass and effective radius for the detected clumps. 
The IR-dark and IR-bright clumps have been plotted with open and filled 
circles, respectively. The dotted line represents the mass-radius threshold 
for massive-star formation proposed by Kauffmann \& Pillai (2010), 
i.e., $M(R)=870$ M$_{\sun}\times(R/{\rm pc})^{1.33}$.}
\label{figure:massradius}
\end{figure}

The estimated masses of the detected IR-dark and IR-bright clumps lie in the 
ranges $\sim25-2\times10^4$ M$_{\sun}$ and $\sim42-9.5\times10^3$ M$_{\sun}$, 
respectively. The clump masses as a function of radius are plotted 
in Fig.~\ref{figure:massradius}. As can be seen, many of the clumps appear to 
lie above the mass-radius threshold for massive-star formation proposed by 
Kauffmann \& Pillai (2010). In particular, 31 IR-dark 
clumps ($\sim78\%$) lie on or above this threshold line, hence being 
potential sites of future high-mass star formation. For example, Beuther 
et al. (2011) estimated that the initial mass of a clump has to be 
$\sim850$ M$_{\sun}$ if it is to form at least one 20 M$_{\sun}$ star; 
similarly, a $\sim1\,900$ M$_{\sun}$ clump is needed to form a 
40 M$_{\sun}$ star. In this regard, our sample contains 28 (22) IR-dark 
clumps, which could be able to give birth to a 20 (40) M$_{\sun}$ star.  

Following Casoli et al. (1986; their Sect.~4.1), we can compute the IR 
luminosities, $L_{\rm IR}$, of the detected \textit{IRAS} sources using the 
sources' distances and flux densities at 12, 25, 60, and 100 $\mu$m. The 
obtained $L_{\rm IR}$ values of IRAS 17474-2704, 18112-1720, 18114-1718, and 
18117-1738 are $4.4\times10^4$, $1.3\times10^4$, $7.4\times10^3$, and 
$1.9\times10^5$ L$_{\sun}$, respectively. These should be interpreted as lower 
limits to the bolometric luminosity, $L_{\rm bol}$, because other wavelength 
data (e.g., near-IR and submm) were not employed. These high luminosities are 
indicative of massive-star formation in the cor\-responding clumps. 
The clump G2.11-SMM 5, associated with IRAS 17474-2704, also shows Class 
{\scriptsize II} CH$_3$OH maser emission at 6.7 GHz (\cite{caswell1995}). 
Because this maser transition results from radiatively pumped population 
inversion by IR emission from warm dust, it can only occur near MYSOs 
(e.g., \cite{cragg1992}; \cite{minier2003}; \cite{xu2008}).

Krumholz \& McKee (2008) showed that a clump's mass surface density needs to be 
$\Sigma \gtrsim 1$ g~cm$^{-2}$ if heating is to prevent clump fragmentation 
into smaller units, and hence, enable high-mass star formation. 
In terms of H$_2$ column density, this threshold is 
$N({\rm H_2})=\Sigma/\mu_{\rm H_2}m_{\rm H}\simeq2.1\times10^{23}$ cm$^{-2}$. 
The estimated beam-averaged $N({\rm H_2})$ values are all smaller than this, 
but the (possible) substructure within the clumps, or dense cores, can have 
much higher column densities. Also, as discussed above, some of the clumps 
already show clear signposts of high-mass star formation.

\subsubsection{Are there any EGOs ?}

\begin{figure*}
\begin{center}
\includegraphics[scale=0.4]{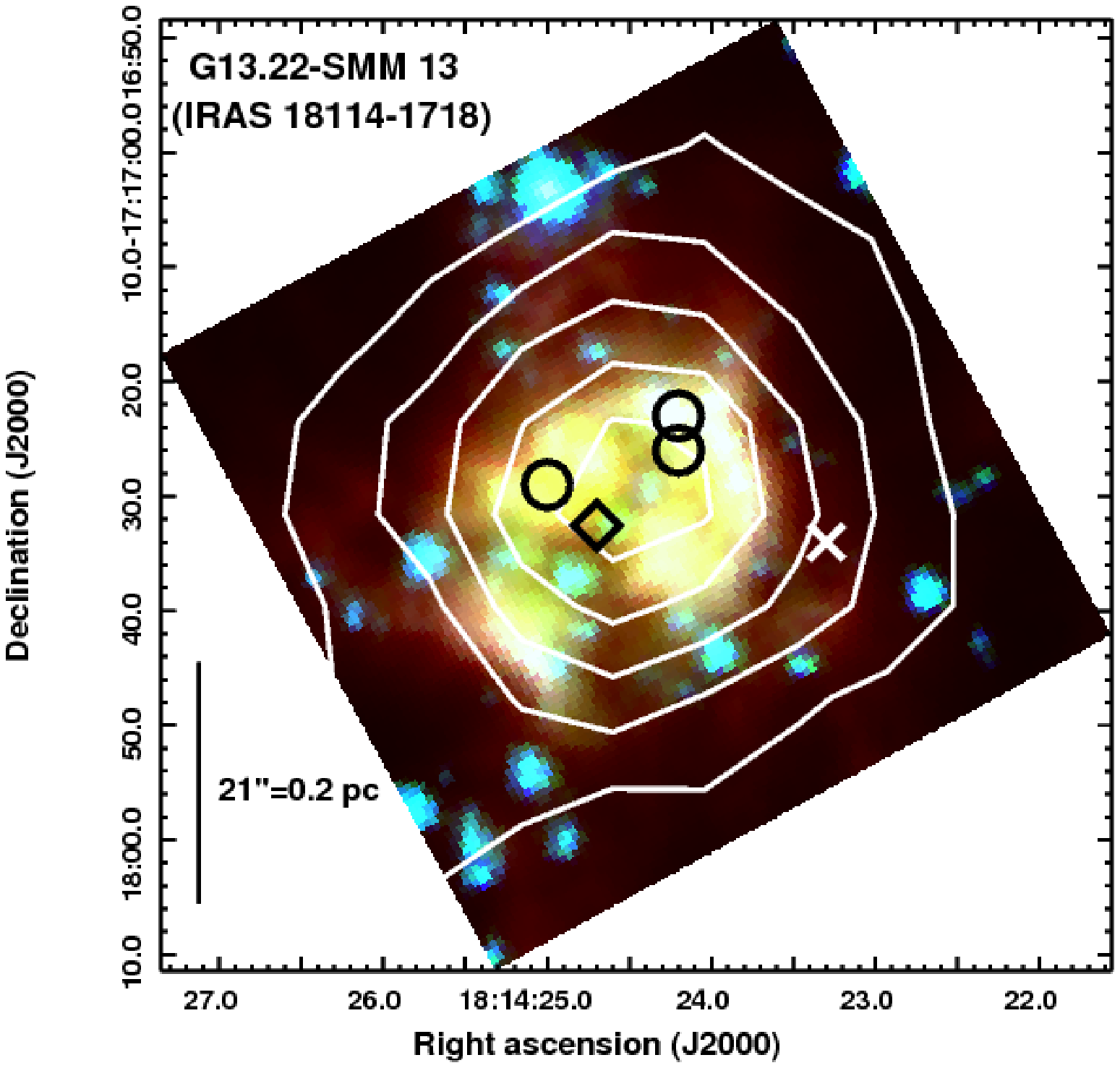}
\includegraphics[scale=0.45]{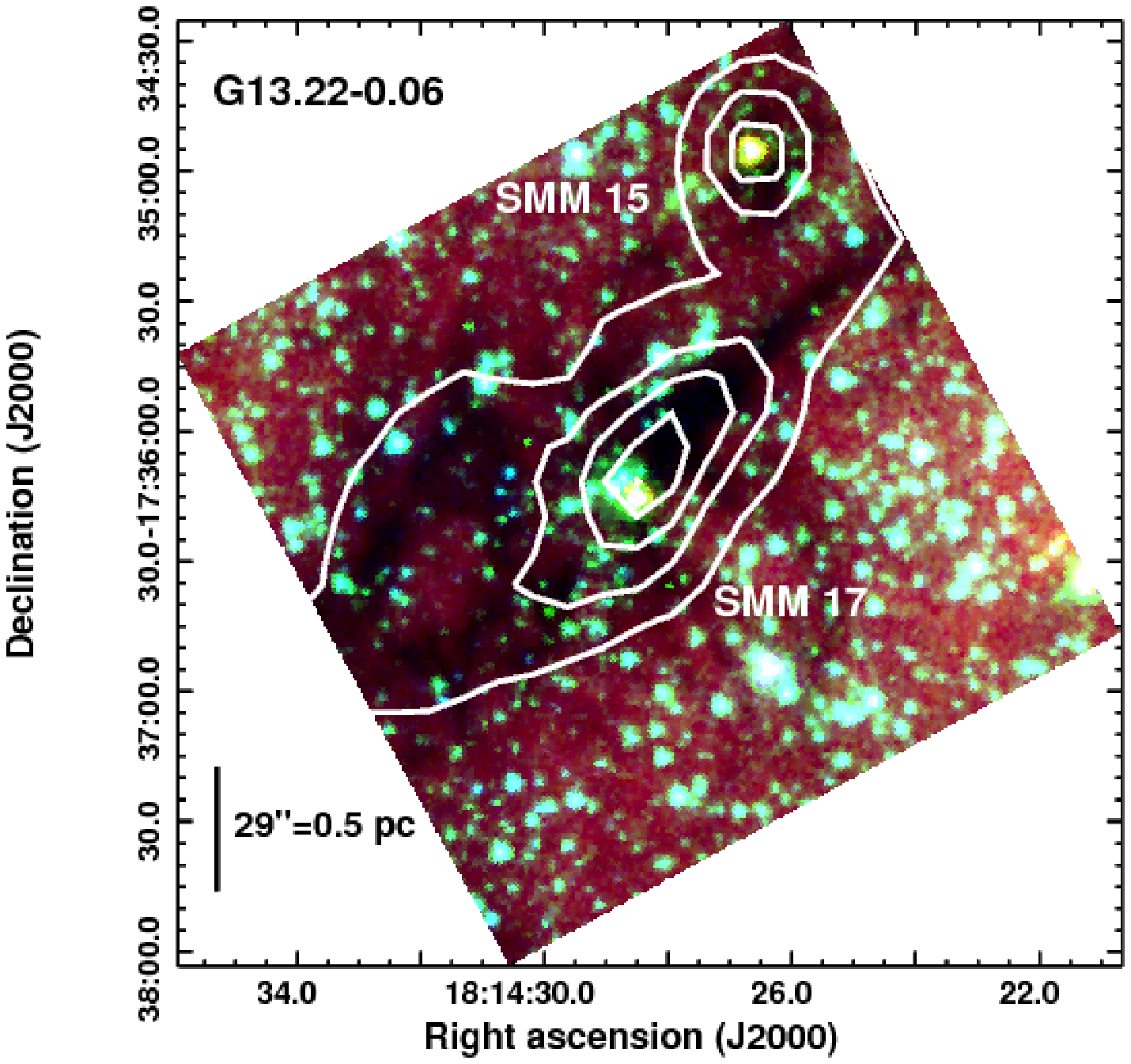}
\includegraphics[scale=0.45]{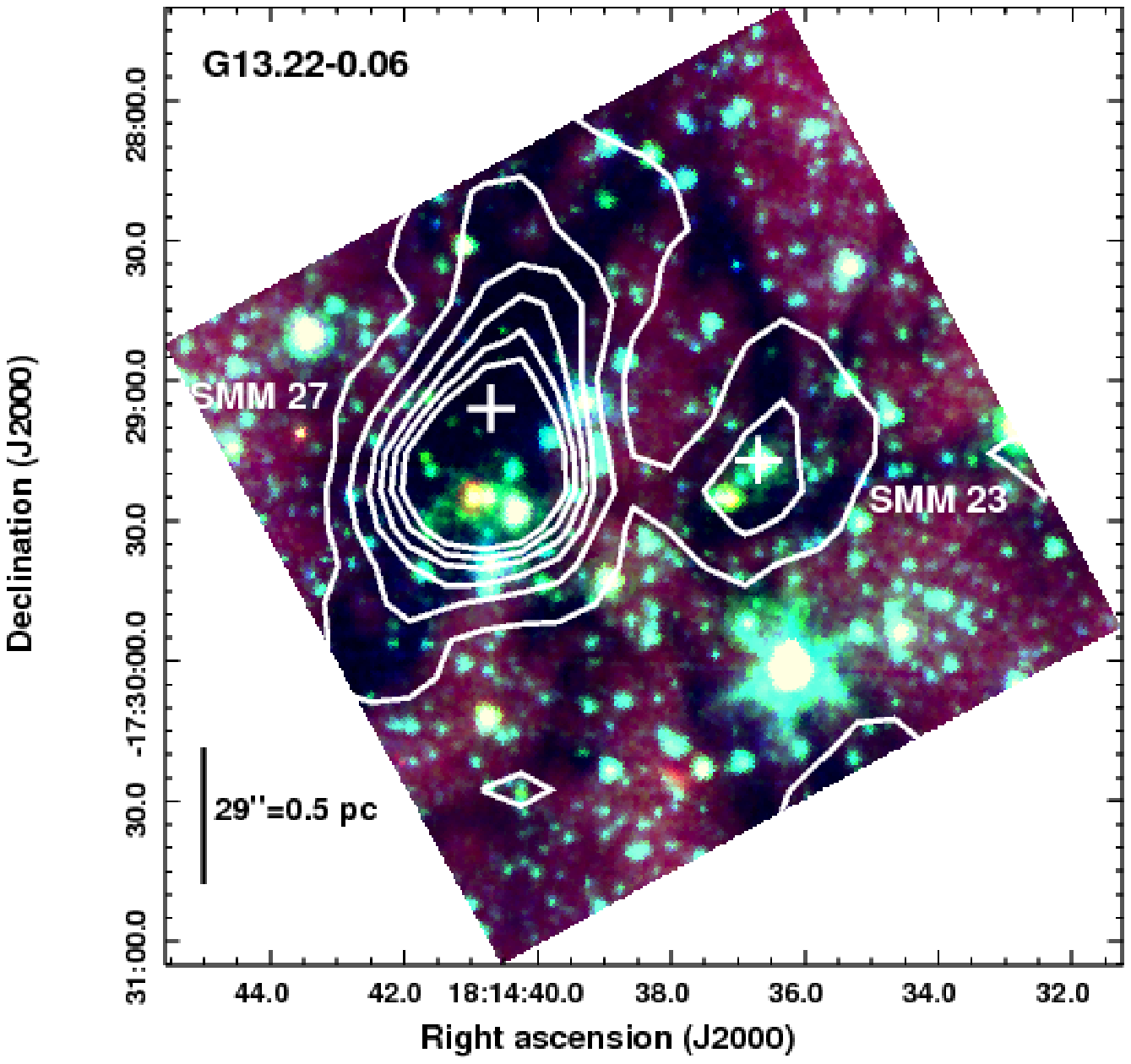}
\includegraphics[scale=0.4]{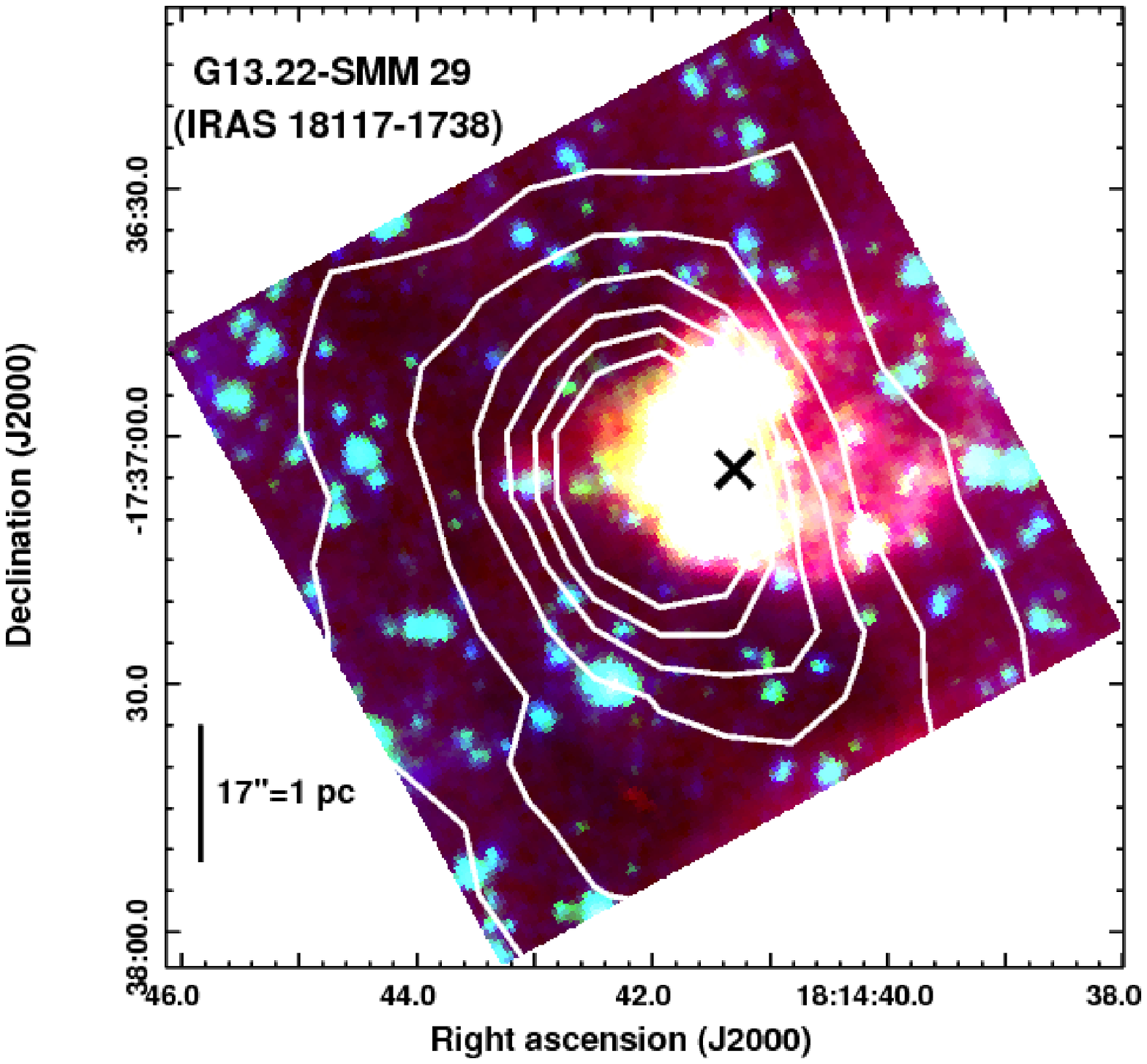}
\includegraphics[scale=0.5]{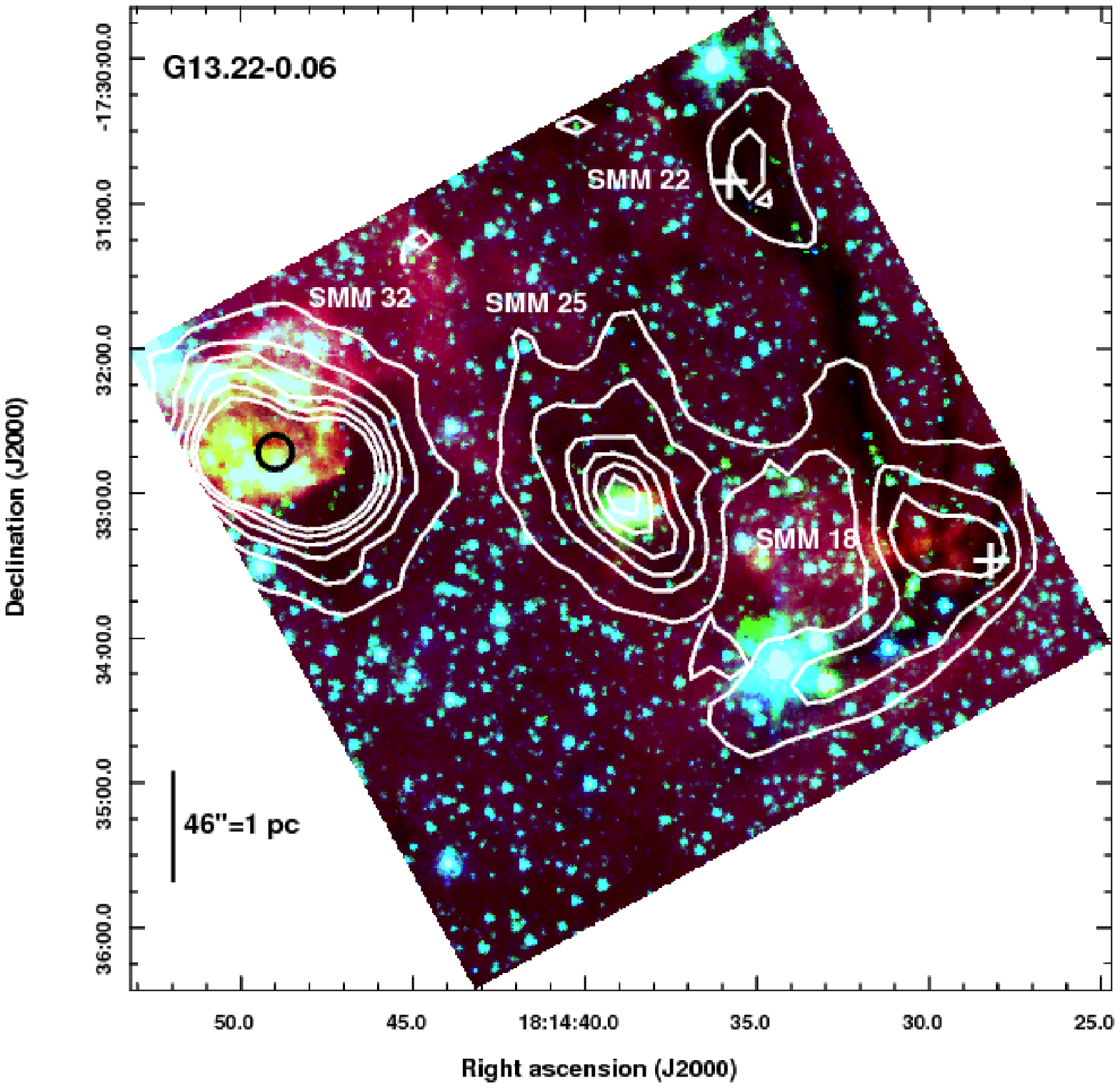}
\caption{\textit{Spitzer}-IRAC three-colour composite images towards EGO 
candidates in G13.22 overlaid with contours of LABOCA dust continuum emission. 
The 3.6, 4.5, and 8.0 $\mu$m emission is coded in blue, green, and red, 
respectively, and the colours are shown in linear scales. The contours 
are as in Fig.~\ref{figure:G1322}. Towards G13.22-SMM 13 (\textit{top left}), 
the black circles indicate the positions of the 
1.5-GHz VLA radio sources from Garwood et al. (1988), the diamond symbol 
shows the position of the 5-GHz VLA radio source from Becker et al. (1994), 
and the white cross shows the nominal catalogue position of IRAS 18114-1718. 
Towards G13.22-SMM 29 (\textit{middle right}), the position of IRAS 18117-1738 
is indicated by a black cross. The black circle towards G13.22-SMM 32 
(\textit{bottom}) denotes the 1.4-GHz radio source 181449-173243 from Condon 
et al. (1998). The white plus signs in the middle left and bottom panels 
represent the positions of our C$^{17}$O observations. We note that in the 
bottom panel the linear scale bar corresponds to the line-of-sight distance 
of SMM 25.}
\label{figure:EGOs}
\end{center}
\end{figure*}

A visual inspection of the \textit{Spitzer} 4.5-$\mu$m images revealed an 
extended-like emission associated with the clumps SMM 13, 17, 23, 25, 27, 29, 
and 32 in G13.22 (see Fig.~\ref{figure:EGOs}). These type of sources are known 
as Extended Green Objects (EGOs; \cite{cyganowski2008}) or ``green fuzzies'' 
(\cite{chambers2009}). The enhanced/extended 4.5-$\mu$m emission is believed to 
be mainly caused by shock-excited H$_2$ lines and/or ro-vibrational lines of 
CO$(\nu=1-0)$, implying the presence of outflow shocks in the source 
(\cite{marston2004}; \cite{smith2005}; \cite{smith2006}; \cite{yabarra2009}; 
\cite{debuizer2010}). However, Takami et al. (2012) discussed the possibility 
that the dominant emission mechanism responsible for EGOs might be scattered 
continuum in outflow cavities. In any case, EGOs appear to be related to 
outflow activity. Finding EGOs among our sources is unsurprising, because 
they often appear to be associated with IRDCs and 6.7-GHz Class 
{\scriptsize II} CH$_3$OH masers 
(\cite{chen2010}).

\subsubsection{The bubbles N10 and N11}

\begin{figure*}
\begin{center}
\includegraphics[scale=0.6]{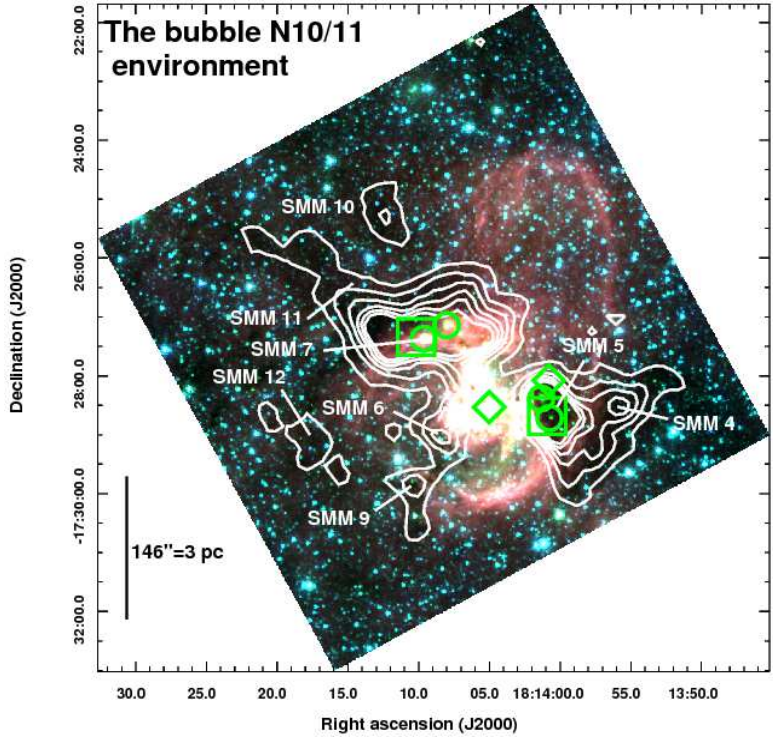}
\caption{\textit{Spitzer}-IRAC three-colour composite image towards the 
bubbles N10 and N11 overlaid with contours of LABOCA dust continuum emission. 
The 3.6, 4.5, and 8.0 $\mu$m emission is coded in blue, green, and red, 
respectively, and the colours are shown in linear scales. The contours are 
as in Fig.~\ref{figure:G1322}. The squares show the positions of the two 
870-$\mu$m condensations found by Deharveng et al. (2010); their 
condensation 1 is associated with our clump SMM 5, and condensation 2 is 
associated with SMM 7. The circles show the positions of the candidate 
embedded MYSOs on the rims of N10 (\cite{watson2008}). The diamond symbols 
indicate the positions of the 6.7-GHz Class {\scriptsize II} methanol masers 
from Szymczak et al. (2002; right) and Pandian et al. (2008; left).}
\label{figure:bubbles}
\end{center}
\end{figure*}

A zoom-in view towards the N10/11 bubble environment in G13.22 is shown in 
Fig.~\ref{figure:bubbles}. Our LABOCA clumps SMM 4, 5, 7, and 11 appear 
to form a $7\farcm4$ or $\sim9.1$ pc long ``ridge'' between these two IR 
bubbles from Churchwell et al. (2006). Moreover, the clumps SMM 6 and 9 
form a filamentary structure, extending perpendicularly to the southeast 
from the above mentioned ridge. 

Churchwell et al. (2006) classified N10 and N11 as complete (or closed) 
bubbles, with N10 enclosing a star cluster (see also \cite{watson2008}; their 
Fig.~14). Moreover, the N10/11 
system was classified as a bipolar bubble whose lobes are in contact. 
Both bubbles are coincident with H{\scriptsize II} regions 
(\cite{deharveng2010} and references therein), and Watson et al. (2008) 
identified four possible ionising stars located inside N10 in projection.
Watson et al. (2008) also identified four embedded candidate MYSOs on the rims 
of N10, and suggested that the bubble could be associated with triggered 
massive-star formation. The elongated dust emission morphology observed here 
is perhaps consistent with this scenario.
The third observation noted by Watson et al. (2008) was that inside 
the 8-$\mu$m shell of N10, emission at both 24 $\mu$m and 20 cm peak at the 
same position, implying the presence of hot dust inside the H{\scriptsize II} 
region.

Deharveng et al. (2010) reported the detection of two LABOCA condensations 
on the border of N10. As shown in Fig.~\ref{figure:bubbles}, the two 
condensations are coincident with our clumps SMM 5 and 7. 
As discussed by these authors, a Class {\scriptsize II} 
methanol maser associated with the edge of SMM 5 (their condensation 1) 
supports the scenario of triggered massive-star formation. Another CH$_3$OH 
maser is seen close to the centre of N10 in projection (\cite{pandian2008}). 
Deharveng et al. (2010) also speculated that N10 could be in the process of 
opening, because of its elongated shape. They speculated that the observed 
bipolar morphology of the system might result from the expansion of an 
H{\scriptsize II} region simultaneously in two opposite directions through the 
edge of the cloud. The dust filament consisting of SMM 6 and 9 could perhaps 
be related to the bubble expansion. As already noted by Deharveng et al. 
(2010), N11 is not associated with significant LABOCA 870-$\mu$m emission.

\subsection{Lifetime of massive IR-dark clumps}

We can use the relative numbers of IR-dark and IR-bright clumps to estimate 
the statistical lifetime of the former stage. We avoid to use the term 
``starless'' clump here, because some of the IR-dark clumps may well harbour 
faint YSOs, which cannot be detected with the current detection limit. 

Following Chambers et al. (2009), we also adopt as a re\-presentative YSO 
lifetime the accretion timescale $\sim2\times10^5$ yr from Zinnecker \& Yorke 
(2007). If we further assume that \textit{i)} all our IR-bright clumps, 
including those which are only partly associated with diffuse-like emission, 
host YSOs, \textit{ii)} the star-formation rate is constant as a 
function of time, and \textit{iii)} the clump's lifetime does not depend on 
its mass, we estimate the duration of the IR-dark phase of clump evolution to 
be $\tau_{\rm IR-dark}\sim40/51\times 2\times10^5\sim1.6\times10^5$ 
yr. Conserning the third assumption here, Clark et al. (2007) noted that the 
clumps of different mass are expected to have different lifetimes, because 
their free-fall timescales, $\tau_{\rm ff}\propto 1/\sqrt{\rho}$, can be 
different. They pointed out that this timescale problem may not be a problem 
if the studied clumps have roughly the same density, i.e., the same 
$\tau_{\rm ff}$. The estimated average densities of our IR-dark and IR-bright 
clumps vary by factors of $\sim7$ and 15, respectively, so their evolutionary 
rates can also vary. This makes our third assumption rather a rough one. 

For comparison, Chambers et al. (2009), by considering only their quiescent 
(no embedded IR emission) and active (green fuzzies and 24 $\mu$m emission) 
sources (69 and 37 sources, respectively), estimated a factor of $\sim2$ 
longer duration for the quiescent phase, i.e., $\sim3.7\times10^5$ yr. 
However, by comparing the relative numbers of their quiescent
and 24-$\mu$m sources only ('intermediate cores'), $69/98\simeq0.7$, the 
quiescent-phase timescale becomes $\sim1.4\times10^5$ yr, which is very 
similar to our estimate. Parsons et al. (2009), adopting the embedded YSO 
duration of $10^4-10^5$ yr, deduced that the IR-dark phase of massive clumps 
lasts a few times $10^3-10^4$ yr. With the aid of SiO spectral-line data, 
Russeil et al. (2010) estimated that the combined lifetime of starless clumps 
(no IR emission/high-velocity SiO emission) and IR-quiet MYSOs is 
$\sim1\times10^4+6\times10^4=7\times10^4$ yr in the NGC 6334/6357 complex, 
which is a factor of $\sim2$ less than our $\tau_{\rm IR-dark}$ value. 
Tackenberg et al. (2012) inferred the lifetime of massive IR-dark ('starless' 
in their nomenclature) clumps to be $\sim(6\pm5)\times10^4$ yr, the upper 
limit being roughly comparable to our estimate. 
We also note that in their \textit{Herschel}/Hi-GAL study of IRDCs, 
Wilcock et al. (2012) inferred the lifetime $\sim2\times10^5$ yr for the 
IR-quiet phase, which is close to our estimate. We note that Wilcock et al. 
(2012) adopted the same YSO lifetime as we ( $\sim2\times10^5$ yr).

Inspecting the diverse and somewhat heterogeneous lifetime estimates for 
IR-dark clumps, it seems that the true value is somewhere between 
$\sim10^4-10^5$ yr, although values as low as $<10^3$ yr have also been 
proposed (\cite{motte2007}). Extremely short lifetime of massive IR-dark clumps 
would, however, mean that it would be rather unlikely to see them in a large 
number in observational surveys.

\subsection{Clump mass distribution}

\begin{figure*}
\begin{center}
\includegraphics[scale=0.3]{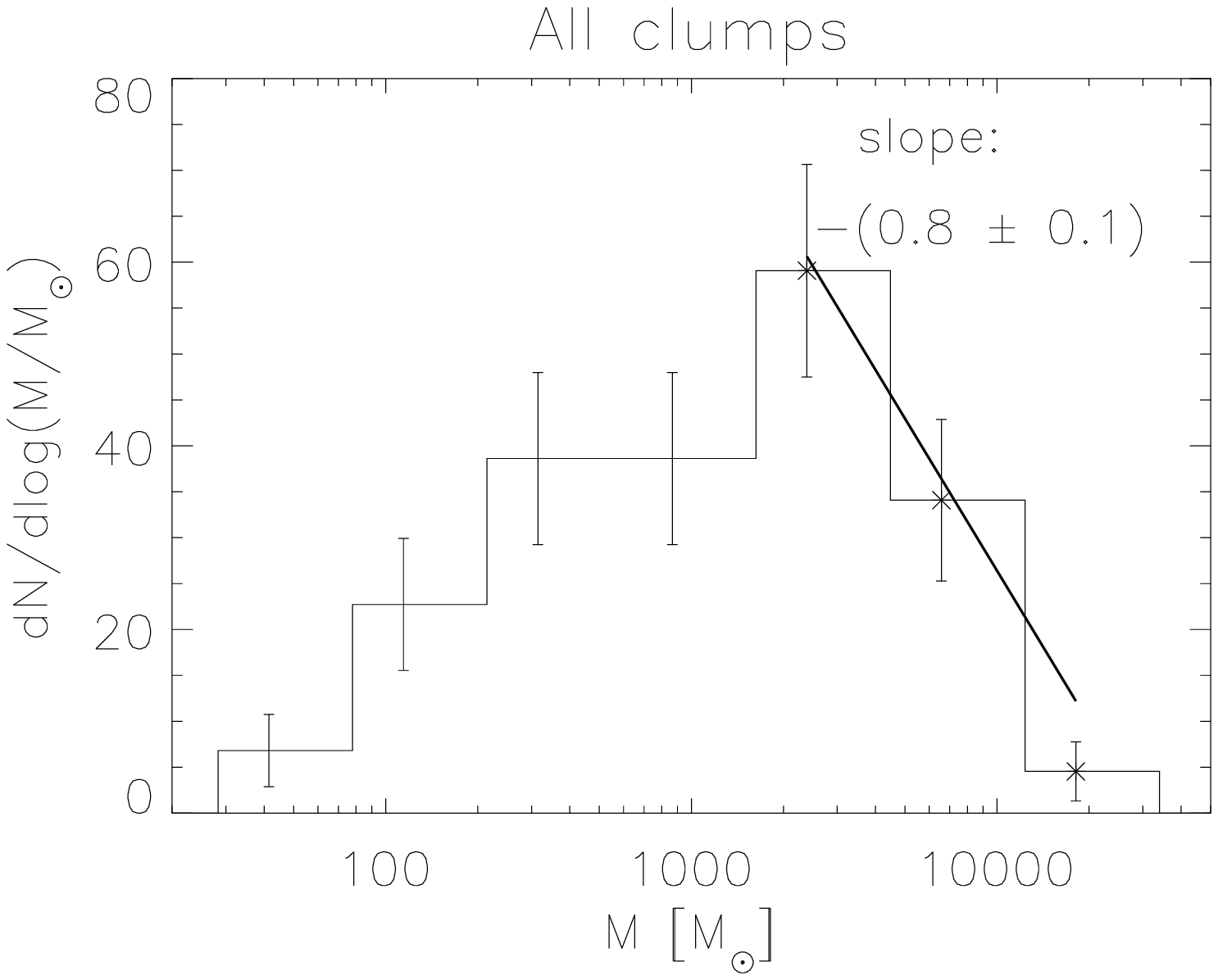}
\includegraphics[scale=0.3]{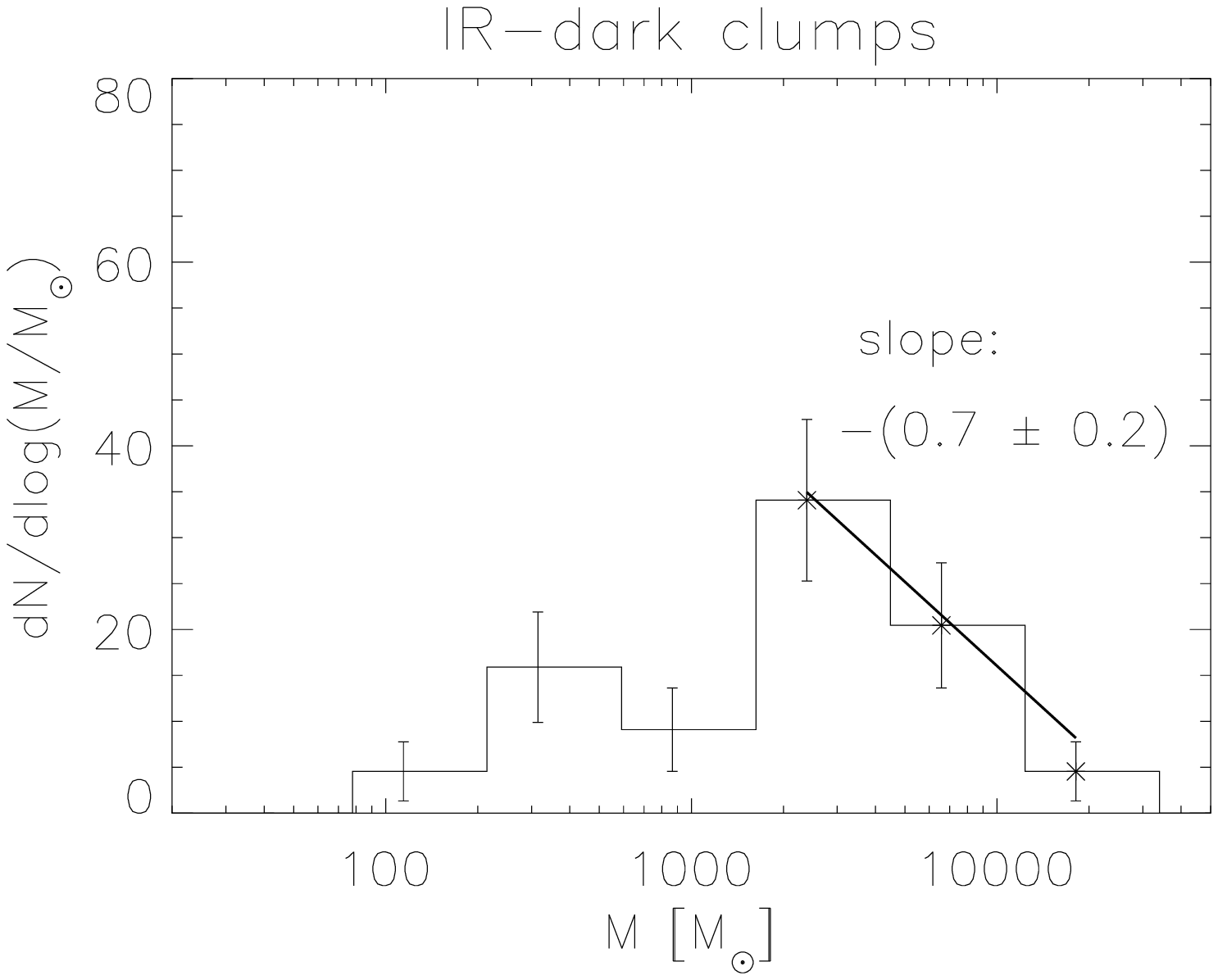}
\includegraphics[scale=0.3]{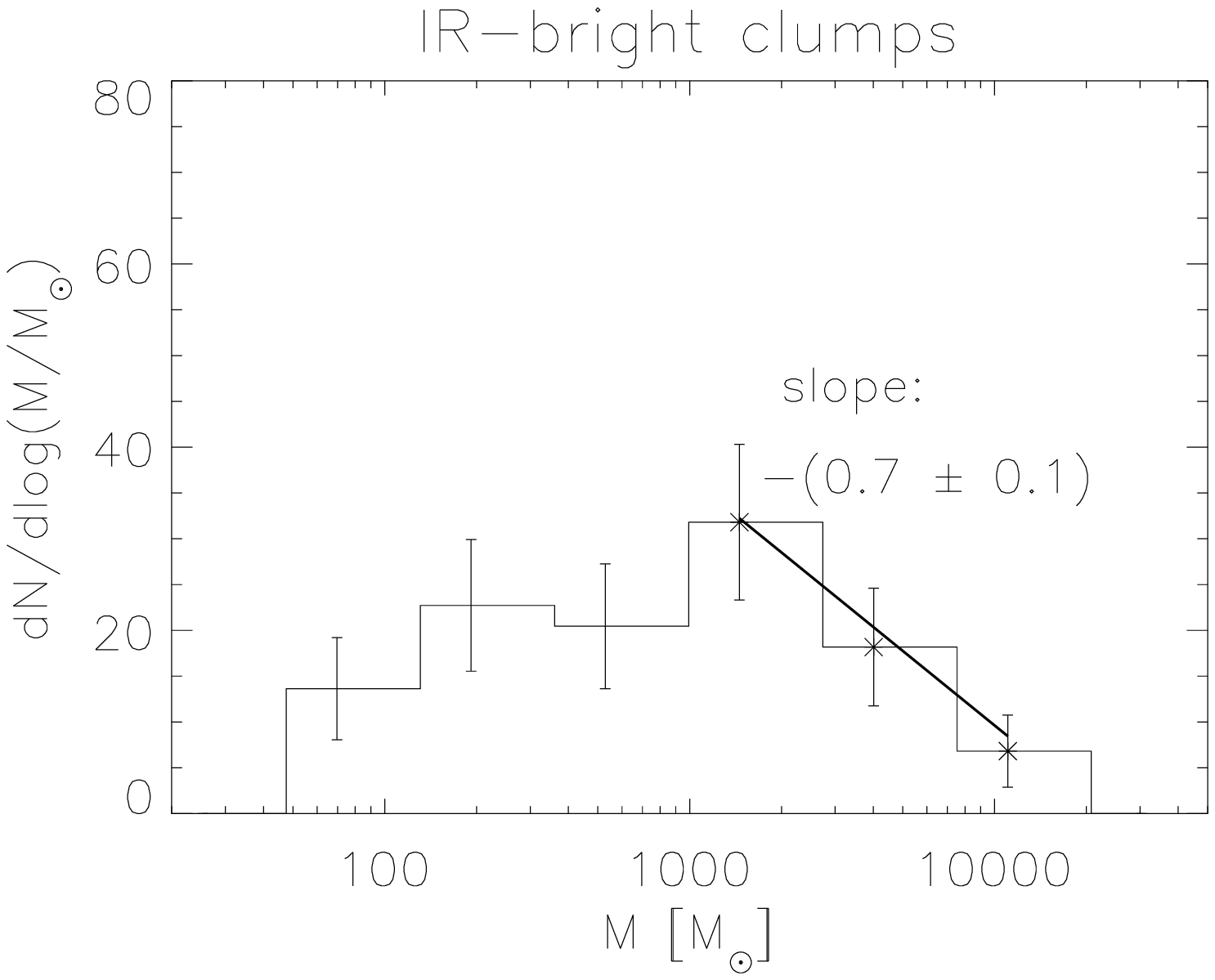}
\caption{From left to right are shown the differential 
(${\rm d}N/{\rm d}\log M$) clump mass distributions of the 
entire sample, IR-dark clumps, and IR-bright clumps, respectively. In all 
panels the mass bin size is $\Delta \log (M/{\rm M}_{\sun})\simeq0.44$. 
The error bars correspond to the standard Poisson $\sqrt{N}$ counting 
uncertainties. The solid lines indicate the best-fit power-law functions of 
the form ${\rm d}N/{\rm d}\log M \propto M^{-\Gamma}$, where the slope is given 
in each panel. The fits refer to masses above $\sim2\,400$ M$_{\sun}$ 
for both the entire sample and IR-dark clumps, and above 
$\sim1\,500$ M$_{\sun}$ for the IR-bright clumps. We note that the Salpeter 
IMF has a slope of $\Gamma=1.35$.}
\label{figure:massdistributions}
\end{center}
\end{figure*}

The clump mass distributions (CMDs) for the entire sample (composed of 
both IR-dark and IR-bright clumps), and separately for the IR-dark and 
IR-bright clumps are shown in Fig.~\ref{figure:massdistributions}. 
The CMDs are plotted 
as ${\rm d}N/{\rm d}\log M$ versus $M$, where the first term is approximated as 
the number of clumps in each bin divided by the logarithmic mass interval, 
i.e., $\Delta N/\Delta \log M$. Following L{\'o}pez et al. (2011), we kept the 
histogram bin size $\Delta \log M$ at a constant value of about 0.44. The 
CMDs were fitted with power-laws of the form 
${\rm d}N/{\rm d}\log M \propto M^{-\Gamma}$, with the slopes 
$\Gamma=0.8\pm0.1$, $0.7\pm0.2$, and $0.7\pm0.1$ for the entire sample, 
IR-dark clumps, and IR-bright clumps, respectively. The errors of the slopes 
were determined by considering only the statistical Poisson $\sqrt{N}$ 
uncertainty of the data, and the mass uncertainties were not taken into 
account. The mass ranges used in the fits were 
$\sim2.4\times10^3-1.8\times10^4$ M$_{\sun}$ for both the entire sample and 
IR-dark clumps, and $\sim1.5\times10^3-1.1\times10^4$ M$_{\sun}$ for the 
IR-bright clumps. Even the CMD slopes for the IR-dark and IR-bright clumps are 
similar, the CMDs themselves appear to be very different from each other; a 
two-sample Kolmogorov-Smirnov (K-S) test gives a probability of only 
0.48\% that they are drawn from the same underlying parent distribution.
We note that the CMDs can also be expressed in the differential form of
${\rm d}N/{\rm d}M \propto M^{-\alpha}$, where $\alpha = \Gamma + 1$. 

The clumps studied here lie at different distances, and the sensitivities of 
the four maps are also different (median being $\sim70$ mJy~beam$^{-1}$). 
Therefore, the mass detection limit varies, and is not straightforward to 
determine. For example, at the median distance of our sources, $\sim4.4$ kpc, 
a source of 1 Jy cor\-responds to a 15-K mass of $\sim225$ M$_{\sun}$. Our 
completeness limit is, however, likely to be at a much higher mass, as 
suggested by the apparent peak near $\sim2\,000$ M$_{\sun}$. 
Because our sample is very heterogeneous in nature, it is not ideal for 
the CMD study. However, it is interesting to compare the 
derived slopes with those from the literature.

The slope of the Salpeter (1955) initial mass function (IMF) is 
$\Gamma=1.35$ or $\alpha=2.35$ for stars with masses in the range 
0.4 M$_{\sun}$ $\lesssim M \lesssim$ 10 M$_{\sun}$. Our slopes are clearly 
shallower than the Salpeter value. On the other hand, our results are similar 
to the CMDs of diffuse CO clumps, which are found to be well described 
by power-law forms with $\alpha$ between 1.6 and 1.8 (e.g., \cite{stutzki1990}; 
\cite{kramer1998}; \cite{simon2001}). These slopes also resemble those of 
stellar clusters' mass function, implying that massive clumps are not 
the direct progenitors of individual stars, but will instead presumably 
fragment to form groups/clusters of stars (e.g., \cite{elmegreen2000}). 

Rathborne et al. (2006) found a slope of 
$\alpha=2.1\pm0.4$ above $\sim100$ M$_{\sun}$ for their sample of 
\textit{MSX} 8-$\mu$m dark clumps, which is quite close to the 
Salpeter power-law IMF, but also consistent with our values within the error 
bars. Interestingly, a two-sample K-S test yields practically a zero 
probabi\-lity ($\sim10^{-12}$) that our IR-dark clump masses and those from 
Rathborne et al. (2006) are drawn from the common underlying distribution. 
We note that for this test the clump masses from Rathborne et al. (2006) were 
multiplied by 1.214 to be consistent with the dust model we have adopted 
(their 1.2-mm dust opacity of 1.0 cm$^2$~g$^{-1}$ was replaced by the value 
$\kappa_{\rm 1.2\, mm}\simeq0.82$ cm$^2$~g$^{-1}$). Ragan et al. (2009) also 
built a CMD for their entire sample of IR-dark and IR-bright clumps (cf. 
left panel of our Fig.~\ref{figure:massdistributions}), and derived the slope 
$\alpha=1.76\pm0.05$ (from 30 to 3\,000 M$_{\sun}$), which is very similar to 
our corresponding value of $1.8\pm0.1$. Tackenberg et al. (2012) 
found a Salpeter-like slope of $\alpha=2.2$ for the CMD of candidate massive 
starless clumps. 
We note that Salpeter-like slopes have also been determined for the mass 
distributions of more evolved massive clumps (e.g., \cite{reid2005}; 
\cite{beltran2006}; \cite{bally2010}), although the uncertainties are often 
too large to say whether they are actually flatter (or steeper) than the 
exact Salpeter slope. Indeed, in addition to our study, some authors 
have found the mass functions of evolved massive clumps to be similar to 
those of CO clumps (e.g., \cite{beuther2011}; \cite{lopez2011}). 

Mass distributions of whole IRDCs have also been stu\-died by Simon et al. 
(2006), Marshall et al. (2009), and Peretto \& Fuller (2010), who determined 
the slopes $\alpha=1.97\pm0.09$, $\alpha=1.75\pm0.06$, and 
$\alpha=1.85\pm0.07$, respectively. These are also comparable to our results 
and to the CMD slopes of the CO clumps. Marshall et al. (2009) suggested 
that the similarity bet\-ween the mass distributions of IRDCs and CO clumps 
perhaps indicates that IRDCs are the result of density fluctuations caused by 
interstellar turbulence. The analytical theory by Hennebelle \& Chabrier 
(2008) is able to explain the flat mass spectra of CO clumps formed by 
supersonic turbulent flows. In the case of large-scale turbulent flows, the 
slope of the CMD's power-law tail is determined by the spectral index of the 
turbulent power spectrum. In our notation, the CMD slope $\Gamma$ in the 
Hennebelle-Chabrier theory is given by $\Gamma=2-n^{,}/3$, where $n^{,}$ is 
the three-dimensional power-spectrum index of the logarithmic density field. 
As discussed by Hennebelle \& Chabrier (2008), $n^{,}$ appears to be 
$\simeq11/3$, i.e., similar to the Kolmogorov power spectrum index 11/3 
of the velocity field in incompressible turbulence. This leads to the value 
$\Gamma \simeq 7/9 \simeq 0.8$, which is remarkably close to the CMD slopes 
of CO clumps and the values derived in the present study. 

\subsection{On the origin and hierarchical fragmentation of filamentary 
IRDCs}

The formation of filamentary structures by colliding shock fronts is clearly 
seen in numerical simulations (e.g., \cite{banerjee2006}).
Large-scale colliding supersonic turbulent flows offer an intriguing mechanism 
for the formation of filamentary IRDCs. As discussed in the previous section, 
the IRDCs' mass functions can be understood in terms of turbulent flows. 
Moreover, Jim{\'e}nez-Serra et al. (2010) found extended SiO emission along 
the filamentary IRDC G035.39-00.33, which might have been produced in shocks
caused by converging flows. Hernandez \& Tan (2011) and M12 found 
evidence that the surface pressure may play an important role in the dynamics 
of filamentary IRDCs. This could also be related to the cloud formation in 
gas overdensities resulting from colliding flows.

Although the formation of IRDCs could be caused by interstellar turbulent 
flows, their further fragmentation may have its origin in some other 
mechanism(s). As was shown in the present study, the fragmentation of the 
filamentary IRDC G11.36 into clumps can be explained by the ``sausage''-type 
fluid instability. In an IRDC, this was first found to be the case in the 
Nessie Nebula by Jackson et al. (2010). More recently, M12 inferred that 
this is also the case in the filamentary IRDC G304.74+01.32. Hence, to our 
knowledge there are three different filamentary IRDCs where the predictions 
of the sausage-instability theory have been tested so far. In all cases, the 
theory and observations are in very good agreement with each other. One 
remarkable feature that should be raised is that in all the three above 
mentioned cases the fragmentaion length scale corresponds to the wavelength 
of the fastest growing mode only when the non-thermal motions are taken into 
account. Non-thermal ('turbulent') motions are therefore still important in the 
fragmentation process, on least at the scale of clumps. However, this can also 
continue in a hierarchical way down to the scale of cores, as found by Wang 
et al. (2011) in the massive IRDC clump G28.34+0.06-P1. A Jeans-type 
gravitational instability may start to dominate inside dense cores, i.e., 
cause them to fragment into still smaller condensations 
(cf. \cite{miettinenetal2012}).

\subsection{CO depletion and its implications on the age of the 
G11.36+0.80 filament}

The CO depletion factors we estimate towards 25 target 
positions lie in the range $\sim1.0\pm0.2-20.7\pm0.8$. Some of the values are 
very large, taking that the beam size $27\farcs8$ is probing the scale of 
clumps at the source distances. In contrast, using the same kind of 
APEX/C$^{17}$O data as here, Miettinen et al. (2011) found depletion factors 
of only $\sim0.6-2.7$ towards a sample of seven clumps in IRDCs and similar 
values of $\sim0.3-2.3$ were found by M12 towards the clumps in the IRDC 
G304.74. On the other hand, Chen et al. (2011), who used $34\arcsec$ C$^{18}$O 
observations, reported $f_{\rm D}$ values as high as $\sim19$ in the IRDC 
G34.43+0.24, which resembles our largest $f_{\rm D}$ values. Pillai et al. 
(2007) derived the CO depletion factor of 9.3 towards one of our clumps, 
namely G13.22-SMM 5 (their source G13.18+00.06).

Some of the largest $f_{\rm D}$ values in the present study are found 
towards the positions C, D, E, and F in the filamentary IRDC in G1.87, and 
position A in G13.22. These might be potential sites to search for high levels 
of molecular deuteration, since the depletion of CO plays a key role in the 
deuterium fractionation. Interestingly, the largest $f_{\rm D}$ value of 
$\sim20.7$ is seen towards the edge of G2.11-SMM 5 (position H). This raises 
the question whether the high depletion factor is the result of molecular 
freeze-out onto dust grains, or possibly due to photodissociation by FUV 
photons at the edge of the clump (\cite{tielens1985}; \cite{visser2009}) ? 
Photodissociation could also play a role in reducing the C$^{17}$O abundance 
in some other positions of our line observations.

The $f_{\rm D}$ values in the G11.36 filament (excluding positions F and G), 
where we best probe the submm peak positions among our sources, are 
$\sim1.6\pm0.2-3.9\pm0.6$. This implies that non-negligible CO 
depletion might be present in some parts of the filament. Hernandez et al. 
(2011) inferred depletion factors of $\sim3-4$ for the thinnest part of the 
filamentary IRDC G035.30-00.33, and used the CO depletion timescale to 
estimate the cloud's age. Here, we do the same exercise for G11.36.
The CO depletion timescale can be computed as the inverse of the 
freezing rate (\cite{rawlings1992}) as

\begin{equation}
\tau_{\rm dep}=\frac{1}{k_{\rm freeze}}=\frac{1}{n_{\rm g}\pi a_{\rm g}^2{\rm v}_{\rm CO}S}\, ,
\end{equation}
where $n_{\rm g}$ is the grain number density, $a_{\rm g}$ is the average grain 
radius ($\pi a_{\rm g}^2$ is the mean cross section of the dust grains, and 
$n_{\rm g}\pi a_{\rm g}^2$ is the grain opacity), ${\rm v}_{\rm CO}$ is the 
thermal speed of the CO molecules, and $S$ is their sticking coefficient.
We can write the grain number density as $n_{\rm g}=x_{\rm g}n({\rm H_2})$, where 
$x_{\rm g}$ is the fractional grain abundance given by

\begin{equation}
x_{\rm g}=\frac{m_{\rm H_2}}{m_{\rm g}}\times R_{\rm d}=\frac{m_{\rm H_2}R_{\rm d}}{\frac{4\pi}{3}a_{\rm g}^3\rho_{\rm g}} \,.
\end{equation}
In the above formula, $m_{\rm H_2}$ is the H$_2$ molecule's mass, $m_{\rm g}$ 
($\rho_{\rm g}$) is the mass (density) of the grain particle, and 
$R_{\rm d}=1/100$ is again the dust-to-gas mass ratio. By adopting the 
``standard'' values $a_{\rm g}=0.1$ $\mu$m and $\rho_{\rm g}=3.0$ g~cm$^{-3}$, 
we obtain $x_{\rm g}=2.7\times10^{-12}$. The Maxwellian mean thermal speed of the 
CO molecules ($m_{\rm CO}=28m_{\rm H}$) at $T_{\rm kin}=15$ K is 
${\rm v}_{\rm CO}=\left(8k_{\rm B}T_{\rm kin}/\pi m_{\rm CO} \right)^{1/2}\simeq106$ 
m~s$^{-1}$. Assuming that $S=1$, i.e., that the CO molecules stick to the 
dust grains in each collision, we can write 
$\tau_{\rm dep}\sim 4.4\times10^{9}/n({\rm H_2})\, [{\rm cm^{-3}}]$ 
yr\footnote{We note that the formula used by Hernandez et al. (2011; their 
Sect.~4.2) gives a factor of 5.5 times shorter CO depletion timescale.}. 
The volume-average density in the G11.36 clumps is $\sim10^4$ cm$^{-3}$ 
(Table~\ref{clumps}), which gives $\tau_{\rm dep}\sim 4.4\times10^5$ yr. 
Interestingly, this is very close to the estimated fragmentation timescale in 
G11.36 ($\sim 5.4\times10^{5}$ yr). The age of the filament may therefore be 
similar to its fragmentation time. In this case, the cloud's fragmentation 
would have started at the same time as it was formed. Perhaps the 
cloud-forming flows also triggered the sausage-type fluid instability in 
the cloud.

\section{Summary and conclusions}

Four selected regions in the Galactic plane, all containing IRDCs, were mapped 
with APEX/LABOCA at 870 $\mu$m. Moreover, selected positions in the fields 
were observed in C$^{17}$O$(2-1)$ with APEX. Our main results and conclusions 
can be summarised as follows:

\begin{enumerate}
\item The total number of clumps identified in this survey is 91. The number 
of \textit{Spitzer} 8- and 24-$\mu$m dark clumps was estimated to be 40. 
The remaining 51 clumps were found to be associated with both 8 and 
24 $\mu$m emission, or only with 24 $\mu$m emission (embedded point sources, 
group of point sources, diffuse/extended emission).
\item Many of the clumps appear to be potential sites of (future) high-mass 
star formation, as implied by the comparison with the mass-radius threshold 
proposed by Kauffmann \& Pillai (2010). Some of the clumps already show clear 
signposts of ongoing high-mass star formation, such as Class {\scriptsize II} 
CH$_3$OH maser emission.
\item Seven clumps show extended-like 4.5 $\mu$m emission, hence are 
classified as candidate EGOs. 
\item One of the mapped fields (G13.22) contains the \textit{Spitzer} 
MIR-bubbles N10/11 found by Churchwell et al. (2006). The bubble region 
is believed to be associated with triggered massive-star formation, and the 
observed dust emission morphology might have been created by the bubble 
shell-cloud interaction process. 
\item The relative numbers of IR-dark and IR-bright clumps implies the 
duration of the IR-dark stage to be $\sim1.6\times10^5$ yr. Although this 
estimate is likely to suffer from the relatively low number statistics, it is 
comparable with some earlier lifetime estimates (\cite{chambers2009}; 
\cite{wilcock2012}). Surveys with higher number of sources typically indicate 
shorter lifetimes for massive IR-dark clumps (e.g., \cite{tackenberg2012}).
\item The clump mass distributions were constructed for the total clump 
sample, and for the IR-dark and IR-bright clumps separately. In all cases, 
the high-mass tail could be fitted with the power-law function of the form 
${\rm d}N/{\rm d}\log M \propto M^{-\Gamma}$, with 
$\Gamma \simeq 0.7\ldots 0.8$. This is very similar to the mass functions 
found for the diffuse CO clumps, and can be understood in terms of 
supersonic-turbulence induced cloud formation (\cite{hennebelle2008}).
\item The C$^{17}$O observations revealed potential targets of strong CO 
depletion. In some cases, however, our line observations probe more dilute gas, 
where the low C$^{17}$O abundances could be caused by photodissociation by FUV 
photons.
\item The filamentary IRDC G11.36+0.80 was studied in more detail, because 
there our line observation positions were well matched with the submm peak 
positions. The filament's fragmentation into clumps can be well explained by 
a ``sausage''-type fluid instability. In particular, the observed projected 
clump separations are in excellent agreement with the theoretical prediction. 
This is in agreement with the results for other IRDCs 
(\cite{jackson2010}; M12). Most clumps in the filament appear to be 
gravitationally bound, and the filament as a whole appers to be close to 
virial equilibrium. Interestingly, the estimated fragmentation timescale of 
the filament and the CO depletion timescale were inferred to be comparable to 
each other ($\sim 5\times10^5$ yr). The perturbations responsible for the 
filament fragmentation into clumps might have been excited by the 
cloud-forming process, i.e., by converging turbulent flows.  
\end{enumerate}

\begin{acknowledgements}

I thank the anonymous referee for his/her constructive comments and 
suggestions. I am grateful to the staff at the APEX telescope for performing 
the service-mode observations presented in this paper. In particular, I would 
like to thank the APEX operator Felipe Mac-Auliffe for providing helpful 
information on the observations. The Academy of Finland is acknowledged for 
the financial support through grant 132291. This work is based in part on 
observations made with the Spitzer Space Telescope, which is operated by the 
Jet Propulsion Laboratory, California Institute of Technology, under contract 
with NASA. This research has made use of NASA's Astrophysics Data System 
and the NASA/IPAC Infrared Science Archive, which is operated by the JPL, 
California Institute of Technology, under contract with the NASA. This 
research has also made use of the SIMBAD database, operated at CDS, 
Strasbourg, France.

\end{acknowledgements}

\longtabL{3}{
\begin{landscape}
{\scriptsize
\renewcommand{\footnoterule}{}
\begin{longtable}{c c c c c c c c c c c c}
\caption{\label{clumps} Characteristics of the clumps identified from the 
LABOCA maps.}\\
\hline\hline
Source & $\alpha_{2000.0}$ & $\delta_{2000.0}$ & $I_{\rm 870\, \mu m}^{\rm peak}$ & $S_{\rm 870\, \mu m}$ & \multicolumn{2}{c}{$R_{\rm eff}$} & $M$ & $N({\rm H_2})$ & $\langle n({\rm H_2}) \rangle$ & Associations\tablefootmark{a} & 8/24 $\mu$m\\
       & [h:m:s] & [$\degr$:$\arcmin$:$\arcsec$] & [Jy~beam$^{-1}$] & [Jy] & [$\arcsec$] & [pc] & [M$_{\sun}$] & [$10^{22}$ cm$^{-2}$] & [$10^4$ cm$^{-3}$] & &\\
\hline 
\hline
\endfirsthead
\caption{continued.}\\
\hline\hline
Source & $\alpha_{2000.0}$ & $\delta_{2000.0}$ & $I_{\rm 870\, \mu m}^{\rm peak}$ & $S_{\rm 870\, \mu m}$ & \multicolumn{2}{c}{$R_{\rm eff}$} & $M$ & $N({\rm H_2})$ & $\langle n({\rm H_2}) \rangle$ & Associations\tablefootmark{a} & 8/24 $\mu$m\\
       & [h:m:s] & [$\degr$:$\arcmin$:$\arcsec$] & [Jy~beam$^{-1}$] & [Jy] & [$\arcsec$] & [pc] & [M$_{\sun}$] & [$10^{22}$ cm$^{-2}$] & [$10^4$ cm$^{-3}$] & &\\
\hline
\endhead
\hline
\endfoot
{\bf G1.87-0.14} & \\
SMM 1 \ldots & 17 49 44.0 & -27 33 28 & $2.39\pm0.26$ & $15.21\pm2.32$ & 41.4 & $2.12\pm0.10$ & $19\,725\pm3\,540$ & $11.8\pm1.3$ & $1.0\pm0.2$ & JCMTSE J174944.3-273330\tablefootmark{1} ($4\farcs4$); & dark/dark\\ 
 & & & & & & & & & & BGPS G1.652-0.066\tablefootmark{2} ($21\farcs4$) & \\
SMM 2 \ldots & 17 49 46.5 & -27 25 12 & $0.53\pm0.10$ & $1.97\pm0.21$ & 26.5 & $1.36\pm0.06$ & $1\,633\pm233$ & $1.7\pm0.3$ & $0.3\pm0.1$ & JCMTSE J174946.2-272512\tablefootmark{1} ($6\farcs8$) & point/point \\
SMM 3 \ldots & 17 49 47.2 & -27 15 48 & $0.65\pm0.11$ & $2.44\pm0.25$ & 29.5 & $1.51\pm0.07$ & $3\,164\pm441$ & $3.2\pm0.5$ & $0.4\pm0.1$ & BGPS G1.908+0.082\tablefootmark{2} ($24\farcs3$) & dark/dark \\
SMM 4 \ldots & 17 49 52.5 & -27 32 56 & $0.55\pm0.11$ & $1.23\pm0.14$ & 20.2 & $1.04\pm0.05$ & $1\,595\pm236$ & $2.7\pm0.5$ & $0.7\pm0.1$ & JCMTSE J174952.4-273307\tablefootmark{1} ($11\farcs0$); & dark/dark \\
 & & & & & & & & & & BGPS G1.680-0.094\tablefootmark{2} ($18\farcs1$)\\ 
SMM 5 \ldots & 17 49 54.5 & -27 32 40 & $1.07\pm0.14$ & $3.20\pm0.33$ & 26.3 & $1.35\pm0.06$ & $4\,150\pm581$ & $5.3\pm0.7$ & $0.8\pm0.1$ & JCMTSE J174955.1-273237\tablefootmark{1} ($14\farcs7$) & dark/dark \\
SMM 6 \ldots & 17 49 56.0 & -27 31 56 & $0.86\pm0.12$ & $1.42\pm0.15$ & 17.3 & $0.89\pm0.04$ & $1\,842\pm261$ & $4.2\pm0.6$ & $1.2\pm0.2$ & JCMTSE J174956.5-273143\tablefootmark{1} ($14\farcs7$); & dark/dark  \\
 & & & & & & & & & & BGPS G1.696-0.092\tablefootmark{2} ($23\farcs5$)\\
SMM 7 \ldots & 17 49 56.7 & -27 31 44 & $0.96\pm0.13$ & $2.95\pm0.30$ & 24.8 & $1.27\pm0.06$ & $3\,826\pm531$ & $4.7\pm0.6$ & $0.9\pm0.1$ & JCMTSE J174956.5-273143\tablefootmark{1} ($2\farcs5$) & dark/dark  \\
SMM 8 \ldots & 17 49 59.4 & -27 29 24 & $0.74\pm0.12$ & $3.08\pm0.31$ & 30.8 & $1.58\pm0.07$ & $3\,994\pm552$ & $3.6\pm0.6$ & $0.5\pm0.1$ & BGPS G1.738-0.080\tablefootmark{2} ($10\farcs6$) & dark/dark  \\
SMM 9 \ldots & 17 49 59.4 & -27 34 28 & $0.59\pm0.11$ & $0.97\pm0.11$ & 17.2 & $0.88\pm0.04$ & $804\pm119$ & $1.9\pm0.3$ & $0.5\pm0.1$ & JCMTSE J174959.2-273431\tablefootmark{1} ($3\farcs5$) & group/group  \\
SMM 10 \ldots & 17 49 59.7 & -27 30 36 & $0.71\pm0.11$ & $1.56\pm0.17$ & 22.1 & $1.13\pm0.06$ & $1\,293\pm187$ & $2.2\pm0.3$ & $0.4\pm0.1$ & JCMTSE J174959.7-273037\tablefootmark{1} ($0\farcs8$) & point/point  \\
SMM 11 \ldots & 17 50 02.1 & -27 34 28 & $0.85\pm0.12$ & $1.56\pm0.17$ & 18.5 & $0.95\pm0.04$ & $1\,293\pm187$ & $2.7\pm0.4$ & $0.7\pm0.1$ & \ldots & point/point \\
SMM 12 \ldots & 17 50 03.0 & -27 33 56 & $1.47\pm0.17$ & $5.74\pm0.58$ & 33.2 & $1.70\pm0.08$ & $4\,759\pm659$ & $4.6\pm0.5$ & $0.4\pm0.1$ & JCMTSE J175002.8-273401\tablefootmark{1} ($5\farcs3$) & point/point \\
SMM 13 \ldots & 17 50 04.2 & -27 34 44 & $1.16\pm0.15$ & $2.95\pm0.30$ & 24.6 & $1.26\pm0.06$ & $3\,826\pm531$ & $5.7\pm0.7$ & $0.9\pm0.1$ & JCMTSE J175004.1-273455\tablefootmark{1} ($10\farcs6$) & dark/dark \\
SMM 14 \ldots & 17 50 04.2 & -27 34 57 & $1.12\pm0.14$ & $2.92\pm0.30$ & 26.0 & $1.33\pm0.06$ & $3\,787\pm529$ & $5.5\pm0.7$ & $0.7\pm0.1$ & JCMTSE J175004.1-273455\tablefootmark{1} ($1\farcs9$) & dark/dark  \\
SMM 15 \ldots & 17 50 05.1 & -27 28 32 & $0.76\pm0.12$ & $3.10\pm0.32$ & 29.1 & $1.49\pm0.05$ & $2\,570\pm360$ & $2.4\pm0.4$ & $0.4\pm0.1$ & \ldots & group/point  \\
SMM 16 \ldots & 17 50 09.0 & -27 33 36 & $0.79\pm0.12$ & $3.70\pm0.37$ & 32.9  &$1.69\pm0.08$ & $4\,798\pm661$ & $3.9\pm0.6$ & $0.5\pm0.1$ & JCMTSE J175009.1-273331\tablefootmark{1} ($5\farcs5$) & dark/dark \\
SMM 17 \ldots & 17 50 12.3 & -27 36 48 & $0.82\pm0.12$ & $1.77\pm0.19$ & 20.9 & $1.07\pm0.05$ & $2\,295\pm328$ & $4.0\pm0.6$ & $0.9\pm0.1$ & BGPS G1.658-0.180\tablefootmark{2} ($11\farcs8$) & dark/dark \\ 
SMM 18 \ldots & 17 50 12.3 & -27 35 28 & $0.59\pm0.11$ & $1.20\pm0.13$ & 19.3 & $0.99\pm0.05$ & $1\,556\pm224$ & $2.9\pm0.5$ & $0.7\pm0.1$ & \ldots & dark/dark \\
SMM 19 \ldots & 17 50 12.6 & -27 36 04 & $0.92\pm0.13$ & $3.38\pm0.34$ & 27.7 & $1.42\pm0.07$ & $2\,802\pm387$ & $2.9\pm0.4$ & $0.5\pm0.1$ & BGPS G1.660-0.178\tablefootmark{2} ($10\farcs0$); & point/point \\ 
 & & & & & & & & & & JCMTSE J175011.8-273607\tablefootmark{1} ($10\farcs4$)\\
SMM 20 \ldots & 17 50 13.5 & -27 20 40 & $0.86\pm0.12$ & $8.13\pm0.82$ & 47.0 & $2.41\pm0.11$ & $6\,740\pm932$ & $2.7\pm0.4$ & $0.2\pm0.1$ & [foreground star HD 316367 ($14\farcs2$)] & point/point \\
SMM 21 \ldots & 17 50 13.8 & -27 35 24 & $0.67\pm0.11$ & $1.50\pm0.16$ & 20.6 & $1.06\pm0.05$ & $1\,244\pm177$ & $2.1\pm0.3$ & $0.5\pm0.2$ & \ldots & point/point  \\
SMM 22 \ldots & 17 50 14.7 & -27 27 48 & $0.54\pm0.10$ & $1.57\pm0.17$ & 22.6 & $1.16\pm0.05$ & $2\,036\pm293$ & $2.7\pm0.5$ & $0.6\pm0.1$ & \ldots & dark/dark \\
SMM 23 \ldots & 17 50 15.0 & -27 21 33 & $1.07\pm0.14$ & $10.05\pm1.01$ & 44.2 &$2.27\pm0.11$ & $13\,033\pm1\,799$ & $5.3\pm0.7$ & $0.5\pm0.1$ & \ldots & dark/dark \\
SMM 24 \ldots & 17 50 15.3 & -27 34 24 & $0.76\pm0.12$ & $4.07\pm0.41$ & 32.7 & $1.68\pm0.08$ & $5\,278\pm729$ & $3.7\pm0.6$ & $0.5\pm0.1$ & BGPS G1.686-0.172\tablefootmark{2} ($0\farcs3$); & dark/dark \\ 
 & & & & & & & & & & JCMTSE J175015.0-273437\tablefootmark{1} ($13\farcs1$)\\
SMM 25 \ldots & 17 50 15.3 & -27 27 52 & $0.58\pm0.11$ & $2.69\pm0.28$ & 29.8 & $1.53\pm0.07$ & $3\,489\pm491$ & $2.9\pm0.5$ & $0.4\pm0.1$ & \ldots & dark/dark \\ 
SMM 26 \ldots & 17 50 18.6 & -27 26 49 & $0.59\pm0.11$ & $1.87\pm0.20$ & 23.7 & $1.21\pm0.06$ & $1\,550\pm221$ & $1.9\pm0.3$ & $0.4\pm0.1$ & \ldots & point/point \\
SMM 27 \ldots & 17 50 18.9 & -27 26 32 & $0.70\pm0.11$ & $1.48\pm0.16$ & 20.7 & $1.06\pm0.05$ & $1\,919\pm276$ & $3.4\pm0.5$ & $0.7\pm0.1$ & \ldots & point(fg)\tablefootmark{c}/dark \\
SMM 28 \ldots & 17 50 19.2 & -27 27 53 & $0.87\pm0.13$ & $6.17\pm0.62$ & 37.7 & $1.93\pm0.09$ & $5\,115\pm706$ & $4.3\pm0.6$ & $0.3\pm0.1$ & BGPS G1.798-0.124\tablefootmark{2} ($23\farcs3$) & group/point \\
SMM 29 \ldots & 17 50 19.5 & -27 21 48 & $0.55\pm0.11$ & $4.49\pm0.45$ & 37.6 & $1.93\pm0.09$ & $5\,823\pm803$ & $2.7\pm0.5$ & $0.4\pm0.1$ & \ldots & dark/dark \\
SMM 30 \ldots & 17 50 22.8 & -27 21 28 & $0.59\pm0.11$ & $4.56\pm0.46$ & 37.9 & $1.94\pm0.09$ & $5\,914\pm818$ & $2.9\pm0.5$ & $0.4\pm0.1$ & \ldots & dark/dark \\
SMM 31 \ldots & 17 50 24.3 & -27 28 29 & $0.43\pm0.10$ & $1.09\pm0.14$ & 20.1 & $1.03\pm0.05$ & $904\pm144$ & $1.4\pm0.3$ & $0.4\pm0.1$ & BGPS G1.798-0.152\tablefootmark{2} ($15\farcs2$) & point/point \\
SMM 32 \ldots & 17 50 25.5 & -27 35 36 & $0.53\pm0.10$ & $0.64\pm0.09$ & 14.6 & $0.75\pm0.04$ & $531\pm90$ & $1.7\pm0.3$ & $0.6\pm0.1$ & BGPS G1.692-0.212\tablefootmark{2} ($1\farcs5$) & dark/point \\ 
SMM 33 \ldots & 17 50 31.8 & -27 22 12 & $0.55\pm0.11$ & $3.96\pm0.40$ & 37.0 & $1.90\pm0.09$ & $5\,136\pm711$ & $2.7\pm0.5$ & $0.3\pm0.1$ & BGPS G1.894-0.118\tablefootmark{2} ($16\farcs9$) & dark/dark \\
SMM 34 \ldots & 17 50 31.9 & -27 24 29 & $0.47\pm0.10$ & $1.90\pm0.20$ & 26.6 & $1.36\pm0.06$ & $2\,464\pm349$ & $2.3\pm0.5$ & $0.5\pm0.1$ & \ldots & dark/dark \\
SMM 35(C, D)\tablefootmark{b} \ldots & 17 50 36.7 & -27 24 12 & $0.64\pm0.11$ & $2.69\pm0.28$ & 29.9 & $1.53\pm0.07$ & $2\,230\pm314$ & $2.0\pm0.3$ & $0.3\pm0.1$ & SSTGLMC G001.8803-00.1521\tablefootmark{3} ($7\farcs4$); & group/point \\
 & & & & & & & & & & BGPS G1.878-0.154\tablefootmark{2} ($11\farcs9$) \\
SMM 36 \ldots & 17 50 37.5 & -27 20 49 & $0.56\pm0.11$ & $1.68\pm0.18$ & 23.7 & $1.21\pm0.06$ & $2\,179\pm311$ & $2.8\pm0.5$ & $0.6\pm0.1$ & BGPS G1.932-0.126\tablefootmark{2} ($6\farcs8$) & point(fg)\tablefootmark{d}/dark \\
SMM 37 \ldots & 17 50 41.5 & -27 18 57 & $0.51\pm0.10$ & $1.39\pm0.15$ & 23.0 & $1.18\pm0.06$ & $1\,152\pm165$ & $1.6\pm0.3$ & $0.3\pm0.1$ & \ldots & point/point \\
SMM 38 \ldots & 17 50 46.3 & -27 24 32 & $0.76\pm0.12$ & $4.42\pm0.45$ & 38.8 & $1.99\pm0.09$ & $5\,732\pm797$ & $3.7\pm0.6$ & $0.3\pm0.1$ & BGPS G1.894-0.186\tablefootmark{2} ($6\farcs9$) & dark/dark \\
SMM 39 \ldots & 17 50 46.6 & -27 19 09 & $0.56\pm0.11$ & $2.24\pm0.23$ & 27.4 & $1.40\pm0.07$ & $2\,905\pm406$ & $2.8\pm0.5$ & $0.5\pm0.1$ & BGPS G1.972-0.140\tablefootmark{2} ($22\farcs7$) & point(fg)\tablefootmark{e}/dark \\
SMM 40 \ldots & 17 50 48.7 & -27 22 01 & $0.52\pm0.10$ & $0.43\pm0.07$ & 12.2 & $0.63\pm0.03$ & $356\pm67$ & $1.6\pm0.3$ & $0.7\pm0.1$ & SSTGLMC G001.9357-00.1696\tablefootmark{3} ($4\farcs1$); & group/point \\ 
 & & & & & & & & & & SSTGLMC G001.9323-00.1700\tablefootmark{3} ($8\farcs7$);\\
 & & & & & & & & & & BGPS G1.934-0.172\tablefootmark{2} ($10\farcs2$) \\

{\bf G2.11+0.00} & \\
SMM 1 \ldots & 17 50 26.8 & -27 02 16 & $0.37\pm0.07$ & $2.41\pm0.25$ & 36.8 & $1.11\pm0.09$ & $846\pm291$ & $1.8\pm0.3$ & $0.3\pm0.1$ & \ldots & dark/dark \\
SMM 2(D)\tablefootmark{b} \ldots & 17 50 30.6 & -27 07 20 & $0.30\pm0.07$ & $0.28\pm0.07$ & 12.4 & $0.37\pm0.03$ & $105\pm42$ & $1.5\pm0.3$ & $1.0\pm0.4$ & \ldots & dark/dark \\
SMM 3 \ldots & 17 50 30.9 & -27 04 12 & $0.47\pm0.08$ & $3.17\pm0.32$ & 37.2 & $1.12\pm0.09$ & $1\,113\pm381$ & $2.3\pm0.4$ & $0.4\pm0.1$ & BGPS G2.154+0.036\tablefootmark{2} ($6\farcs2$) & dark/dark \\
SMM 4 \ldots & 17 50 32.1 & -27 08 48 & $0.37\pm0.07$ & $1.50\pm0.16$ & 29.0 & $0.87\pm0.07$ & $527\pm181$ & $1.8\pm0.3$ & $0.4\pm0.1$ & BGPS G2.090-0.008\tablefootmark{2} ($16\farcs0$) & point(fg)\tablefootmark{f}/dark \\
SMM 5 \ldots & 17 50 36.0 & -27 05 44 & $2.74\pm0.28$ & $6.27\pm0.63$ & 34.4 & $1.23\pm0.02$ & $1\,457\pm152$ & $4.9\pm0.5$ & $0.4\pm0.1$ & GPSR5 2.143+0.010\tablefootmark{4} ($1\farcs6$); & extended/point \\
 & & & & & & & & & & G002.143+00.010\tablefootmark{5} ($2\farcs5$); \\
 & & & & & & & & & & G002.14+0.01 A\tablefootmark{6} ($3\farcs0$);  \\
 & & & & & & & & & & IRAS 17474-2704\tablefootmark{7} ($3\farcs7$);  \\
 & & & & & & & & & & 2MASS J17503612-2705483 ($4\farcs4$);  \\
 & & & & & & & & & & 002.14+00.01\tablefootmark{8} ($5\farcs9$);  \\
 & & & & & & & & & & MSX5C G002.1419+00.0099\tablefootmark{9} ($6\farcs6$);  \\
 & & & & & & & & & & BGPS G2.144+0.006\tablefootmark{2} ($13\farcs8$);  \\
 & & & & & & & & & & 002.14+00.01\tablefootmark{10} ($16\farcs0$)  \\
SMM 6 \ldots & 17 50 36.9 & -27 06 32 & $0.38\pm0.07$ & $1.69\pm0.18$ & 28.8 & $0.87\pm0.07$ & $379\pm131$ & $1.2\pm0.2$ & $0.3\pm0.1$ & \ldots & point/point \\
SMM 7 \ldots & 17 50 41.7 & -26 59 28 & $0.33\pm0.07$ & $1.46\pm0.16$ & 28.8 & $0.87\pm0.07$ & $513\pm177$ & $1.6\pm0.3$ & $0.4\pm0.1$ & BGPS G2.240+0.046\tablefootmark{2} ($6\farcs2$) & point(fg)\tablefootmark{g}/dark \\
SMM 8 \ldots & 17 51 03.3 & -27 11 24 & $0.39\pm0.07$ & $1.47\pm0.16$ & 26.8 & $0.81\pm0.06$ & $516\pm178$ & $1.9\pm0.3$ & $0.4\pm0.2$ & BGPS G2.114-0.130\tablefootmark{2} ($11\farcs2$) & dark/dark \\
SMM 9 \ldots & 17 51 15.3 & -27 03 19 & $0.32\pm0.07$ & $1.64\pm0.17$ & 30.4 & $0.91\pm0.07$ & $576\pm198$ & $1.6\pm0.3$ & $0.4\pm0.1$ & \ldots & weak/diffuse \\
SMM 10 \ldots & 17 51 16.5 & -27 05 39 & $0.38\pm0.07$ & $2.57\pm0.26$ & 36.6 & $1.10\pm0.09$ & $902\pm309$ & $1.9\pm0.3$ & $0.3\pm0.1$ & \ldots & dark/dark \\

{\bf G11.36+0.80} & \\
SMM 1(A)\tablefootmark{b} \ldots & 18 07 35.0 & -18 43 46 & $0.62\pm0.07$ & $1.58\pm0.16$ & 23.9 & $0.38\pm0.06$ & $128\pm41$ & $2.0\pm0.2$ & $1.1\pm0.3$ & \ldots & point/point \\
SMM 2(C)\tablefootmark{b} \ldots & 18 07 35.6 & -18 43 22 & $0.48\pm0.06$ & $1.24\pm0.13$ & 23.0  & $0.37\pm0.06$ & $157\pm50$ & $2.4\pm0.3$ & $1.4\pm0.5$ & \ldots & dark/dark \\
SMM 3(B)\tablefootmark{b} \ldots & 18 07 35.8 & -18 42 42 & $0.75\pm0.09$ & $2.28\pm0.23$ & 29.6 & $0.47\pm0.07$ & $184\pm59$ & $2.4\pm0.3$ & $0.8\pm0.3$ & SDC G11.360+0.800\tablefootmark{11} ($3\farcs1$) & point/point \\
SMM 4 \ldots & 18 07 36.1 & -18 44 26 & $0.60\pm0.07$ & $1.25\pm0.13$ & 23.5 & $0.38\pm0.06$ & $101\pm32$ & $1.9\pm0.2$ & $0.8\pm0.3$ & SSTGLMC G011.3364+00.7861\tablefootmark{3} ($2\farcs2$) & group/group \\
SMM 5(E)\tablefootmark{b} \ldots & 18 07 36.7 & -18 41 14 & $0.62\pm0.07$ & $2.37\pm0.24$ & 32.5 & $0.52\pm0.08$ & $192\pm61$ & $2.0\pm0.2$ & $0.6\pm0.2$ & SDC G11.380+0.809\tablefootmark{11} ($10\farcs1$) & point/point \\ 
SMM 6(F)\tablefootmark{b} \ldots & 18 07 39.0 & -18 42 10 & $0.20\pm0.04$ & $0.15\pm0.04$ & 11.1 & $0.18\pm0.03$ & $25\pm9$ & $1.0\pm0.2$ & $2.0\pm0.7$ & SDC G11.374+0.792\tablefootmark{11} ($9\farcs1$) & dark/dark \\ 
SMM 7(G)\tablefootmark{b} \ldots & 18 07 40.4 & -18 43 18 & $0.36\pm0.05$ & $0.72\pm0.08$ & 21.2 & $0.34\pm0.05$ & $58\pm19$ & $1.1\pm0.2$ & $0.7\pm0.2$ & SDC G11.361+0.777\tablefootmark{11} ($12\farcs8$) & point/point \\ 

{\bf G13.22-0.06} & \\
SMM 1 \ldots & 18 13 47.0 & -17 22 09 & $0.49\pm0.09$ & $0.62\pm0.10$ & 15.2 & $0.32\pm0.02$ & $85\pm18$ & $1.5\pm0.3$ & $1.2\pm0.3$ & BGPS G13.245+0.158\tablefootmark{2} ($5\farcs0$) & point/point \\ 
SMM 2 \ldots & 18 13 53.4 & -17 31 10 & $0.44\pm0.09$ & $0.54\pm0.10$ & 14.8 & $0.30\pm0.02$ & $72\pm17$ & $1.4\pm0.3$ & $1.2\pm0.3$ & JCMTSF J181353.2-173119\tablefootmark{1} ($14\farcs7$); & diffuse/weak \\ 
& & & & & & & & & & SDC G13.125+0.072\tablefootmark{11} ($20\farcs5$) & \\
SMM 3 \ldots & 18 13 54.6 & -17 20 42 & $0.44\pm0.09$ & $1.09\pm0.14$ & 21.4 & $0.45\pm0.03$ & $150\pm28$ & $1.4\pm0.3$ & $0.8\pm0.1$ & \ldots & point/point \\ 
SMM 4 \ldots & 18 13 55.9 & -17 28 34 & $1.21\pm0.15$ & $9.12\pm0.92$ & 45.2 & $0.93\pm0.07$ & $1\,310\pm222$ & $3.8\pm0.5$ & $0.1\pm0.1$ & JCMTSF J181356.1-172837\tablefootmark{1} ($4\farcs3$); & point/point \\ 
& & & & & & & & & & N10-4\tablefootmark{12} ($5\farcs1$); & \\
& & & & & & & & & & N10-3\tablefootmark{12} ($7\farcs8$) & \\
SMM 5 \ldots & 18 14 00.7 & -17 28 38 & $5.61\pm0.57$ & $25.92\pm2.59$ & 52.7 & $1.08\pm0.08$ &$3\,335\pm704$ & $17.0\pm1.7$ & $1.2\pm0.3$ & SDC G13.177+0.061\tablefootmark{11} ($2\farcs1$); & point/extended \\ 
 & & & & & & & & & & outflow\tablefootmark{13} ($4\farcs7$); & \\
 & & & & & & & & & & SCAMPS G13.18+0.06\tablefootmark{14} ($5\farcs0$); & \\
 & & & & & & & & & & JCMTSF J181400.7-172843\tablefootmark{1} ($5\farcs5$); & \\
 & & & & & & & & & & N10-7\tablefootmark{12}  ($5\farcs6$); & \\
 & & & & & & & & & & G013.177+00.059SMM\tablefootmark{15} ($9\farcs1$); & \\
 & & & & & & & & & & BGPS G13.179+0.060\tablefootmark{2} ($16\farcs5$); & \\
 & & & & & & & & & & SSTGLMC G013.1818+00.0610\tablefootmark{3}/N10-9\tablefootmark{12} ($17\farcs4$) & \\
SMM 6 \ldots & 18 14 08.8 & -17 28 57 & $0.83\pm0.12$ & $4.89\pm0.50$ & 36.2 & $0.74\pm0.05$ & $652\pm114$ & $2.5\pm0.4$ & $0.7\pm0.1$ & BGPS G13.191+0.034\tablefootmark{2} ($4\farcs8$); & extended/extended \\ 
& & & & & & & & & & JCMTSF J181408.7-172907\tablefootmark{1} ($9\farcs6$); & \\
& & & & & & & & & & G013.19+00.04 b\tablefootmark{15} ($10\farcs4$) & \\
SMM 7 \ldots & 18 14 09.4 & -17 27 21 & $3.02\pm0.31$ & $48.22\pm4.82$ & 75.2 & $1.55\pm0.11$ & $6\,433\pm1\,114$ & $9.5\pm1.0$ & $0.8\pm0.1$ & 2MASS J18140960-1727219\tablefootmark{16} ($3\farcs6$); & extended/extended \\ 
 & & & & & & & & & & JCMTSF J181409.5-172725\tablefootmark{1} ($5\farcs0$) & \\
SMM 8 \ldots & 18 14 09.9 & -17 19 53 & $0.53\pm0.10$ & $0.41\pm0.09$ & 11.9 & $0.25\pm0.02$ & $42\pm10$ & $1.8\pm0.3$ & $1.2\pm0.3$ & IRAS 18112-1720 ($5\farcs1$); & extended/extended \\ 
& & & & & & & & & & GPSR5 13.322+0.096\tablefootmark{18} ($8\farcs2$) & \\
SMM 9 \ldots & 18 14 10.5 & -17 29 50 & $0.78\pm0.11$ & $2.12\pm0.23$ & 26.1 & $0.54\pm0.04$ & $295\pm52$ & $2.5\pm0.3$ & $0.9\pm0.2$ & SDC G13.177+0.017\tablefootmark{11} ($3\farcs0$); & extended/extended \\ 
 & & & & & & & & & & JCMTSF J181410.4-172955\tablefootmark{1} ($15\farcs4$) & \\
SMM 10 \ldots & 18 14 12.2 & -17 25 14 & $0.54\pm0.10$ & $1.82\pm0.20$ & 25.6 & $0.53\pm0.04$ & $392\pm70$ & $2.7\pm0.5$ & $1.2\pm0.2$ & JCMTSF J181412.5-172525 \tablefootmark{1} ($12\farcs2$); & dark/dark \\ 
& & & & & & & & & & BGPS G13.253+0.044\tablefootmark{2} ($20\farcs2$); & \\
SMM 11 \ldots & 18 14 14.4 & -17 26 30 & $0.49\pm0.09$ & $4.08\pm0.42$ & 39.3 & $0.81\pm0.06$ & $544\pm95$ & $1.5\pm0.3$ & $0.5\pm0.1$ & \ldots & diffuse/diffuse \\   
SMM 12 \ldots & 18 14 17.7 & -17 28 58 & $0.45\pm0.09$ & $1.99\pm0.21$ & 29.0 & $0.60\pm0.04$ & $265\pm47$ & $1.4\pm0.3$ & $0.6\pm0.1$ & Unknown object\tablefootmark{17} ($4\farcs4$); & dark/extended \\ 
& & & & & & & & & & JCMTSF J181417.9-172908\tablefootmark{1} ($10\farcs6$) & \\
SMM 13 \ldots & 18 14 24.5 & -17 17 30 & $1.41\pm0.16$ & $4.01\pm0.41$ & 29.0 & $0.27\pm0.10$ & $65\pm47$ & $2.0\pm0.2$ & $1.5\pm1.1$ & GPSR5 13.385+0.066\tablefootmark{18} ($4\farcs8$); & extended/extended \\ 
& & & & & & & & & & 013.385+0.069\tablefootmark{19} ($5\farcs2$); & \\
& & & & & & & & & & 013.386+0.069\tablefootmark{19} ($7\farcs6$); & \\
& & & & & & & & & & BGPS G13.387+0.066\tablefootmark{2} ($7\farcs8$); & \\
& & & & & & & & & & 013.386+0.065\tablefootmark{19} ($7\farcs9$); & \\
& & & & & & & & & & IRAS 18114-1718 ($17\farcs0$) & \\
SMM 14 \ldots & 18 14 25.6 & -17 22 46 & $0.51\pm0.09$ & $1.50\pm0.17$ & 23.0 & $0.50\pm0.03$ & $221\pm39$ & $1.6\pm0.3$ & $0.8\pm0.1$ & SDC G13.310+0.021\tablefootmark{11} ($2\farcs2$); & dark/point \\ 
SMM 15 \ldots & 18 14 26.7 & -17 34 58 & $0.91\pm0.12$ & $2.58\pm0.27$ & 28.2 & $0.49\pm0.05$ & $243\pm60$ & $2.9\pm0.4$ & $0.9\pm0.2$ & SSTGLMC G013.1343-00.0794\tablefootmark{3} ($2\farcs5$); & point/point  \\ 
SMM 16 \ldots & 18 14 27.8 & -17 22 38 & $0.54\pm0.10$ & $2.57\pm0.27$ & 29.3 & $0.63\pm0.04$ & $379\pm65$ & $1.7\pm0.3$ & $0.7\pm0.1$ & \ldots & point/point \\ 
SMM 17 \ldots & 18 14 28.3 & -17 36 14 & $1.23\pm0.15$ & $5.16\pm0.52$ & 33.6 & $0.58\pm0.07$ & $485\pm120$ & $3.9\pm0.5$ & $1.1\pm0.3$ & SSTGLMC G013.1182-00.0966\tablefootmark{3} ($3\farcs7$); & point/point \\ 
 & & & & & & & & & & BGPS G13.121-0.094\tablefootmark{2} ($19\farcs0$); & \\
 & & & & & & & & & & SDC G13.121-0.091\tablefootmark{11} ($19\farcs9$) & \\
SMM 18(A)\tablefootmark{b} \ldots & 18 14 28.6 & -17 33 26 & $1.00\pm0.13$ & $19.29\pm1.93$ & 49.2 & $1.07\pm0.07$ & $2\,860\pm479$ & $3.1\pm0.4$ & $1.1\pm0.2$ & SDC G13.158-0.073\tablefootmark{11} ($5\farcs6$) & extended/extended \\ 
SMM 19 \ldots & 18 14 28.9 & -17 21 30 & $0.40\pm0.09$ & $1.08\pm0.13$ & 21.5 & $0.46\pm0.03$ & $249\pm45$ & $2.0\pm0.4$ & $1.2\pm0.2$ & \ldots & dark/dark \\ 
SMM 20 \ldots & 18 14 31.4 & -17 36 06 & $0.48\pm0.09$ & $3.01\pm0.31$ & 33.4 & $0.58\pm0.06$ & $443\pm109$ & $2.4\pm0.4$ & $1.0\pm0.3$ & \ldots & dark/dark \\ 
SMM 21 \ldots & 18 14 34.5 & -17 34 34 & $0.48\pm0.09$ & $2.09\pm0.22$ & 28.2 & $0.61\pm0.04$ & $310\pm53$ & $1.5\pm0.3$ & $0.6\pm0.1$ & \ldots & point/point \\ 
SMM 22(C)\tablefootmark{b} \ldots & 18 14 35.4 & -17 30 42 & $0.55\pm0.10$ & $1.87\pm0.20$ & 24.5 & $0.42\pm0.05$ & $275\pm68$ & $2.7\pm0.5$ & $1.7\pm0.4$ & BGPS G13.213-0.076\tablefootmark{2} ($5\farcs9$) & dark/dark \\ 
SMM 23(E)\tablefootmark{b} \ldots & 18 14 36.8 & -17 29 22 & $0.67\pm0.10$ & $1.90\pm0.21$ & 25.1 & $0.43\pm0.05$ & $179\pm45$ & $2.1\pm0.3$ & $1.0\pm0.3$ & \ldots & point/point \\ 
SMM 24 \ldots & 18 14 37.0 & -17 38 50 & $1.61\pm0.18$ & $8.35\pm0.70$ & 43.6 & $2.62\pm0.08$ & $9\,496\pm1\,005$ & $5.1\pm0.6$ & $0.2\pm0.1$ & SSTGLMC G013.0970-00.1447\tablefootmark{20} ($7\farcs7$); & point/point \\ 
 & & & & & & & & & & BGPS G13.097-0.146\tablefootmark{2} ($18\farcs3$) & \\
SMM 25 \ldots & 18 14 39.0 & -17 33 02 & $1.69\pm0.19$ & $12.70\pm1.27$ & 53.1 &$1.15\pm0.08$ & $1\,883\pm315$ & $5.3\pm0.6$ & $0.6\pm0.1$ & BGPS G13.186-0.108\tablefootmark{2} ($9\farcs2$); & extended/point \\ 
 & & & & & & & & & & SDC G13.190-0.105\tablefootmark{11} ($19\farcs8$) & \\
SMM 26 \ldots & 18 14 39.6 & -17 41 10 & $0.64\pm0.10$ & $2.51\pm0.26$ & 27.8 & $1.67\pm0.05$ & $4\,465\pm545$ & $3.2\pm0.5$ & $0.4\pm0.1$ & SDC G13.067-0.172\tablefootmark{11} ($4\farcs4$) & dark/dark \\ 
SMM 27(F)\tablefootmark{b} \ldots & 18 14 40.7 & -17 29 22 & $3.16\pm0.33$ & $13.25\pm1.33$ & 44.3 & $0.76\pm0.09$ & $1\,246\pm307$ & $10.0\pm1.0$ & $1.3\pm0.3$ & SDC G13.246-0.081\tablefootmark{11} ($18\farcs1$); & group/point \\ 
 & & & & & & & & & &  BGPS G13.245-0.084\tablefootmark{2} ($19\farcs9$) & \\
SMM 28 \ldots & 18 14 41.2 & -17 23 18 & $1.53\pm0.17$ & $8.21\pm0.82$ & 38.8 & $0.84\pm0.06$ & $1\,212\pm203$ & $4.8\pm0.5$ & $0.9\pm0.2$ & BGPS G13.333-0.038\tablefootmark{2} ($16\farcs6$) & group/group \\ 
SMM 29 \ldots & 18 14 42.1 & -17 37 06 & $2.42\pm0.25$ & $11.27\pm1.13$ & 46.3 &$2.77\pm0.09$ & $7\,269\pm868$ & $8.3\pm0.9$ & $0.2\pm0.1$ & BGPS G13.133-0.150\tablefootmark{2} ($2\farcs9$); & extended/point \\ 
& & & & & & & & & & IRAS 18117-1738\tablefootmark{21} ($11\farcs2$) & \\
SMM 30 \ldots & 18 14 42.1 & -17 21 45 & $0.43\pm0.09$ & $0.66\pm0.10$ & 16.3 & $0.35\pm0.02$ & $97\pm20$ & $1.4\pm0.3$ & $1.0\pm0.2$ & SDC G13.357-0.029\tablefootmark{11} ($6\farcs3$); & point/point \\ 
& & & & & & & & & & BGPS G13.359-0.030\tablefootmark{2} ($13\farcs1$) & \\
SMM 31 \ldots & 18 14 46.0 & -17 23 53 & $0.58\pm0.10$ & $1.87\pm0.20$ & 25.8 & $0.56\pm0.04$ & $432\pm74$ & $2.1\pm0.4$ & $1.1\pm0.2$ & \ldots & dark/dark \\  
SMM 32\tablefootmark{h} \ldots & 18 14 49.9 & -17 32 45 & $4.28\pm0.44$ & $31.75\pm3.18$ & 62.3 & $1.33\pm0.09$ & $2\,620\pm443$ & $7.7\pm0.8$ & $0.5\pm0.1$ & G013.210-0.144\tablefootmark{22} ($1\farcs4$); & extended/extended \\ 
 & & & & & & & & & & MSX6C G013.2097-00.1436\tablefootmark{23} ($1\farcs6$); & \\
 & & & & & & & & & & PMN J1814-1732\tablefootmark{24} ($1\farcs8$); & \\
 & & & & & & & & & & 181154.9-173348\tablefootmark{25} ($5\farcs4$); & \\
 & & & & & & & & & & 013.210-0.145\tablefootmark{26} ($6\farcs7$); & \\
 & & & & & & & & & & BGPS G13.211-0.142\tablefootmark{2} ($11\farcs0$) & \\
SMM 33 \ldots & 18 14 52.4 & -17 41 14 & $0.55\pm0.10$ & $0.98\pm0.13$ & 18.2 & $1.09\pm0.04$ & $1\,743\pm257$ & $2.7\pm0.5$ & $0.6\pm0.1$ & BGPS G13.091-0.218\tablefootmark{2} ($12\farcs1$); & dark/dark \\ 
SMM 34 \ldots & 18 14 53.6 & -17 40 06 & $0.84\pm0.12$ & $1.75\pm0.19$ & 22.7 & $1.36\pm0.04$ & $1\,990\pm251$ & $2.6\pm0.4$ & $0.4\pm0.1$ & BGPS G13.109-0.216\tablefootmark{2} ($8\farcs7$) & point/point \\ 
\hline
\end{longtable}
\tablefoot{\tablefoottext{a}{Remarks on the associated sources. Offset [$\arcsec$] from the 870-$\mu$m peak position is indicated in parenthesis.}\tablefoottext{b}{The clump is associated with the marked C$^{17}$O$(2-1)$-observation target position.}\tablefoottext{c}{The 8-$\mu$m point source is associated with SSTGLMC G001.8134-00.1136, which has the $[3.6]-[4.5]$, $[4.5]-[5.8]$, and $[4.5]-[8.0]$ colours of 0.098, 0.245, and 0.031, respectively, and is likely to be a foreground star.}\tablefoottext{d}{The 8-$\mu$m point source corresponds to SSTGLMC G001.9301-00.1234 with the colours $[3.6]-[4.5]=0.081$, $[4.5]-[5.8]=0.158$, and $[4.5]-[8.0]=0.167$, and is likely a foreground star.}\tablefoottext{e}{The 8-$\mu$m point source corresponds to SSTGLMC G001.9716-00.1375 with the colours $[3.6]-[4.5]=-0.033$, $[4.5]-[5.8]=0.294$, and $[4.5]-[8.0]=0.141$, and is likely a foreground star.}\tablefoottext{f}{The 8-$\mu$m point source is associated with SSTGLMC G002.0927-00.0039, which has the $[3.6]-[4.5]$, $[4.5]-[5.8]$, and $[4.5]-[8.0]$ colours of 0.802, 0.775, and 0.290, respectively, and is likely to be 
a foreground star.}\tablefoottext{g}{The 8-$\mu$m point source corresponds to SSTGLMC G002.2440+00.0447. Its $[3.6]-[4.5]$, $[4.5]-[5.8]$, and $[4.5]-[8.0]$ colours are, respectively, 0.523, 0.436, and 0.838, and it is likely a foreground star.}\tablefoottext{h}{The clump can be resolved by eye into two ``subclumps'', but they are dealt as a one source by {\tt clumpfind}.}\tablefoottext{1}{SCUBA clump (\cite{difrancesco2008}).}\tablefoottext{2}{Bolocam clump (\cite{rosolowsky2010}).}\tablefoottext{3}{YSO candidate (\cite{robitaille2008}).}\tablefoottext{4}{UC H{\scriptsize II} region (\cite{becker1994}).}\tablefoottext{5}{OH maser (\cite{argon2000}).}\tablefoottext{6}{UC H{\scriptsize II} region (\cite{forster2000}).}\tablefoottext{7}{MacLeod et al. (1998) search for 6.7-GHz CH$_3$OH maser emission towards this sources, but did not detect it.}\tablefoottext{8}{6.7-GHz CH$_3$OH maser (\cite{caswell1995}).}\tablefoottext{9}{IR source (\cite{egan2001}).}\tablefoottext{10}{OH and H$_2$O maser source (see \cite{caswell1983}).}\tablefoottext{11}{IRDC from the Peretto \& Fuller (2009) catalogue; available at {\tt www.irdarkclouds.org/}.}\tablefoottext{12}{YSO candidate (\cite{watson2008}).}\tablefoottext{13}{CO outflow associated with the bubble N10 (\cite{beaumont2010}).}\tablefoottext{14}{A submm clump studied by Pillai et al. (2007).}\tablefoottext{15}{SCUBA clump (\cite{thompson2006}).}\tablefoottext{16}{Possible AGB star (\cite{cutri2003}).}\tablefoottext{17}{Source nro.~28 in Russeil (2003); its nature is unknown.}\tablefoottext{18}{Radio source (\cite{becker1994}).}\tablefoottext{19}{1.5-GHz radio source (\cite{garwood1988}).}\tablefoottext{20}{Possible AGB star (\cite{robitaille2008}).}\tablefoottext{21}{The distance ambiguity of the IRAS source was resolved by Sewilo et al. (2004).}\tablefoottext{22}{H{\scriptsize II} region (\cite{white2005}; \cite{urquhart2009}).}\tablefoottext{23}{YSO (\cite{egan2003}).}\tablefoottext{24}{1.4-GHz radio source (\cite{condon1998}).}\tablefoottext{25}{H{\scriptsize II} region (\cite{chini1987}).}\tablefoottext{26}{H{\scriptsize II} region (\cite{wink1982}).}   }
}
\end{landscape}
}


\begin{thebibliography}{}

\bibitem[Aguirre et al. 2011]{aguirre2011} Aguirre, J.~E., Ginsburg, A.~G., 
Dunham, M.~K., et al. 2011, \apjs, 192, 4 

\bibitem[Argon et al. 2000]{argon2000} Argon, A.~L., Reid, 
M.~J., and Menten, K.~M. 2000, \apjs, 129, 159 

\bibitem[Bally et al. 2010]{bally2010} Bally, J., Aguirre, J., Battersby, C., 
et al. 2010, \apj, 721, 137 

\bibitem[Banerjee et al. 2006]{banerjee2006} Banerjee, R., Pudritz, 
R.~E., and Anderson, D.~W. 2006, \mnras, 373, 1091 

\bibitem[Battersby et al. 2010]{battersby2010} Battersby, C., Bally, 
J., Jackson, J.~M., et al. 2010, \apj, 721, 222 

\bibitem[Beaumont \& Williams 2010]{beaumont2010} Beaumont, C.~N., and 
Williams, J.~P. 2010, \apj, 709, 791 

\bibitem[Becker et al. 1994]{becker1994} Becker, R.~H., White, 
R.~L., Helfand, D.~J., and Zoonematkermani, S. 1994, \apjs, 91, 347 

\bibitem[Belitsky et al. 2007]{belitsky2007} Belitsky, V., Lapkin, I., 
Vassilev, V., et al. 2007, in \textit{Proceedings of joint 32nd International 
Conference on Infrared Millimeter Waves and 15th International Conference on 
Terahertz Electronics}, September 3-7, 2007, City Hall, Cardiff, Wales, UK, 
pp.~326-328

\bibitem[Beltr{\'a}n et al. 2006]{beltran2006} Beltr{\'a}n, M.~T., Brand, J., 
Cesaroni, R., et al. 2006, \aap, 447, 221 

\bibitem[Benjamin et al. 2003]{benjamin2003} Benjamin, R.~A., Churchwell, E., 
Babler, B.~L., et al. 2003, \pasp, 115, 953 

\bibitem[Bergin \& Tafalla 2007]{bergin2007} Bergin, E.~A., and Tafalla, M. 
2007, \araa, 45, 339 

\bibitem[Bertoldi \& McKee 1992]{bertoldi1992} Bertoldi, F., and McKee, C. F. 
1992, \apj, 395, 140 

\bibitem[Beuther \& Steinacker 2007]{beuther2007} Beuther, H., \& Steinacker, 
J. 2007, \apjl, 656, L85 

\bibitem[Beuther et al. 2002]{beuther2002} Beuther, H., Schilke, 
P., Menten, K.~M., et al. 2002, \apj, 566, 945 

\bibitem[Beuther et al. 2011]{beuther2011} Beuther, H., Linz, H., Henning, T., 
et al. 2011, \aap, 531, A26 

\bibitem[Bohlin et al. 1978]{bohlin1978} Bohlin, R.~C., Savage, B.~D., and 
Drake, J.~F. 1978, \apj, 224, 132 

\bibitem[Carey et al. 1998]{carey1998} Carey, S.~J., Clark, F.~O., 
Egan, M.~P., et al. 1998, \apj, 508, 721

\bibitem[Carey et al. 2009]{carey2009} Carey, S.~J., Noriega-Crespo, A., 
Mizuno, D.~R., et al. 2009, \pasp, 121, 76 

\bibitem[Casoli et al. 1986]{casoli1986} Casoli, F., Combes, F., Dupraz, C., 
et al. 1986, \aap, 169, 281 

\bibitem[Caswell et al. 1983]{caswell1983} Caswell, J.~L., 
Batchelor, R.~A., Forster, J.~R., and Wellington, K.~J. 1983, 
Australian Journal of Physics, 36, 401 

\bibitem[Caswell et al. 1995]{caswell1995} Caswell, J.~L., Vaile, 
R.~A., Ellingsen, S.~P., et al. 1995, \mnras, 272, 96 

\bibitem[Chambers et al. 2009]{chambers2009} Chambers, E.~T., 
Jackson, J.~M., Rathborne, J.~M., and Simon, R. 2009, \apjs, 181, 360 

\bibitem[Chandrasekhar \& Fermi 1953]{chandrasekhar1953} Chandrasekhar, S., 
and Fermi, E. 1953, \apj, 118, 116 

\bibitem[Chen et al. 2010]{chen2010} Chen, X., Shen, Z.-Q., Li, J.-J., 
et al. 2010, \apj, 710, 150 

\bibitem[Chen et al. 2011]{chen2011} Chen, H.-R., Liu, S.-Y., Su, Y.-N., and 
Wang, M.-Y. 2011, \apj, 743, 196 

\bibitem[Chini et al. 1987]{chini1987} Chini, R., Kruegel, E., and 
Wargau, W. 1987, \aap, 181, 378 

\bibitem[Churchwell et al. 2006]{churchwell2006} Churchwell, E., 
Povich, M.~S., Allen, D., et al. 2006, \apj, 649, 759 

\bibitem[Churchwell et al. 2009]{churchwell2009} Churchwell, E., 
Babler, B.~L., Meade, M.~R., et al. 2009, \pasp, 121, 213 

\bibitem[Clark et al. 2007]{clark2007} Clark, P.~C., Klessen, R.~S., and 
Bonnell, I.~A. 2007, \mnras, 379, 57 

\bibitem[Condon et al. 1998]{condon1998} Condon, J.~J., Cotton, 
W.~D., Greisen, E.~W., et al. 1998, \aj, 115, 1693 

\bibitem[Cragg et al. 1992]{cragg1992} Cragg, D.~M., Johns, 
K.~P., Godfrey, P.~D., and Brown, R.~D. 1992, \mnras, 259, 203 

\bibitem[Cutri et al. 2003]{cutri2003} Cutri, R.~M., Skrutskie, 
M.~F., van Dyk, S., et al. 2003, VizieR Online Data Catalog, 2246, 0 

\bibitem[Cyganowski et al. 2008]{cyganowski2008} Cyganowski, C.~J., 
Whitney, B.~A., Holden, E., et al. 2008, \aj, 136, 2391 

\bibitem[Dame \& Thaddeus 2008]{dame2008} Dame, T.~M., and Thaddeus, P. 
2008, \apjl, 683, L143 

\bibitem[De Buizer \& Vacca 2010]{debuizer2010} De Buizer, J.~M., and 
Vacca, W.~D. 2010, \aj, 140, 196 

\bibitem[Deharveng et al. 2010]{deharveng2010} Deharveng, L., Schuller, F., 
Anderson, L.~D., et al. 2010, \aap, 523, A6 

\bibitem[Devine et al. 2011]{devine2011} Devine, K.~E., Chandler, 
C.~J., Brogan, C., et al. 2011, \apj, 733, 44 

\bibitem[Di Francesco et al. 2008]{difrancesco2008} Di Francesco, J., 
Johnstone, D., Kirk, H., et al. 2008, \apjs, 175, 277 

\bibitem[Draine 2003]{draine2003} Draine, B.~T. 2003, \araa, 41, 241 

\bibitem[Egan et al. 1998]{egan1998} Egan, M. P., Shipman, R. F., Price, S. D.,
et al. 1998, \apj, 494, L199

\bibitem[Egan et al. 2001]{egan2001} Egan, M.~P., Price, S.~D., 
Moshir, M.~M., et al.\ 2001, VizieR Online Data Catalog, 5107, 0 

\bibitem[Egan et al. 2003]{egan2003} Egan, M.~P., Price, S.~D., 
Kraemer, K.~E., et al. 2003, VizieR Online Data Catalog, 5114, 0 

\bibitem[Elmegreen et al. 2000]{elmegreen2000} Elmegreen, B.~G., Efremov, Y., 
Pudritz, R.~E., and Zinnecker, H. 2000, in \textit{Protostars and Planets IV}, 
eds. V. Mannings, A.~P. Boss, and S.~S. Russell (Tucson: Univ. of Arizona 
Press), p.~179 

\bibitem[Fazio et al. 2004]{fazio2004} Fazio, G.~G., Hora, 
J.~L., Allen, L.~E., et al. 2004, \apjs, 154, 10 

\bibitem[Fiege \& Pudritz 2000a]{fiege2000a} Fiege, J.~D., and Pudritz, R.~E. 
2000a, \mnras, 311, 85 

\bibitem[Fiege \& Pudritz 2000b]{fiege2000b} Fiege, J.~D., and Pudritz, R.~E. 
2000b, \mnras, 311, 105 

\bibitem[Forster \& Caswell 1989]{forster1989} Forster, J.~R., and 
Caswell, J.~L. 1989, \aap, 213, 339 

\bibitem[Forster \& Caswell 2000]{forster2000} Forster, J.~R., and 
Caswell, J.~L. 2000, \apj, 530, 371 

\bibitem[Galli et al. 2002]{galli2002} Galli, D., Walmsley, M., and 
Gon{\c c}alves, J. 2002, \aap, 394, 275 

\bibitem[Garwood et al. 1988]{garwood1988} Garwood, R.~W., Perley, 
R.~A., Dickey, J.~M., and Murray, M.~A. 1988, \aj, 96, 1655 

\bibitem[Green et al. 2011]{green2011} Green, J.~A., Caswell, 
J.~L., McClure-Griffiths, N.~M., et al. 2011, \apj, 733, 27 

\bibitem[Gutermuth et al. 2008]{gutermuth2008} Gutermuth, R.~A., 
Myers, P.~C., Megeath, S.~T., et al. 2008, \apj, 674, 336 

\bibitem[G{\"u}sten et al. 2006]{gusten2006} G{\"u}sten, R., Nyman, L.~{\AA}., 
Schilke, P., et al. 2006, \aap, 454, L13 

\bibitem[Hatchell et al. 2007]{hatchell2007} Hatchell, J., Fuller, G.~A., 
Richer, J.~S., et al. 2007, \aap, 468, 1009 

\bibitem[Hennebelle \& Chabrier 2008]{hennebelle2008} Hennebelle, P., and 
Chabrier, G. 2008, \apj, 684, 395 

\bibitem[Hennemann et al. 2009]{hennemann2009} Hennemann, M., Birkmann, S.~M., 
Krause, O., et al. 2009, \apj, 693, 1379 

\bibitem[Henning et al. 1990]{henning1990} Henning, T., Pfau, W., and 
Altenhoff, W.~J. 1990, \aap, 227, 542 

\bibitem[Hernandez \& Tan 2011]{hernandezandtan2011} Hernandez, A.~K., and 
Tan, J.~C. 2011, \apj, 730, 44 

\bibitem[Hernandez et al. 2011]{hernandez2011} Hernandez, A.~K., Tan, J.~C., 
Caselli, P., et al. 2011, \apj, 738, 11 

\bibitem[Heyer \& Brunt 2004]{heyer2004} Heyer, M.~H., \& Brunt, C.~M. 
2004, \apjl, 615, L45 

\bibitem[Jackson et al. 2010]{jackson2010} Jackson, J.~M., Finn, 
S.~C., Chambers, E.~T., et al. 2010, \apjl, 719, L185 

\bibitem[Jim{\'e}nez-Serra et al. 2010]{jimenez2010} Jim{\'e}nez-Serra, I., 
Caselli, P., Tan, J.~C., et al. 2010, \mnras, 406, 187 

 \bibitem[Kauffmann \& Pillai 2010]{kauffmann2010} Kauffmann, J., \& Pillai, 
T. 2010, \apjl, 723, L7 

\bibitem[Klein et al. 2006]{klein2006} Klein, B., Philipp, S.~D., Kr{\"a}mer, 
I., et al. 2006, \aap, 454, L29 

\bibitem[Kov{\'a}cs 2008]{kovacs2008} Kov{\'a}cs, A. 2008, \procspie, 7020, 45 

\bibitem[Kramer et al. 1998]{kramer1998} Kramer, C., Stutzki, J., Rohrig, R., 
and Corneliussen, U. 1998, \aap, 329, 249 

\bibitem[Krumholz \& McKee 2008]{krumholz2008} Krumholz, M.~R., and 
McKee, C.~F. 2008, \nat, 451, 1082 

\bibitem[Ladd et al. 1998]{ladd1998} Ladd, E.~F., Fuller, G.~A., and 
Deane, J.~R. 1998, \apj, 495, 871 

\bibitem[Larson 1981]{larson1981} Larson, R.~B. 1981, \mnras, 194, 809 

\bibitem[L{\'o}pez et al. 2011]{lopez2011} L{\'o}pez, C., Bronfman, L., Nyman, 
L.-{\AA}., et al. 2011, \aap, 534, A131 

\bibitem[MacLeod et al. 1998]{macleod1998} MacLeod, G.~C., van der 
Walt, D.~J., North, A., et al. 1998, \aj, 116, 2936 

\bibitem[Marshall et al. 2009]{marshall2009} Marshall, D.~J., Joncas, G., and 
Jones, A.~P. 2009, \apj, 706, 727 

\bibitem[Marston et al. 2004]{marston2004} Marston, A.~P., Reach, 
W.~T., Noriega-Crespo, A., et al. 2004, \apjs, 154, 333 

\bibitem[McKee \& Zweibel 1992]{mckee1992} McKee, C.~F., and Zweibel, E.~G. 
1992, \apj, 399, 551 

\bibitem[Miettinen 2012]{miettinen2012} Miettinen, O. 2012, \aap, 540, A104 
(M12)

\bibitem[Miettinen \& Harju 2010]{miettinenharju2010} Miettinen, O., and 
Harju, J. 2010, \aap, 520, A102 (MH10)

\bibitem[Miettinen et al. 2011]{miettinen2011} Miettinen, O., Hennemann, M., 
and Linz, H. 2011, \aap, 534, A134 

\bibitem[Miettinen et al. 2012]{miettinenetal2012} Miettinen, O., Harju, J., 
Haikala, L.~K., and Juvela, M. 2012, \aap, 538, A137 

\bibitem[Minier et al. 2003]{minier2003} Minier, V., Ellingsen, S.~P., Norris, 
R.~P., and Booth, R.~S. 2003, \aap, 403, 1095 

\bibitem[Motte et al. 2007]{motte2007} Motte, F., Bontemps, S., Schilke, P., 
et al. 2007, \aap, 476, 1243 

\bibitem[Nagasawa 1987]{nagasawa1987} Nagasawa, M. 1987, Prog. Theor. Phys., 
77, 635 

\bibitem[Nakamura et al. 1993]{nakamura1993} Nakamura, F., Hanawa, 
T., and Nakano, T. 1993, \pasj, 45, 551 

\bibitem[Ossenkopf \& Henning 1994]{ossenkopf1994} Ossenkopf, V., and 
Henning, T. 1994, \aap, 291, 943 

\bibitem[Pandian et al. 2008]{pandian2008} Pandian, J.~D., Momjian, E., and 
Goldsmith, P.~F. 2008, \aap, 486, 191 

\bibitem[Parsons et al. 2009]{parsons2009} Parsons, H., Thompson, M.~A., and 
Chrysostomou, A. 2009, \mnras, 399, 1506 

\bibitem[P\'erault et al. 1996]{perault1996} P\'erault, M., Omont, A., 
Simon, G., et al. 1996, \aap, 315, L165 

\bibitem[Peretto \& Fuller 2009]{peretto2009} Peretto, N., and Fuller, G.~A. 
2009, \aap, 505, 405 

\bibitem[Peretto \& Fuller 2010]{peretto2010} Peretto, N., and Fuller, G.~A. 
2010, \apj, 723, 555 

\bibitem[Pillai et al. 2006]{pillai2006} Pillai, T., Wyrowski, F., Carey, 
S.~J. et al. 2006, \aap, 450, 569 

\bibitem[Pillai et al. 2007]{pillai2007} Pillai, T., Wyrowski, F., 
Hatchell, J., et al. 2007, \aap, 467, 207 

\bibitem[Ragan et al. 2009]{ragan2009} Ragan, S.~E., Bergin, E.~A., and 
Gutermuth, R.~A. 2009, \apj, 698, 324 

\bibitem[Ragan et al. 2011]{ragan2011} Ragan, S.~E., Bergin, 
E.~A., and Wilner, D. 2011, \apj, 736, 163 

\bibitem[Rathborne et al. 2006]{rathborne2006} Rathborne, J.~M., 
Jackson, J.~M., and Simon, R. 2006, \apj, 641, 389 

\bibitem[Rathborne et al. 2010]{rathborne2010} Rathborne, J.~M., Jackson, 
J.~M., Chambers, E.~T., et al. 2010, \apj, 715, 310 

\bibitem[Rawlings et al. 1992]{rawlings1992} Rawlings, J.~M.~C., 
Hartquist, T.~W., Menten, K.~M., and Williams, D.~A. 1992, \mnras, 255, 471 

\bibitem[Reid \& Wilson 2005]{reid2005} Reid, M.~A., \& Wilson, C.~D. 2005, 
\apj, 625, 891 

\bibitem[Reid et al. 2009]{reid2009} Reid, M.~J., Menten, K.~M., Zheng, X.~W., 
et al. 2009, ApJ, 700, 137

\bibitem[Rieke et al. 2004]{rieke2004} Rieke, G.~H., Young, 
E.~T., Engelbracht, C.~W., et al. 2004, \apjs, 154, 25 

\bibitem[Robitaille et al. 2008]{robitaille2008} Robitaille, T.~P., 
Meade, M.~R., Babler, B.~L., et al. 2008, \aj, 136, 2413 

\bibitem[Roman-Duval et al. 2009]{romanduval2009} Roman-Duval, J., 
Jackson, J.~M., Heyer, M., et al. 2009, \apj, 699, 1153 

\bibitem[Rosolowsky et al. 2010]{rosolowsky2010} Rosolowsky, E., 
Dunham, M.~K., Ginsburg, A., et al. 2010, \apjs, 188, 123 

\bibitem[Russeil 2003]{russeil2003} Russeil, D. 2003, \aap, 397, 133 

\bibitem[Russeil et al. 2010]{russeil2010} Russeil, D., Zavagno, A., Motte, 
F., et al. 2010, \aap, 515, A55 

\bibitem[Sakai et al. 2008]{sakai2008} Sakai, T., Sakai, N., 
Kamegai, K., et al. 2008, \apj, 678, 1049 

\bibitem[Salpeter 1955]{salpeter1955} Salpeter, E.~E. 1955, \apj, 121, 161 

\bibitem[Schlingman et al. 2011]{schlingman2011} Schlingman, W.~M., 
Shirley, Y.~L., Schenk, D.~E., et al. 2011, \apjs, 195, 14 

\bibitem[Schuller et al. 2009]{schuller2009} Schuller, F., Menten, K.~M., 
Contreras, Y., et al. 2009, \aap, 504, 415 

\bibitem[Sch{\"o}ier et al. 2005]{schoier2005} Sch{\"o}ier, F.~L., van der 
Tak, F.~F.~S., van Dishoeck, E.~F., and Black, J.~H. 2005, \aap, 432, 369 

\bibitem[Sewilo et al. 2004]{sewilo2004} Sewilo, M., Watson, C., 
Araya, E., et al. 2004, \apjs, 154, 553 

\bibitem[Simon et al. 2001]{simon2001} Simon, R., Jackson, 
J.~M., Clemens, D.~P., et al. 2001, \apj, 551, 747 

\bibitem[Simon et al. 2006]{simon2006} Simon, R., Rathborne, J.~M., Shah, 
R.~Y., et al. 2006, \apj, 653, 1325 

\bibitem[Siringo et al. 2009]{siringo2009} Siringo, G., Kreysa, E., 
Kov{\'a}cs, A., et al. 2009, \aap, 497, 945 

\bibitem[Smith \& Rosen 2005]{smith2005} Smith, M.~D., and Rosen, A. 2005, 
\mnras, 357, 1370 

\bibitem[Smith et al. 2006]{smith2006} Smith, H.~A., Hora, J.~L., Marengo, M., 
and Pipher, J.~L. 2006, \apj, 645, 1264 

\bibitem[Spitzer 1978]{spitzer1978} Spitzer, L., Jr. 1978, \textit{Physical 
Processes in the Interstellar Medium} (New York: Wiley Interscience) 

\bibitem[Sreenilayam \& Fich 2011]{sreenilayam2011} Sreenilayam, G., and 
Fich, M. 2011, \aj, 142, 4 

\bibitem[Sridharan et al. 2005]{sridharan2005} Sridharan, T.~K., 
Beuther, H., Saito, M., et al. 2005, \apjl, 634, L57 

\bibitem[Stutzki \& G{\"u}sten 1990]{stutzki1990} Stutzki, J., and G{\"u}sten, 
R. 1990, \apj, 356, 513 

\bibitem[Szymczak et al. 2002]{szymczak2002} Szymczak, M., Kus, A.~J., Hrynek, 
G., et al. 2002, \aap, 392, 277 

\bibitem[Tackenberg et al. 2012]{tackenberg2012} Tackenberg, J., 
Beuther, H., Henning, T., et al. 2012, \aap, \textit{in press}, 
{\tt arXiv:1201.4732} 

\bibitem[Tafalla et al. 2004]{tafalla2004} Tafalla, M., Myers, P.~C., 
Caselli, P., and Walmsley, C.~M. 2004, \aap, 416, 191

\bibitem[Takami et al. 2012]{takami2012} Takami, M., Chen, H.-H., Karr, J.~L., 
et al. 2012, \apj, 748, 8 

\bibitem[Teyssier et al. 2002]{teyssier2002} Teyssier, D., Hennebelle, P., and 
P{\'e}rault, M. 2002, \aap, 382, 624 

\bibitem[Thompson et al. 2006]{thompson2006} Thompson, M.~A., Hatchell, J., 
Walsh, A.~J., et al. 2006, \aap, 453, 1003 

\bibitem[Tielens \& Hollenbach 1985]{tielens1985} Tielens, A.~G.~G.~M., \& 
Hollenbach, D. 1985, \apj, 291, 722 

\bibitem[Urquhart et al. 2009]{urquhart2009} Urquhart, J.~S., Hoare, M.~G., 
Purcell, C.~R., et al. 2009, \aap, 501, 539 

\bibitem[Vassilev et al. 2008a]{vassilev2008a} Vassilev, V., Meledin, D., 
Lapkin, I., et al. 2008a, \aap, 490, 1157 

\bibitem[Vassilev et al. 2008b]{vassilev2008b} Vassilev, V., Henke, D., 
Lapkin, I., et al. 2008b, IEEE Microwave and Wireless Components Letters, 
pp.~55-60, Vol.~18, Number~1

\bibitem[Visser et al. 2009]{visser2009} Visser, R., van Dishoeck, E.~F., and 
Black, J.~H. 2009, \aap, 503, 323 

\bibitem[Wang et al. 2011]{wang2011} Wang, K., Zhang, Q., Wu, Y., and 
Zhang, H. 2011, \apj, 735, 64 

\bibitem[Watson et al. 2008]{watson2008} Watson, C., Povich, 
M.~S., Churchwell, E.~B., et al. 2008, \apj, 681, 1341 

\bibitem[Werner et al. 2004]{werner2004} Werner, M.~W., Roellig, T.~L., 
Low, F.~J., et al. 2004, \apjs, 154, 1 

\bibitem[White et al. 2005]{white2005} White, R.~L., Becker, 
R.~H., and Helfand, D.~J. 2005, \aj, 130, 586 

\bibitem[Wilcock et al. 2012]{wilcock2012} Wilcock, L.~A., 
Ward-Thompson, D., Kirk, J.~M., et al. 2012, \mnras, 2638

\bibitem[Williams et al. 1994]{williams1994} Williams, J. P., de Geus, E. J., 
and Blitz, L.1994, \apj, 428, 693

\bibitem[Wink et al. 1982]{wink1982} Wink, J.~E., Altenhoff, W.~J., and 
Mezger, P.~G. 1982, \aap, 108, 227 

\bibitem[Xu et al. 2008]{xu2008} Xu, Y., Li, J.~J., Hachisuka, K., et al. 
2008, \aap, 485, 729 

\bibitem[Ybarra \& Lada 2009]{yabarra2009} Ybarra, J.~E., and Lada, E.~A. 
2009, \apjl, 695, L120 

\bibitem[Zhang et al. 2011]{zhang2011} Zhang, S.~B., Yang, J., 
Xu, Y., et al. 2011, \apjs, 193, 10 

\bibitem[Zinnecker \& Yorke 2007]{zinnecker2007} Zinnecker, H., and 
Yorke, H.~W. 2007, \araa, 45, 481 


\end{thebibliography}
\end{document}